\documentclass[11pt]{article}
\usepackage{mymacros}
\newcommand{\tT}{\tilde{t}}

\title{Symmetry Sectors in Chord Space and Relational Holography in the DSSYK\\
{\LARGE\rm{Lessons from branes, wormholes, and de Sitter space}}
}
\author[a]{Sergio E. Aguilar-Gutierrez}
\affiliation[a]{Qubits and Spacetime Unit, Okinawa Institute of Science and Technology Graduate University,\footnote{\begin{CJK}{UTF8}{min}沖縄科学技術大学院大学.\end{CJK}} 1919-1 Tancha, Onna, Okinawa 904 0495, Japan}
\emailAdd{sergio.ernesto.aguilar@gmail.com}
\abstract{In holography, gauging symmetries of the boundary theory leads to important modifications in the bulk. In this work, we study constraints to gauge symmetry sectors in the chord Hilbert space of the double-scaled SYK (DSSYK) with matter, and we connect them to different proposals of its bulk dual. These sectors include specific symmetries in chord space, corresponding to End-Of-The-World (ETW) branes and Euclidean wormholes in sine dilaton gravity; and relative time-translations in a doubled DSSYK model (resulting from a single DSSYK with an infinitely heavy matter chord) used in de Sitter holography. We define and evaluate partition functions and thermal correlation functions of the ETW brane and Euclidean wormhole systems in the boundary theory. We deduce the holographic dictionary by matching geodesic lengths in the bulk with the spread complexity of the gauged DSSYK. The Euclidean wormholes of fixed size are perturbatively stable, and their baby universe Hilbert space is non-trivial only when matter is added. We conclude studying the constraints in the path integral of the doubled DSSYK. We derive the gauge invariant operator algebra of one of the DSSYKs dressed to the other one and discuss its holographic interpretation.}

\begin{document}

\maketitle
\section{Introduction}\label{sec:intro}
In the recent years, there has been considerable interest in understanding the holographic correspondence beyond asymptotically anti-de Sitter (AdS) spacetimes. There are different approaches to this problem. In particular, lower dimensional models often have significant simplifications that allow studying quantum gravity in non-AdS spacetimes with analytic control, and in certain cases, to identify a precise holographic dual theory. One important example is the double-scaled SYK (DSSYK) model. It has remarkable features that hint its bulk dual might be a cosmological spacetime toy model \cite{Susskind:2021esx,Narovlansky:2023lfz,Blommaert:2023opb}. For instance, it has drawn a lot of interest for possible applications in de Sitter (dS) space holography and more general toy models for quantum cosmology \cite{Susskind:2021esx,Narovlansky:2023lfz,HVtalks,Okuyama:2025hsd,Blommaert:2023opb,Milekhin:2023bjv,Aguilar-Gutierrez:2024nau,Narovlansky:2025tpb,Blommaert:2025rgw,Blommaert:2024whf}.\footnote{See also interesting developments on a normal matrix model \cite{Betzios:2025eev} which is argued to dual to a gravitational sine-Gordon model transiting between AdS$_2$ and dS$_2$ spacetimes.} Moreover, given that the infrared limit of the DSSYK reproduces the Schwarzian mode of Jackiw-Teitelboim (JT) \cite{JACKIW1985343,TEITELBOIM198341} gravity at the disk level, then by studying this model we can learn how the holographic dictionary work when the gauge symmetry SL$^+(2,\mathbb{R})$ of JT gravity is q-deformed to a SL$_q^+(2,\mathbb{R})$ quantum group of the DSSYK \cite{Belaey:2025ijg}, where $q=\rme^{-\lambda}\in[0,1)$ is a fixed parameter.\footnote{See \cite{Berkooz:2024lgq} for a great introduction to the DSSYK model, or App. {{C}} in our companion work \cite{Aguilar-Gutierrez:2025pqp}.} For instance, there are different examples where Krylov complexity plays a prominent role in the holographic dictionary of the DSSYK model \cite{Heller:2024ldz,Rabinovici:2023yex,Ambrosini:2024sre,Xu:2024gfm,Aguilar-Gutierrez:2025pqp}. However, the holographic correspondence beyond the low energy sector of the DSSYK is still disputed, as there are, seemingly, multiple bulk dual candidates for the same type of boundary theory. For instance, sine dilaton gravity \cite{Blommaert:2023opb,Blommaert:2024whf,Blommaert:2024ydx,Blommaert:2025avl} is a consistent framework that might describe the bulk theory dual DSSYK in terms of a complex geometry that can be mapped to AdS$_2$ black hole spacetimes, and thus {{has}} notable differences with respect to JT gravity, and in fact it can reproduce (A)dS$_2$ or flat space in certain regimes. Other proposals under active development include dS$_3$ space with matter \cite{Verlinde:2024znh,Verlinde:2024zrh,Narovlansky:2023lfz,Gaiotto:2024kze,Tietto:2025oxn}, corresponding to a pair of decoupled DSSYK models; or dS$_2$ JT gravity (in the s-wave reduction of dS$_3$ space), where the DSSYK model is expected to live in the stretched cosmological horizon \cite{Susskind:2022bia,Lin:2022nss,Rahman:2022jsf,Rahman:2023pgt,Rahman:2024iiu,Rahman:2024vyg,Yuan:2024utc}. While there are hints that these proposals are ultimately compatible {{in some aspects}} one to the other \cite{Rahman:2024iiu,Blommaert:2024whf,Narovlansky:2025tpb,Aguilar-Gutierrez:2024oea}, it is desirable to understand the underlying reason behind the similarities and differences between these approaches. For this reason, we ask\footnote{We stress this question is different from the puzzle in \cite{Antonini:2024mci}, where a specific holographic conformal field theory (CFT) state seemly may have more than one-bulk dual description. The issue was later addressed in \cite{Engelhardt:2025vsp}.} 
\begin{quote}
\emph{Why are there so many bulk theories corresponding to, seemingly, the same kind of boundary theory?}
\end{quote}
{{The arguably simplest explanation is that assumptions and/or the physical interpretation in some of the proposals might be inaccurate. However, the above proposals also assume different physical conditions in the boundary theory (such as the constraints in the doubled DSSYK model in \cite{Narovlansky:2023lfz} been different from the other proposals). Thus, it is important to search for a framework within the boundary theory that incorporates the constraints used in the different bulk proposals. Furthermore, gauging boundary symmetries is a topic that should be studied more thoroughly in the AdS/CFT correspondence (see e.g. \cite{Maldacena:2001kr,Maldacena:2018vsr}, which we discuss below); the analytic solvability of the DSSYK model might allow us to make important steps in this direction. For these reasons, we investigate gauging}} different symmetry sectors in the chord (also called auxiliary) Hilbert space of the DSSYK model. These different subspaces are determined in terms of constraint operators (associated to the generators of the symmetry groups) \cite{dirac2013lectures,Henneaux:1994lbw}. Each of them, in principle, results in the different bulk descriptions. This is implemented by considering the set of states $\ket{\psi}$ which are invariant under the action of a set of operators $\qty{\hat{\varphi}_j}$, i.e. $\hvarphi_j\ket{\psi}=0$,
for all states within the symmetry sector of the Hilbert space, $\ket{\psi}\in \mathcal{H}_{\rm phys}\subseteq\mathcal{H}_{m}$, where the $\mathcal{H}_{m}$ is the Hilbert space without the constraints. In the system under consideration, we identify this with the chord Hilbert space with $m$-particles \cite{Lin:2022rbf}. Thus, the main physical interpretation of our work is that there is no unique bulk dual of the DSSYK with matter, instead there is ``a'' dual theory for a given symmetry sector in the chord space, corresponding to a physical bulk Hilbert space. {{This is not surprising from the perspective of the AdS/CFT correspondence. One can recover additional structure in the gravity theory by gauging symmetries in the dual boundary theory. Examples include gauging $\mathbb{Z}_2$ symmetry in a two-copy CFT \cite{Maldacena:2001kr}, or by promoting SU($N$) from a gauge to a global symmetry \cite{Maldacena:2018vsr} in the Banks-Fischler-Shenker-Susskind matrix model \cite{Banks:1996vh,Seiberg:1997ad}. Nevertheless, this needs to be studied carefully in}} the DSSYK since it has a quantum group (SL$_q^+(2,\mathbb{R})$) instead of a gauge or global symmetry group (unless we work in the triple-scaling limit, SL$^+(2,\mathbb{R})$). Nevertheless, this also means that there are different bulk set-ups that can be realized from the same boundary theory by specifying the type of constraints in the chord space. We anticipate that this can be advantageous for model building lower-dimensional bulk geometries dual to a UV finite, solvable boundary theory, possibly with more examples that ones presented in this work.

\paragraph{What are the symmetry sectors?}
In short, we first derive precise conditions to recover End-Of-The-World (ETW) brane Hamiltonians and the associated Euclidean wormhole geometries in sine dilaton gravity by gauging specific symmetries in the DSSYK model, with respect to constraint operators acting on the subspace labeled with respect to the left/right one-particle insertion in the chord diagram (which we review properly in Sec. \ref{sec:constrained quantization general DSSYK}) in the gauged Hilbert space formed by states invariant under the constraints.\footnote{The particle corresponds to two matter operators that are Wick contracted (thus forming a matter chord). Technically, the results extend to arbitrarily many matter chord insertions, albeit in as a composite operator, studied in more detail in \cite{Aguilar-Gutierrez:2025pqp}.} Later, we find a set of conditions to recover the doubled DSSYK models used in dS$_3$ space holography (originally proposed by Narovlansky and Verlinde (NV) \cite{Narovlansky:2023lfz}; followed up with different extensions \cite{Verlinde:2024znh,Verlinde:2024zrh,Gaiotto:2024kze,Tietto:2025oxn} and related models \cite{Narovlansky:2025tpb}). We first impose a constraint to insert a very heavy particle in the DSSYK. In this limiting case, the one-particle Hilbert state factorizes in a tensor product of zero-particle states. The resulting system has a notion of relative Hamiltonian evolution between the pair of DSSYKs. This leads to the double DSSYK models in the literature (see e.g. \cite{Narovlansky:2023lfz,Narovlansky:2025tpb}), where the symmetry sector corresponds to time automorphism due to the relative Hamiltonian evolution.

On the other hand, the bulk interpretation of the DSSYK model with matter and without gauging additional symmetries has been discussed in \cite{Aguilar-Gutierrez:2025pqp}. There, we found, as expected from the holographic dictionary, that matter chord operators in the DSSYK correspond to free massive scalar fields backreacting the effective dual AdS$_2$ black hole geometry. In particular, in some limits this generates shockwave geometries, which are advantageous to describe bulk configurations with arbitrarily many operator insertions \cite{Aguilar-Gutierrez:2025mxf}. All of these cases provide a concrete realization of the boundary-to-bulk map in \cite{Lin:2022rbf}.

Thus, there is a concrete connection between {{at least two}} proposals {{on the bulk}} dual to the DSSYK model. {{We do not attempt to show that the bulk proposals are both correct. For instance, we not provide a bulk calculation supporting the bulk interpretation in \cite{Narovlansky:2023lfz} (and subsequent works) for the doubled DSSYK model. Instead, the goal in this manuscript is to show that there are different symmetry sectors in chord space associated to the bulk proposals in the existing literature, which might play a role in understanding the relations between them.}}
Sine dilaton gravity \cite{Blommaert:2023opb,Blommaert:2024ydx,Blommaert:2024whf,Blommaert:2022ucs} with matter and the dS$_3$ holographic approach in the series of works \cite{Narovlansky:2023lfz,Verlinde:2024znh,HVtalks,Verlinde:2024zrh,Gaiotto:2024kze,Tietto:2025oxn} {{could be}} {different bulk manifestations of the symmetry sectors in chord space}.\footnote{Incorporating the proposal by Susskind \cite{Susskind:2021esx} and several collaborators \cite{Susskind:2022bia,Lin:2022nss,Rahman:2022jsf,Rahman:2023pgt,Rahman:2024iiu,Rahman:2024vyg,Sekino:2025bsc,Miyashita:2025rpt} in this formulation will be presented elsewhere, following up on \cite{Aguilar-Gutierrez:2024oea}.} We proceed with more details about the different cases that we consider.

\paragraph{ETW branes, the holographic dictionary and Euclidean wormholes}
ETW branes are an arguably well-studied case of matter strongly backreacting on AdS space. This can be used to generate higher genus topologies in JT gravity \cite{Jafferis:2022uhu,Jafferis:2022wez} and in sine dilaton gravity \cite{Blommaert:2025avl}. So far, most of the approaches to describe holographically JT or sine dilaton gravity with ETW branes and higher genus topologies require a matrix model completion \cite{Gao:2021uro,Blommaert:2025avl,Jafferis:2022uhu,Jafferis:2022wez}. However, the precise relation between {{ETW branes in the DSSYK model with a matrix model completion, such as}} the ETH matrix model\cite{Jafferis:2022uhu,Jafferis:2022wez} is not entirely settled \cite{Miyaji:2025ucp,Berkooz:2020fvm}.

Moreover, we are also motivated by interesting results showing that ETW brane and trumpet partition functions can be derived from the characters of the SL$^+_q(2,\mathbb{R})$ quantum group of the DSSYK model \cite{Belaey:2025ijg}. However, it is unknown how to interpret the ETW branes and trumpets in terms of chord diagrams. On the other hand, a family of ETW brane Hamiltonians in sine-dilaton gravity \cite{Blommaert:2025avl} was built {{by studying the corresponding}} Wheeler-de-Witt (WDW) \cite{Wheeler:1968iap,DeWitt:1967yk} equation. While the bulk theory is argued to be dual to the DSSYK when there are no branes \cite{Blommaert:2024whf}; there has not been any clear relation between the ETW brane Hamiltonians and the DSSYK Hamiltonian so far. This has also been motivated by a pioneering model in \cite{Okuyama:2023byh} based on a natural generalization of the q-Hermite polynomials. This seems to reproduce an ETW brane configuration in the dual theory. However, their approach lacks a first principles derivation that relates it with the DSSYK model. For these reasons, we ask
\begin{quote}
\emph{Under what circumstances are the ETW brane Hamiltonians in \cite{Okuyama:2023byh,Blommaert:2025avl} dual to the DSSYK model with matter chords?}   
\end{quote}
Here, we show that gauging specific symmetries in the DSSYK model leads to the Al-Salam Chihara (ASC) Hamiltonian in \cite{Blommaert:2025avl}. The constraints generating the symmetries can be interpreted in terms of a particle moving on a surface in phase space; which was previously studied by \cite{Toms:2015lza} in other systems. Importantly, we do not need to assume any connection with the ETH matrix model \cite{Jafferis:2022uhu,Jafferis:2022wez} to generate the wormhole observables associated with the DSSYK model, unlike the previous literature.

Our results also include the evolution of the classical phase space of the DSSYK dual with constraints. Motivated by \cite{Rabinovici:2023yex}, we interpret the solutions for the expectation value of the chord number in terms of spread complexity, which we show is dual to a wormhole minimal geodesic length. This allows us to find match very precise boundary/bulk quantities, from which we deduce the holographic dictionary (summarized in Tab. \ref{tab:holographic_dictionary}).
\begin{table}
    \centering
    \begin{tabular}{|C{7.5cm}|C{7.5cm}|}
    \hline\textbf{Bulk (AdS$_2$ black hole)}\vspace{0.2cm}&\textbf{Boundary (DSSYK)}\\\hline
        \textbf{\footnotesize ETW brane spacetime}\vspace{0.2cm}&\textbf{\footnotesize Constrained HH state,} (\ref{eq:HH ETW})\\
    \footnotesize{Bulk temperature} $\Phi_h/(2\pi)$ \eqref{eq:dictionary 1}\vspace{0.2cm}&\footnotesize{Fake temperature} $J\sin\theta/\pi$\\
    \footnotesize{Asymptotic boundary time} $u_{L/R}$ \eqref{eq:dictionary 1}&\footnotesize{DSSYK Hamiltonian time} $t_{L/R}$ \\
{\footnotesize Wormhole distance with an ETW brane}, $L_{\rm ETW}(t)$ (\ref{eq:length ETW brane}) \vspace{0.2cm}&\multirow{ 2}{*}{\footnotesize Spread complexity, $\lambda~\mathcal{C}(t)$ (\ref{eq:womrhole XY})}\\
{\footnotesize Exponentiated length regulator} $\rme^{L_{\rm reg}}$ \eqref{eq:dictionary 2}\vspace{0.2cm}&${\tfrac{\sqrt{\prod_{X_i=X,Y}(X_i^2-2X_i\cos\theta+1)-(1+XY-(X+Y)\cos\theta)^2}}{2\sin\theta}}$\\
    {\footnotesize Backreaction parameter} $\frac{m_{\rm ETW}}{\sqrt{1-K_{\rm ETW}^2}}$ \eqref{eq:dictionary 3}\vspace{0.2cm}&$\frac{(1+XY-(X+Y)\cos\theta)\sin\theta}{\sqrt{\prod_{X_i=X,Y}(X_i^2-2X_i\cos\theta+1)-(1+XY-(X+Y)\cos\theta)^2}}$\\\hline
    \end{tabular}
    \caption{Holographic dictionary of the ETW brane system/DSSYK with matter. The entries are obtained by equating the spread complexity of the Hartle-Hawking \cite{Hartle:1983ai} (HH) state in the constrained chord space with matter chords and wormhole lengths in AdS$_2$ black hole geometries with an ETW brane. $K_{\rm ETW}$ is the extrinsic curvature of the ETW brane, $m_{\rm ETW}$ is the brane's tension; $X$, $Y$, $\lambda$ are parameters of the theory (\ref{eq:ASC Hamiltonian}), and $\theta$ a parametrization of the energy spectrum (\ref{eq:energy spectrum}). A more detailed summary of the notation is shown in App \ref{app:notation}.}
    \label{tab:holographic_dictionary}
\end{table}

Moreover, we generate Euclidean wormhole partition functions from the ETW brane Hamiltonians {{based on the corresponding bulk partition functions}} in JT gravity \cite{Gao:2021uro} or sine-dilaton gravity \cite{Blommaert:2025avl}. This part of the study is motivated by Euclidean wormholes in higher dimensions (see a modern review in \cite{Hebecker:2018ofv}). Euclidean wormholes naturally arise from the sum over topologies in the Euclidean path integral, as well as in certain cases from the Lorentzian path integral (see e.g. \cite{Loges:2022nuw}). Very interesting developments show that (up to technical details still debated) Euclidean wormholes are perturbatively stable \cite{Hertog:2024nys,Aguilar-Gutierrez:2023ril,Loges:2022nuw,Marolf:2021kjc,Jonas:2023ipa,Jonas:2023qle,Marolf:2025evo}. However, they lead to the appearance of puzzling features. For instance, the factorization puzzle in AdS/CFT \cite{Maldacena:2001kr,Witten:1999xp} (see also \cite{Maldacena:2004rf,Arkani-Hamed:2007cpn}), which has been treated through several approaches, e.g. \cite{Saad:2021uzi,Saad:2021rcu,Saad:2019pqd,Gesteau:2024gzf,McNamara:2020uza,Blommaert:2021fob,Blommaert:2022ucs,Marolf:2020rpm,Marolf:2020xie}. They can also generate baby universes \cite{Coleman:1988cy,Coleman:1988tj,Giddings:1988cx,Preskill:1988na,McNamara:2019rup,McNamara:2020uza,Gesteau:2024gzf,Blommaert:2021fob,Blommaert:2022ucs,Marolf:2020rpm,Marolf:2020xie,Marolf:2021ghr}. In Coleman's interpretation \cite{Coleman:1988tj}, the effect of Euclidean wormholes in the Euclidean path integral is to generate an effective bilocal action (or generalized notions for multi-boundary Euclidean wormholes) which is substituted by an effective local action where the coupling constants of the original theory get shifted by the so-called alpha parameters (which are spacetime independent). This suggests that there might be superpositions of universes (i.e. a multiverse) where the alpha parameters take different specific values in each universe. This would mean that there is no way to predict the fundamental coupling constants that we observe (although they may have a statistical interpretation \cite{Preskill:1988na}). There have been several arguments against non-trivial baby universes in different contexts, e.g. \cite{Engelhardt:2025vsp,Harlow:2025pvj,Marolf:2020xie,Marolf:2020rpm,McNamara:2020uza,Dong:2024tjx,Usatyuk:2024isz,Usatyuk:2024mzs}.

While we do not expect that {{developing wormhole observables in the DSSYK with a gauged symmetry}} may directly address the previous core questions in more realistic settings, it still mimics specific features of Euclidean wormholes in higher dimensional models. Thus, this model may be used as an analytically trackable and UV finite testground to develop technical methods and concepts in other holographic spacetimes. We will find that the Euclidean wormholes in this model turn out to be perturbatively stable (including one-loop corrections in the on-shell action), and they have a trivial baby universe Hilbert space (which we construct explicitly). Moreover, our work on wormhole partition functions also provides the first steps towards higher genus topology entirely from the boundary side, without assuming an underlying matrix model completion at finite $N$. We also discuss interesting new developments connecting closed universes with Euclidean wormholes in relation with our results in Sec. \ref{ssec:outlook}.

\paragraph{Doubled DSSYK models and their gauge invariant algebra}\cite{Narovlansky:2023lfz} showed that a pair of decoupled DSSYK model with the same energy spectrum and gauge-fixed charge, parity, and time reversal (CPT) symmetry reproduces the same semiclassical two-point correlation functions of dS$_3$ space. This connection was further polished with substantial results beyond the semiclassical limit and different extensions/modifications by \cite{Verlinde:2024znh,Verlinde:2024zrh,Gaiotto:2024kze,Tietto:2025oxn,Narovlansky:2025tpb}.

After recovering the doubled DSSYK by constraining the one-particle chord space, we investigate how to implement the path integral to compare our formulation of the system with the original formulation as a pair of finite $N$ SYK systems which is then double-scaled \cite{Narovlansky:2023lfz}. We find that the path integrals take a very similar form between these two perspectives.

We further investigate the gauge invariant algebra of observables, defined in \cite{DeVuyst:2024uvd,DeVuyst:2024pop}. This is a *-algebra of relational observables in the perspective neutral framework to {\textcolor{blue}{quantum reference frames (QRFs)}} (see e.g. \cite{Krumm:2020fws,Hohn:2017cpr,Hoehn:2019fsy,Hoehn:2020epv,Hoehn:2023ehz,Vanrietvelde:2018pgb,Vanrietvelde:2018dit,delaHamette:2021oex,Hohn:2018toe,Hohn:2018iwn,Hoehn:2021flk,Giacomini:2021gei,Yang_2020} among many others), i.e. the corresponding operators are dressed with respect to any possible internal frame in the system. 
Our motivation for carrying this out is to develop a relational notion of the holographic correspondence, which we expect to be a crucial concept to study holography in spatially compact universes, among other situations. The framework of relational quantum dynamics describes gauge invariant notions of operators  dressed with respect to the QRF. This is the role taken by the asymptotic AdS boundary in more conventional settings in AdS/CFT holography, so the notions of QRF transformations are usually not required. This means that all the gauge-invariant observables are dressed with respect to the dual CFT(s). It is worth {{further}} development to describe an observer falling into a AdS black hole \cite{Jafferis:2020ora,deBoer:2022zps,Gao:2021tzr,Stanford:2022fdt,Blommaert:2024ftn,Iliesiu:2024cnh} (and more recently realized in spread complexity \cite{Li:2025fqz}). However, holography for spatially compact universes (or possibly even in flat space holography) indeed allows for different kinds of QRFs and gauge-invariant observables, {{which would have}} dual description {{if the corr{\textcolor{blue}{e}}sponding gravity theories were indeed holographic}}. For this reason, it is rather natural that the doubled DSSYK model in dS holography is ``relational'' in the sense of having associated gauge invariant observables.\footnote{This is also relevant to describe the non-trivial evolution of an observer in a closed universe, the problem of time \cite{Isham:1992ms} (possibly with more reference frames), which is addressed naturally in the model in Sec. \ref{sec:NV model} {{if it were indeed holographic}}. {{We also expect this also applies}} in the boundary construction of Euclidean wormholes, as we discuss in Sec. \ref{ssec:outlook}.} In the specific case we consider, the reference is either one of the DSSYK models, acting as a clock observer that performs measurements of the system (its pair DSSYK clock). We find a type I algebra in the kinematical Hilbert space. We argue that the semiclassical and thermodynamical limit of this algebra similarly may lead to the same type of algebra for dS space (e.g. \cite{Chandrasekaran:2022cip,Witten:2023qsv,Jensen:2023yxy,Kudler-Flam:2023qfl,Witten:2023xze,AliAhmad:2024wja})\footnote{There are technical details that have been recently put in question \cite{Geng:2025bcb}.} in a similar realization as \cite{Kolchmeyer:2024fly}. In deriving these results, we also show that Krylov complexity in the DSSYK model is a relational observable, meaning that it depends on the observers making the measurement. This formalizes some of the recent discussions about the role of observers in spread complexity by \cite{Li:2025fqz}.\footnote{We completed the calculations regarding this point before their work appeared on arXiv.} Furthermore, we uncover a natural extension of the clock states employed in the perspective neutral approach to QRFs (see e.g. \cite{DeVuyst:2024uvd,DeVuyst:2024pop}).\footnote{See \cite{AliAhmad:2024qrf,AliAhmad:2024vdw,AliAhmad:2024wja,Fewster:2024pur} for other related approaches.}
 This calls for further investigations bridging the conceptual frameworks and technical tools in holography and QRFs.

\paragraph{Bulk quantization}To recover a full equivalence between the boundary and bulk theories, one would like to implement the constraints in the quantization of the bulk theory. The two main ways to quantize constrained Hamiltonian systems are reduced phase space quantization \cite{faddeev1969feynman} and Dirac quantization \cite{dirac2013lectures}. These approaches might result in different quantum theories due to operator order ambiguity \cite{Plyushchay:1994pk,Loll:1990rx,Kuchar:1983rc,Kuchar:1986ji,Kuchar:1986jj,Kuchar:1987py,McMullan:1988ty,McMullan:1988tw,Dolan:1990zv,Ashtekar:1982wv,Schleich:1990gd,Shimizu:1996vf,Romano:1989zb} in the canonical quantization. However, given that we rely on the bulk interpretation of chord Hilbert space \cite{Lin:2022rbf}, this can be most naturally realized in Dirac quantization.\footnote{See also App. \ref{app:FJ approach} for a Faddeev-Jackiw path integral approach.} This is a useful framework to organize what are the redundant and physical degrees of freedom in the quantization of constrained systems. The former ones can then be dealt with by introducing gauge invariant observables, such as in relational quantum dynamics (see \cite{Hoehn:2019fsy,Hoehn:2020epv,Hoehn:2023ehz,Casali:2021ewu}). The constraints are imposed in the Hilbert space rather than at the operator level. Quantum gravity in higher dimensions has second-class constraints (due to structure functions in the phase space variables, e.g. \cite{Esposito1994}) that make it difficult to work with \cite{Sundermeyer:1982gv}.\footnote{This might be a reason that the techniques of constraint quantization in holographic systems in higher dimensions are not under continuous active development; although there are interesting developments in holography of information \cite{Chakravarty:2023cll}.} However, these problems are avoided in lower dimensions. For this reason, it turns out that constraint quantization {{methods are}} well-adapted to explore the symmetry sectors within the chord Hilbert space of the DSSYK model, which is argued to describe its bulk dual Hilbert space \cite{Lin:2022rbf}. Dirac quantization often has ambiguities besides the operator ordering, such as in defining the Hilbert space before solving constraints and on the choice of inner product (see the relevant discussion about this topic in \cite{Marolf:2000iq,Giulini:1998rk,Marolf:1995cn} and references within). These issues can be naturally dealt with by adopting the chord Hilbert space \cite{Lin:2022rbf} where the constraints can be imposed. One specifically demands discreteness of the lengths and {{other conditions on the}} energy spectrum of the bulk theory. This also avoids resorting to the finite N SYK \cite{Narovlansky:2023lfz}, where the constraints in the dual bulk theory might not have a clear interpretation. We briefly mention how to employ this approach while studying the constraint quantization of the ETW brane systems, which has been carried out (albeit without using this language) in \cite{Blommaert:2024whf}, while the reader is referred to \cite{Verlinde:2024znh} for the canonical quantization of dS$_3$ space as a DSSYK Hamiltonian. 
 
\paragraph{Plan of the paper:}In \textbf{Sec.} \ref{sec:constrained quantization general DSSYK} we illustrate how to apply our procedure in later sections to gauge symmetries in chord Hilbert space of the DSSYK. In \textbf{Sec.} \ref{sec:ASC from DSSYK}, we discuss the specific constraints in the chord space that recovers ETW brane Hamiltonians in JT \cite{Kourkoulou:2017zaj,Gao:2021uro} and sine dilaton gravity \cite{Blommaert:2025avl}. We interpret them as either one or two-sided branes with similar characteristics as in \cite{Okuyama:2023byh,Xu:2024hoc}. We also study the classical phase space solutions from the path integral, where the classical expectation value of the chord number turns out to describe spread complexity with respect to the HH state prepared by the path integral. We match the result to a Lorentzian wormhole distance. This allows us to spell out the holographic dictionary resulting from this match. Later, in \textbf{Sec.} \ref{sec:wormhole topology} we use the ETW brane interpretation of the constrained chord space to define wormhole partition and correlation functions (i.e. beyond the disk level). We build Euclidean wormhole partition functions with one and two disk boundaries, and also {{deduce}} cylinder two-point functions by implementing appropriate constraints. Moreover, we analyze the (one-loop corrected) perturbative stability of the saddle points in this theory. We then explore the baby universe of the Euclidean wormhole, which we later connect with recent discussion on observers in closed universes \cite{Balasubramanian:2023xyd,Harlow:2025pvj,Abdalla:2025gzn,Chen:2025fwp,Nomura:2025whc}. In \textbf{Sec.} \ref{sec:NV model} we describe the doubled DSSYK model used in dS$_3$ holographic constructions \cite{Narovlansky:2023lfz,Narovlansky:2025tpb} in terms of the one-particle chord space of a single DSSYK. We present both the operator constraints in the bulk Hilbert space, and a corresponding path integral analysis. This allows us to study the gauge invariant algebra of observables \cite{DeVuyst:2024uvd,DeVuyst:2024pop} in this model, where the QRFs are the pair of DSSYK models. In \textbf{Sec.} \ref{sec:disc} we close with a summary and discussion of our work, and future directions.

For the convenience of the reader, App. \ref{app:notation} contains a summary of the notation we used. In App. \ref{app:ETW brane partition function} we {{evaluate}} the ETW brane partition function in the semiclassical limit; {{and present the corresponding}} effective entropy, temperate and the conditions for thermal stability (from the heat capacity).
{App.} \ref{app:Morse from ASC} contains another consistency check, where we analyze the triple-scaling limit of the ETW brane Hamiltonian of the ASC form (\ref{eq:ASC Hamiltonian}). Next, {{in}} App. \ref{app:FJ approach} {{we explain how to use}} an alternative formulation from the one we followed in the main text, Faddeev-Jackiw (FJ) constraint quantization \ref{app:FJ approach}, {{to study the ETW brane systems}}. At last, {App.} \ref{app:alternative derivation}, describes an alternative derivation of (\ref{eq:sp theta}).

\section{Background on constraints and the chord Hilbert space}\label{sec:constrained quantization general DSSYK}
In this short section, we review how to gauge symmetries by implementing constraints, and some details about the chord Hilbert space of the DSSYK with matter chords that are used in the rest of the work. 

We define symmetries in a general Hilbert space as transformations, $\hat{U}_i$, that leaves states invariant,
\begin{equation}\label{eq:LR transformation}
    \hat{U}_i:=\rme^{\alpha_i\hat{\varphi}_i}~:\quad\ket{\psi}\rightarrow\ket{\psi}~,
\end{equation}
where $\alpha_i$ are constants, and $\qty{\hat{\varphi}_j}$ are a set of generators that annihilate states, which are then used to build the gauged Hilbert space, denoted $\mH_{\rm phys}$. This means
\begin{equation}\label{eq:symmetries general2}
    \hat{\varphi}_i\ket{\psi}=0~,\quad \ket{\psi}\in{\mathcal{H}}_{\rm phys}~.
\end{equation}
The additional states in the original Hilbert space that do not obey (\ref{eq:symmetries general2}) have to be modded out to build the Hilbert space above. One can also construct \emph{observables} within the theory, i.e. maps between physical states within the same Hilbert space ${\mathcal{H}}_{\rm phys}$. These have to commute with all the constraints above,
\begin{equation}
    \qty[\hat{\varphi}_i,~\hO_\Delta]=0~, \quad\forall i,~j~.
\end{equation}
The discussion above can be implemented in the boundary or bulk theories. In the latter case, we interpret the chord Hilbert space of the DSSYK model (denoted $\mathcal{H}_m$, see (\ref{eq:Fock space with matter}) below), as the corresponding bulk Hilbert space according to the boundary-to-bulk map in \cite{Lin:2022rbf}. In terms of Dirac quantization \cite{dirac2013lectures}, this eliminates the ambiguity of how to define the linear space where $\hat{\varphi}_i$ act. Moreover, since we work with a dual boundary description, with an explicit Hamiltonian, this also eliminates the operator ordering ambiguity in the canonical quantization of the bulk theory, since we can act with the constraints to generate the bulk physical Hilbert space that reproduces the chord diagram Hamiltonian. 

The constraints are generators of (gauge or global) symmetry groups depending on whether they are \emph{first}- or \emph{second-class} in Dirac's classification \cite{dirac2013lectures}. First-class (generating gauge symmetries) means that all the constraints commute with each other, i.e.
\begin{equation}
    {\rm First-class}~:\quad\qty[\hvarphi_i,~\hvarphi_j]=0\forall i,j~.
\end{equation}
Meanwhile, second-class corresponds to anything that is not first-class.

All the states $\ket{\psi}\in\mathcal{H}_{m}$ such that $\hat{\phi}_i\ket{\psi}\neq0$ might belong to a different symmetry sector and they need to be modded out from $\mathcal{H}_{\rm m}$ to build $\mathcal{H}_{\rm phys}$. In the holographic context, if there are multiple symmetry sectors in the boundary theory each of them are associated with a different bulk dual description.\footnote{I thank Gonçalo Araujo-Regado for pointing out this possible interpretation.}

We now focus on the DSSYK model. Given that the DSSYK model has a SL$_q^+(2,\mathbb{R})$ quantum group instead of the SL$^+(2,\mathbb{R})$ gauge group of JT gravity; the procedure above involves second-class constraints generically (we study an exception in Sec. \ref{sec:NV model}). {In order to discuss the quantization of the bulk dual, one can implement Dirac quantization with specific constraints (originally explored in \cite{Blommaert:2024whf} for sine dilaton gravity). In this work, we proceed in a different approach from \cite{Blommaert:2024whf}. We work directly in the chord Hilbert space of the DSSYK without assuming a specific bulk dual, since there might be several depending on the constraints that we impose, and we also incorporate matter in the bulk Hilbert space.} We will present the specific constraints $\hat{\varphi}_i$ in the different examples explored in Secs. \ref{sec:ASC from DSSYK}, \ref{sec:NV model}. 

We now briefly explain the general set-up. The DSSYK model with matter chords can be expressed through a chord Hilbert space as in \cite{Lin:2022rbf} 
\begin{equation}\label{eq:Fock space with matter}
    \mathcal{H}_{m}=\bigoplus_{n_0,n_1,\cdots,n_m=0}^\infty\mathbb{C}\ket{\tilde{\Delta},n_0,n_1,\cdots,n_m}~.
\end{equation}
Here, $\tilde{\Delta}=\qty{\Delta_1,\dots,\Delta_m}$ is a string of conformal dimensions, where each $\Delta_i$ corresponds to a double-scaled operator $\hat{\mathcal{O}}_{\Delta_i}$. Also, $n_0$ is identified with the number of open (H-)chords to the left of all particle chords; $n_1$ is the number between the first two particles; until we count for all the $m$ particles. The chord Hilbert space expressed in the chord number basis \eqref{eq:Fock space with matter} has a natural bulk interpretation, emphasized in \cite{Lin:2022rbf}, where the parameter $n$ is related a wormhole bulk length (although the basis where the lengths have this interpretation has to be carefully constructed through the Gram-Schmidt procedure).

It was also found through chord diagram combinatorics that the Hamiltonian that describes this system can be expressed as \cite{Lin:2022rbf}
\begin{align}\label{eq:two-sided new}
    &\hH_{L/R}=\frac{J}{\sqrt{\lambda}}\qty(\hat{a}_{L/R}+\hat{a}^\dagger_{L/R})~,\quad \text{where}\\
&\hat{a}^\dagger_L=\hat{a}^\dagger_0~,\quad \hat{a}_L=\sum_{i=0}^m\hat{\alpha}_i\qty(\frac{1-q^{\hat{n}_i}}{1-q})q^{\hat{n}_i^<}~,\quad\text{with}\quad\hat{n}_i^<=\sum_{j=0}^{i-1}\qty(\hat{n}_j+\Delta_{j+1})~,\label{eq:aLdagger,aL}\\
&\hat{a}^\dagger_R=\hat{a}^\dagger_m~,\quad\hat{a}_R= \sum_{i=0}^m\hat{\alpha}_i\qty(\frac{1-q^{\hat{n}_i}}{1-q})q^{\hat{n}_i^>}~,\quad\text{with}\quad\hat{n}_i^>=\sum_{j={i+1}}^{m}\qty(\hat{n}_j+\Delta_{j})~,\label{eq:aRdagger,aR}
\end{align}
where $J$ is a coupling constant (resulting from Gaussian ensemble averaging in Hamiltonian moments of the SYK model). The division of the system into left/right chord sectors is due to a particle chord insertion (see Fig. \ref{fig:brane_from_DSSYK} for examples). The operators in \eqref{eq:two-sided new} act as follows
\begin{subequations}\label{eq:Fock Hm}
    \begin{align}\label{eq:Fock Hm 1}
    \hat{a}^\dagger_{i}\ket{\tilde{\Delta};n_0,\dots n_i,\dots, n_m}&=\ket{\tilde{\Delta};n_0,\dots, n_i+1,\dots n_m}~,\\\label{eq:Fock Hm 2}
    \hat{\alpha}_{i}\ket{\tilde{\Delta};n_0,\dots n_i,\dots, n_m}&=\ket{\tilde{\Delta};n_0,\dots, n_i-1,\dots n_m}~.
\end{align}
\end{subequations}
Similar to \cite{Aguilar-Gutierrez:2025pqp}, we can perform a canonical change of variables
\begin{equation}\label{eq:non-conjugate ops many particles}
\hat{a}_{i}^\dagger=\frac{\rme^{-\rmi \hat{P}_{i}}}{\sqrt{1-q}}~,\quad \hat{\alpha}_{i}=\sqrt{1-q}~\rme^{\rmi \hat{P}_{i}}~,\quad q^{\hat{n}_{i}}=\rme^{-\hat{\ell}_{i}}~.
\end{equation}
For the particular case with only one-particle insertion\footnote{We discuss generalization with more particles chords in Sec. \ref{ssec:outlook}.} (\ref{eq:two-sided new}) becomes
\begin{equation}\label{eq:pair DSSYK Hamiltonians 1 particle}
\begin{aligned}
    \hH_{L/R}=\frac{J}{\sqrt{\lambda(1-q)}}\qty(\rme^{-\rmi \hat{P}_{L/R}}+\rme^{\rmi \hat{P}_{L/R}}\qty(1-\rme^{-\hat{\ell}_{L/R}})+q^{\Delta}\rme^{\rmi \hat{P}_{R/L}}\rme^{-\hat{\ell}_{L/R}}\qty(1-\rme^{-\hat{\ell}_{R/L}}))~.
    \end{aligned}
\end{equation}
The canonical operators in the two-sided Hamiltonian  $\hat{\ell}_{L/R}$ correspond to length variables reaching between the left/right asymptotic boundaries of the dual theory reaching a bulk matter insertion in the boundary-to-bulk map of \cite{Lin:2022rbf}, and their corresponding conjugate momenta $\hat{P}_{L/R}$.

We provide more details about the inner product and the Hamiltonian for the states where constraints are imposed in Secs. \ref{sec:ASC from DSSYK}, \ref{sec:NV model}.

\section{End-of-the-world (ETW) branes from matter chords, and the holographic dictionary}\label{sec:ASC from DSSYK}
In this section, we study the constraints in the chord space that generate ETW branes in the bulk and the holographic dictionary of the DSSYK model after gauging specific symmetries {{(introduced in Sec. \ref{ssec:constraints ETW})}}, whose Hamiltonian acting on the physical Hilbert space takes the form
\begin{equation}\label{eq:ASC Hamiltonian}
    \hH_{\rm ASC}=\frac{J}{\sqrt{\lambda(1-q)}}\qty(\rme^{-\rmi \hat{P}}+(X+Y)\rme^{-\hat{\ell}}+\qty(1-XY\rme^{-\hat{\ell}})\rme^{\rmi \hat{P}}\qty(1-\rme^{-\hat{\ell}}))~,
\end{equation}
where $X$, $Y$ are constant parameters of the theory, and the overall coefficient is fixed by the original DSSYK model Hamiltonian (\ref{eq:pair DSSYK Hamiltonians 1 particle}). Later, we show that the expectation value of the chord number in the constrained system (with the HH state as the reference) matches the Lorentzian wormhole distance in an AdS$_2$ black hole background with an ETW brane. This allows us to derive its holographic dictionary. Similar to the bulk-to-boundary map in \cite{Lin:2022rbf}, fixed chord number states (after the Gram–Schmidt process) form a Krylov basis {{that allows us to evaluate the corresponding}} spread complexity {{for a reference state \cite{Balasubramanian:2022tpr}}}. We then relate a concrete quantum mechanical observable with the bulk geometry.\footnote{See \cite{Rabinovici:2023yex,Xu:2024gfm,Heller:2024ldz,Aguilar-Gutierrez:2025pqp,Aguilar-Gutierrez:2025mxf} for similar approaches in this direction.} Thus, spread complexity takes a prominent role in the holographic duality of this model (as also emphasized in \cite{Heller:2024ldz,Rabinovici:2023yex,Ambrosini:2024sre,Xu:2024gfm}).

Additionally, we show that the triple scaling limit of (\ref{eq:ASC Hamiltonian}) in App. \ref{app:Morse from ASC}, reproduces the JT gravity ETW brane Hamiltonians in \cite{Jafferis:2022wez}. We will keep working in the double-scaling limit everywhere else.

\paragraph{Outline}In Sec. \ref{ssec:constraints ETW} we show how to recover two families of ETW brane Hamiltonians of the form (\ref{eq:ASC Hamiltonian}) from constraints in the one-particle chord space of the DSSYK model. {{In Sec. \ref{ssec:path integral HH} we investigate the HH state of the theory (\ref{eq:ASC Hamiltonian}) and the evolution of operators in the Heisenberg picture.}} In Sec. \ref{ssec:saddle points} we study the evolution of the saddle point solutions of the path integral that prepares the HH state in the constrained theory. Later, in Sec. \ref{ssec:correlation functions}, we discuss about the semiclassical thermal two-point functions at the disk level. Meanwhile, in Sec. \ref{ssec:spread} we determine the Krylov basis, Lanczos coefficients and the spread complexity of the HH state in the constrained model, which builds on the saddle point solutions in the previous section. At last, we show in Sec. \ref{ssec:spread bulk picture from ETW} that spread complexity of the HH state corresponds to a Lorentzian wormhole distance in the bulk theory, which leads us to the holographic dictionary of the DSSYK model under the constraints in the bulk Hilbert space.

\paragraph{Connection with the Al-Salam Chihara (ASC) polynomials}
We first explain how to recover the ASC recurrence relation from (\ref{eq:ASC Hamiltonian}).

The eigenfunctions of the Hamiltonian (\ref{eq:ASC Hamiltonian}) are the ASC polynomials \cite{al1976convolutions}. This can be seen by defining a basis $\qty{\ket{H_n}}$, where we have
\begin{equation}
\begin{aligned}
&\hat{a}_{\rm ASC}^\dagger={\rme^{-\rmi \hat{P}}}~,\quad &&\hat{a}_{\rm ASC}=\rme^{\rmi \hat{P}}~,\quad &&q^{\hat{n}}=\rme^{-\hat{\ell}}~,\\
&\hat{a}_{\rm ASC}^\dagger\ket{H_n}=\ket{H_{n+1}}~,\quad &&\hat{a}_{\rm ASC}\ket{H_n}=\ket{H_{n-1}}~,\quad &&\hat{n}\ket{H_n}=n\ket{H_n}~.\label{eq:number op ASC}
\end{aligned}
\end{equation}
Here we have introduced a variable $q$ where $q^{\hat{n}}=\rme^{-\hat{\ell}}$ and $\hat{n}$ is the so-called chord number operator (e.g. \cite{Berkooz:2018jqr}), which in this case corresponds to (\ref{eq:number op ASC}). The Hamiltonian (\ref{eq:ASC Hamiltonian}) can then be expressed\footnote{Note that despite \eqref{eq:ASC H2} having seemingly non-Hermitian form, it is Hermitian, just as the chord Hamiltonian \eqref{eq:pair DSSYK Hamiltonians 1 particle} for the one-particle sector chord inner product (Sec 2.6 \cite{Lin:2023trc}) since we are considering a subspace of states in the one-particle Hilbert space.}
\begin{equation}\label{eq:ASC H2}
    \hH_{\rm ASC}=\frac{J}{\sqrt{\lambda(1-q)}}\qty(\rme^{-\rmi \hat{P}}+(X+Y)q^{\hat{n}}+\rme^{\rmi \hat{P}}\qty(1-XYq^{\hat{n}-1})\qty(1-q^{\hat{n}}))~,
\end{equation}
where we have used commutation relations
\begin{equation}\label{eq:recurrence ASC n}
\hat{a}_{\rm ASC}~\hat{n}=[\hat{a}_{\rm ASC},\hat{n}]+\hat{n}~\hat{a}_{\rm ASC}=(1+\hat{n})\hat{a}_{\rm ASC}~.
\end{equation}
The eigenvalue problem of (\ref{eq:ASC H2}) is therefore equivalent to the ASC recurrence relation:
\begin{equation}\label{eq:ASC recurrence relation}
\begin{aligned}
    2\cos\theta ~Q_n=Q_{n+1}+(X+Y)q^n Q_n+(1-XY q^{n-1})(1-q^n)Q_{n-1}~,
\end{aligned}
\end{equation}
where $\frac{2J}{\sqrt{\lambda(1-q)}}\cos\theta$ is the eigenvalue of $\hH_{\rm ASC}$ in (\ref{eq:ASC Hamiltonian}); $Q_n(\cos\theta|X,Y;q)\equiv\braket{\theta}{H_n}$ (with $\qty{\ket{\theta}}$ the energy basis) are the ASC polynomials \cite{al1976convolutions}, which are given by
\begin{equation}\label{eq:ASC pol}
    Q_l(\cos\theta|X,Y;q)=\frac{(XY;q)_l}{X^l}\sum_{n=0}^\infty\frac{(q^{-l},X \rme^{\pm\rmi\theta};q)_n}{(XY,q;q)_n}q^n~,
\end{equation}
and we introduced the q-Pochhammer symbol and its products
\begin{subequations}
    \begin{align}
        (a;~q)_n&:=\prod_{k=0}^{n-1}(1-aq^k)~,\label{eq:q-Pochhammer symbol}\\
    \label{eq:combine Pochhammer}
    (a_0,\dots, a_N;q)_n&:=\prod_{i=1}^N(a_i;~q)_n~,\\
    (x^{\pm a_1\pm a_2};q)_n&:=(x^{a_1+ a_2};q)_n(x^{- a_1+a_2};q)_n(x^{-a_1+ a_2};q)_n(x^{- -a_1-a_2};q)_n~.
    \end{align}
    \end{subequations}
The recurrence relation is initiated with $Q_{-1}=0$, $Q_{0}=1$ in (\ref{eq:number op ASC}). (\ref{eq:ASC pol}) also satisfies an orthogonality relation of the ASC polynomials, 
\begin{equation}\label{eq:ortho relations ASC}
\int_0^\pi\rmd\theta ~\tilde{\mu}(\theta)\frac{Q_{l_1}(\cos\theta,X,Y|q)}{\sqrt{(q,XY;q)_{l_1}}}\frac{Q_{l_2}(\cos\theta,X,Y|q)}{\sqrt{(q,XY;q)_{l_2}}}=\delta_{l_1,l_2}~,
\end{equation}
where
\begin{equation}\label{eq:completeness relation}
    \tilde{\mu}(\theta)=\frac{(q,XY,\rme^{\pm2 \rmi\theta};q)_\infty}{2\pi(X\rme^{\pm \rmi\theta},Y\rme^{\pm \rmi\theta};q)_\infty}~,\quad \int_0^\pi\rmd\theta~\tilde{\mu}(\theta)\ket{\theta}\bra{\theta}=\mo~.
\end{equation}

\subsection{Chord ETW branes}\label{ssec:constraints ETW}
We now apply the techniques in Sec. \ref{sec:constrained quantization general DSSYK} for the specific case where the second-class constraint operators are
\begin{equation}
    \hat{\varphi}_A=A_0\mo+A_L\hat{\ell}_L+A_R\hat{\ell}_R~, \quad \label{eq:chi ETW}
    \hat{\varphi}_B=B_0\mo+B_L\hat{P}_L+B_R\hat{P}_R~,
\end{equation}
and where $A_{0/L/R}$, $B_{0/L/R}$ are constants. Note that (\ref{eq:chi ETW}) corresponds to fixing length scales in the bulk dual theory, according to the bulk-to-boundary map \cite{Lin:2022rbf}. Below, we find two types of transformations on the two-sided DSSYK Hamiltonian (\ref{eq:pair DSSYK Hamiltonians 1 particle}) that lead to two different boundary dual ETW brane Hamiltonians (\ref{eq:ASC Hamiltonian}).

\paragraph{Types of solutions}
We will consider two types of constraints:
\begin{itemize}
\item \textbf{One-sided branes}: 
\begin{equation}\label{eq:constraints Okuyama}
\begin{aligned}    
    &\hvarphi_A=\hat{\ell}_{L/R}-\tilde{N}\mathbb{1}~,\quad \hvarphi_B=\hat{P}_{L/R}~,
\end{aligned}
\end{equation}
where we take a limit $\tilde{N}\rightarrow\infty$.\footnote{As an alternative to \eqref{eq:constraints Okuyama}, consider the constraints $\hat{\varphi}_A=:\hat{\ell}_{L/R}-\hat{a}_{L/R}/\delta$ and $\hat{\varphi}_B=:\hat{P}_{L/R}-\delta~\hat{b}_{L/R}$, and then take the limit $\delta\rightarrow0$ with $\hat{a}_{L/R}$ and $\hat{b}_{L/R}$ being finite in this limit.} Thus, this limit eliminates the dynamical propagation of either of the left/right chord sectors, and we relabel the remaining sector 
\begin{equation}
    \hat{P}_{R/L}=\hat{P},~ \text{and}\quad\hat{\ell}_{R/L}=\hat{\ell}~.
\end{equation}
Note that the resulting two-sided Hamiltonian $\hH_{L/R}$ (\ref{eq:pair DSSYK Hamiltonians 1 particle}) acting on states in $\mathcal{H}_{\rm phys}$,
\begin{equation}\label{eq:true Okuyama}
    \hH^{(\rm 1s)}_{\rm ASC}=\frac{J}{\sqrt{\lambda(1-q)}}\qty(\rme^{-\rmi\hat{P}}+\rme^{\rmi\hat{P}}\qty(1-\rme^{-\hat{\ell}})+q^\Delta\rme^{-\hat{\ell}})~.
\end{equation}
The result is precisely the model proposed by K.~Okuyama \cite{Okuyama:2023byh}, as we can see by acting with the basis $\ket{H_n}$ \eqref{eq:number op ASC} on \eqref{eq:true Okuyama}
\begin{equation}
    \hH_{\rm ASC}^{(\rm 1s)}\ket{H_n}=\frac{J}{\sqrt{\lambda(1-q)}}\qty(\ket{H_{n+1}}+q^{\Delta+n}\ket{H_n}+(1-q^n)\ket{H_{n-1}})~,
\end{equation}
which indeed corresponds to (4.1, 31) in \cite{Okuyama:2023byh}; and by projecting onto $\ket{\theta}$ as in \eqref{eq:ASC recurrence relation}, this reduces to the recurrence relation of the big q-Hermite polynomials used in \cite{Okuyama:2023byh}. Their model considers a generalization of the auxiliary DSSYK Hamiltonian without matter chords using the big q-Hermite polynomials. The triple-scaling limit of the model by K.~Okuyama \cite{Okuyama:2023byh} also reproduces the ETW brane Hamiltonian of \cite{Gao:2021uro}; see App.~\ref{app:Morse from ASC} for details of the triple-scaling limit of the ASC Hamiltonian in \eqref{eq:ASC Hamiltonian} (which is not used in our evaluations).

Moreover, we see that the integration measure \eqref{eq:completeness relation} in this case reduces to
\begin{equation}\label{eq:integration measure coherent}
    \tilde{\mu}(\theta)=\mu(\theta)\frac{1}{(q^\Delta\rme^{\pm \rmi\theta};q)_\infty}=:\bra{\theta}\ket{B_\Delta}~,\quad\mu(\theta)=\frac{(q,\rme^{\pm2\rmi\theta};q)_\infty}{2\pi}~,
\end{equation}
where $\ket{B_\Delta}$ is the coherent state previously identified by K.~Okuyama \cite{Okuyama:2023byh}
\begin{equation}\label{eq:B Delta}
    \ket{B_\Delta}=\sum_{n=0}^\infty\frac{q^{\Delta n}}{(q;q)_n}\ket{n}~,\quad \hat{a}\ket{B_\Delta}=q^\Delta\ket{B_\Delta}~,
\end{equation}
and we express above state in the zero-particle chord number basis (i.e.~\eqref{eq:Fock space with matter} with $m=0$) where $\hat{a}_L=\hat{a}_R:=\hat{a}$ in (\ref{eq:aLdagger,aL}, \ref{eq:aRdagger,aR}). From the integration measure \eqref{eq:integration measure coherent} in the partition function (see App. \ref{sapp:comparison 1p partition function}), one can associate $\ket{B_\Delta}$ with the boundary state of an ETW brane \cite{Okuyama:2023byh}.

Now, we would like to interpret the {{procedure leading to \eqref{eq:true Okuyama} in view of the}} previous work on partially entangled thermal states \cite{Goel:2018ubv} {{(see below their (5.4))}} in the finite $N$ SYK model.\footnote{{{This has also been an extended with double-scaled chord matter operators in \cite{Aguilar-Gutierrez:2025pqp}.}}} ETW branes {{in the bulk}} can be described by SYK systems with two {{operator}} insertions. {{In our case,}} one of the chord sectors is removed {{ for a heavy enough operator insertion. Similar to \cite{Goel:2018ubv}, when $\beta_R\rightarrow\infty$ (i.e. the zero temperature limit), the right-chord subsystem reaches its ground state, which suppresses the dynamics of the corresponding chord number $n_R$. This limit eliminates the corresponding terms in the two-sided DSSYK Hamiltonian \eqref{eq:pair DSSYK Hamiltonians 1 particle} that lead to \eqref{eq:true Okuyama}.}} This describes strong backreaction in the bulk due to a heavy one-particle insertion in the DSSYK model which removes a large portion of the geometry, as illustrated in Fig. \ref{fig:brane_from_DSSYK} (a). 
\begin{figure}
    \centering
    \subfloat[One-sided brane]{\includegraphics[height=0.5\textwidth]{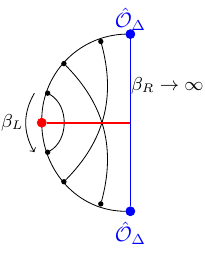}}\hfill\subfloat[Two-sided brane]{\includegraphics[height=0.5\textwidth]{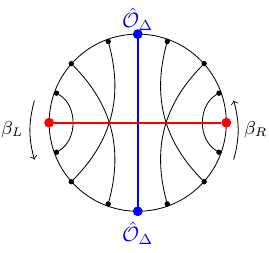}}
    \caption{Chord diagram of the DSSYK model with a one-particle chord insertion (blue), and a probe matter insertion (red). (a) One-sided constrained configuration where $\beta_R\rightarrow\infty$, while $\beta_L$ remains finite. The right chord sector is integrated out in the path integral (\ref{eq:part integral ETW}). (b) We impose $\beta_L=\beta_R$ in the path integral. The two-sided Hamiltonian theory in both of these configurations is invariant under an exchange of the left/right chord sectors (which we denote chord parity invariance).}
    \label{fig:brane_from_DSSYK}
\end{figure}
Here, the thermal circle is bisected into left and right sectors with respect to a matter chord insertion, {{and it}} is split apart in limit where the {{one of the}} subregions becomes flat and only the left or right side remains. {{To confirm our interpretation,}} in App. \ref{sapp:comparison 1p partition function} we study the corresponding partition function $Z_\Delta(\beta_L,\beta_R)$ in the {{$\beta_R\rightarrow\infty$}} limit.\footnote{{{We remark that the global state remains pure; the evaluation of $Z(\beta_L,\beta_R\rightarrow\infty)$ (App. \ref{app:ETW brane partition function}) corresponds to setting $\beta_L$ in the HH state of the one-particle chord space, so the system we describe does not originate from tracing the left or chord sector of a state in chord space. Meanwhile, from the bulk perspective, it is argued in \cite{Goel:2018ubv} (see below (5.4)) that the ETW brane itself has an associated entropy that purifies the rest of the geometry. It would be interesting to confirm their interpretation using sine dilaton gravity with ETW branes \cite{Blommaert:2025avl}.}}} By integrating out one of the two chord sectors (e.g. $n_L$ from (\ref{eq:constraints Okuyama})) we {{recover}} a Hamiltonian of the form (\ref{eq:ASC Hamiltonian}) with
\begin{equation}\label{eq:XY 1s}
    X=0~,\quad Y=q^{\Delta}~,
\end{equation}
{{which thus confirms our interpretation that setting either $\beta_{L/R}\rightarrow\infty$ (while keeping $\beta_{R/L}$ finite) corresponds to the system described by the one-sided ETW brane Hamiltonian \eqref{eq:true Okuyama}. To discuss this further, let us consider again $\beta_R\rightarrow\infty$, which geometrically corresponds to a planar limit for the would-be right half-circle, as in Fig. \ref{fig:brane_from_DSSYK} (a). This might seem in tension with $n_R\rightarrow\infty$; where do the right-sided chords go as we remove the right half-circle? Our interpretation is that this should be understood as a limiting case where the right side of the Euclidean circle approaches the ETW brane, and in the singular limit $\beta_R\rightarrow\infty$ one fits arbitrarily many chords behind the ETW brane. Note however, that the system remains well-defined in this limit as we show in App. \ref{app:ETW brane partition function}. It would be interesting to explore this point further.}}

Thus, our procedure both in this section and App. \ref{app:ETW brane partition function} shows that the ETW brane model by K. Okuyama \cite{Okuyama:2023byh} turns out to have a chord diagram origin in the one-particle chord space.
    \item \textbf{Two-sided branes}: Motivated by the triple-scaling case studied in \cite{Xu:2024hoc})\footnote{We thank Jiuci Xu for pointing out a connection between our results with his previous work \cite{Xu:2024hoc} which motivated this subsection.}, we consider the constraints:
    \begin{equation}\label{eq:constraints Xu case}
    \begin{aligned}
    &\hvarphi_A=\hat{\ell}_L-\hat{\ell}_R~,\quad &&\hvarphi_B=\hP_R-\hP_L~,
    \end{aligned}
    \end{equation}
    which is shown in Fig. \ref{fig:brane_from_DSSYK} (b).

We recover a representation of the Hamiltonian acting on the physical states is Hamiltonian:
    \begin{equation}\label{eq:H plus}
\hH^{(\rm 2s)}_{\rm ASC}=\frac{J}{\sqrt{\lambda(1-q)}}\qty(\rme^{-\rmi \hat{P}}+\rme^{\rmi \hat{P}}\qty(1-\rme^{-\hat{\ell}})+q^{\Delta}\rme^{\rmi \hat{P}}\rme^{-\hat{\ell}}\qty(1-\rme^{-\hat{\ell}}))~.
\end{equation}
We can rewrite the above expression using (\ref{eq:recurrence ASC n}) with
\begin{equation}
    \rme^{-\hat{\ell}}=q^{\hat{n}}~,\quad \hat{a}_{\rm ASC}=\frac{1}{\sqrt{1-q}}\rme^{\rmi\hat{P}}~,
\end{equation}
which implies
\begin{equation}\label{eq:comm q}
\rme^{\rmi \hat{P}}\rme^{-\hat{\ell}}=q^{\hat{n}+1}\rme^{\rmi \hat{P}}~.
\end{equation}
The Hamiltonian (\ref{eq:ASC Hamiltonian}) becomes
\begin{equation}\label{eq:H + ASC}
\hH_{\rm ASC}^{(\rm 2s)}=\frac{J}{\sqrt{\lambda(1-q)}}\qty(\rme^{-\rmi \hat{P}}+\qty(1+q^{\Delta+1}\rme^{-\hat{\ell}})\rme^{\rmi \hat{P}}\qty(1-\rme^{-\hat{\ell}}))~.
\end{equation}
This form of the {{theory}} manifestly matches (\ref{eq:ASC Hamiltonian}) when 
\begin{equation}\label{eq:Z2 sym condition}
    X+Y=0~,\quad XY=q^{\Delta+1}~.
\end{equation}
We see that the constraint imposes an equal time flow direction generated by the left and right Hamiltonians. We will interpret the dual description of the DSSYK models as an AdS$_2$ black hole geometry which has been bisected by an ETW brane. The two copies are then glued back together. It follows that the Israel junction conditions in the AdS$_2$ black hole geometry \cite{Engelhardt:2022qts} are trivially satisfied. We illustrate the configuration in Fig. \ref{fig:brane_from_DSSYK} (a). 
\end{itemize}
We discuss about the bulk interpretation for both of the above cases in Sec. \ref{ssec:spread bulk picture from ETW}. In App. \ref{app:FJ approach} we also recover these results through FJ quantization \cite{Faddeev:1988qp}, which has a more intuitive interpretation of the previous constraints in terms of a particle moving in a surface in phase space.

Following the interpretation of the chord Hilbert space being the bulk Hilbert space of the holographic dual of the DSSYK, the implementation of constraints allows us to study the \emph{symmetry sectors} within the model {{generated by the constraints (\ref{eq:constraints Okuyama}, \ref{eq:constraints Xu case})}}. Each one is associated with a different bulk theory, since the Hamiltonians in the physical Hilbert space (\ref{eq:H + ASC}, \ref{eq:true Okuyama}) are inequivalent.

\paragraph{Constraints with more particle chords}While our analysis of the constrained path integral focuses on the one-particle chord space of the DSSYK model, there is a straightforward extension of the previous analysis for the composite operators, where we introduce constraints of the form
\begin{equation}\label{eq:more general conditions}
    \hvarphi_0=\hvarphi_0(\hat{\ell}_{0},\hat{\ell}_{m})~,\quad\hvarphi_{1\leq j\leq m-1}=\hat{\ell}_j~,\quad \hvarphi_m=\hvarphi_m(\hat{P}_{0},\hat{P}_{m})~.
\end{equation}
This reduces a composite operator insertion acting on $\mathcal{H}_m$ (the chord space with m-particle insertions) {{which reduces}} to its $\mathcal{H}_1$ irrep. \cite{Aguilar-Gutierrez:2025pqp}.

In the more general cases, where we analyze non-composite operators (which have the form $\hO_{\Delta_1}\rme^{-\tau\hH} \hO_{\Delta_2}\cdots$ \cite{Aguilar-Gutierrez:2025pqp}), we expect that one can extend the techniques developed here for arbitrarily many particles propagating in the bulk; {{which}} might describe other theories. See \ref{ssec:outlook} for additional comments.

\paragraph{Are there more general ways to derive (\ref{eq:ASC Hamiltonian})?}
We now search for a more general way of reproducing the ASC Hamiltonian (\ref{eq:ASC Hamiltonian}). A naive implementation is
\begin{equation}
\begin{aligned}
&\hvarphi_A^L=\hP_R-\hP_L-d_L\mo~,\quad &&\hvarphi_B^L=\hel_L-\hel_R~;\\
&\hvarphi_A^R=\hP_L-d_R\mo~,\quad &&\hvarphi_B^R=\hel_L-\tilde{N}\mo~,
\end{aligned}
\end{equation}
where $\tilde{N}\rightarrow\infty$. We could try relabeling $\hP_L\rightarrow \hP$ and $\hel_L\rightarrow \hel$ in $\hH_L$ acting on $\ket{\psi}\in\mH_{\rm phys}$; while $\hP_R\rightarrow \hP$, $\hel_L\rightarrow\hel$ in $\hH_R$ acting on $\ket{\psi}$.
This could generate a total Hamiltonian of the type (\ref{eq:ASC Hamiltonian}), namely
\begin{equation}
\hH_L+\hH_R=\frac{2J}{\sqrt{\lambda(1-q)}}\qty(\rme^{-\rmi \hP}+\frac{q^{\Delta}\rme^{-\rmi d_L}}{2}\rme^{-\hel}+\rme^{\rmi \hP}\rme^{-\rmi d_R}\qty(1+\frac{q^\Delta\rme^{-\rmi d_R}}{2}\rme^{-\hel})(1-\rme^{-\hel}))~.
\end{equation}
However, this would be wrong since the constraints are global, and thus they cannot be imposed simultaneously in the different sectors of the two-sided Hamiltonian. In the bulk, this would require generating both geometries  in Fig. \ref{fig:ETW brane interpretation} at once.

\subsection{The Hartle-Hawking state}\label{ssec:path integral HH}
We would like to learn about dynamical aspects of the boundary theory after gauging symmetries in the left/right chord sectors. To do this, we consider the HH state for the ASC Hamiltonian (\ref{eq:ASC Hamiltonian}), which we define as 
\begin{equation}\label{eq:HH ETW state}
\ket{\Psi(\tau)}=\rme^{-\tau\hH_{\rm ASC}}\ket{H_0}  ~,  
\end{equation}
where $\tau=\frac{\beta}{2}+\rmi t$. We can therefore construct the corresponding partition function from (\ref{eq:HH ETW state}), which we evaluate in App. \ref{app:ETW brane partition function}.

(\ref{eq:HH ETW state}) allows us to construct the Heisenberg picture rescaled chord number $\ell\equiv\lambda\hat{n}$, and momentum conjugate operators that similarly commute with the constraint operators in (\ref{eq:further constraints}), i.e.
\begin{equation}
    \hat{\ell}(\tau)=\rme^{-\tau^*\hH_{\rm ASC}}\hat{\ell}\rme^{-\tau\hH_{\rm ASC}}~,\quad \hat{P}(\tau)=\rme^{-\tau^*\hH_{\rm ASC}}\hat{P}\rme^{-\tau\hH_{\rm ASC}}~,
\end{equation}
so that expectation values agree with the Schrodinger picture, i.e. 
\begin{equation}
\bra{\Psi(\tau)} \hat{\ell}\ket{\Psi(\tau)}=\bra{H_0} \hat{\ell}(\tau)\ket{H_0}~.    
\end{equation}
These operators obey the usual Heisenberg equation 
\begin{equation}\label{eq:heisenberg 1}
    \partial_t\hat{\ell}(\tau)=\rmi [\hH_{\rm ASC},\hat{\ell}(\tau)]~,\quad \partial_t\hat{P}(\tau)=\rmi [\hH_{\rm ASC},\hat{P}(\tau)]~.
\end{equation}
Next, by setting $\tau=\beta/2+\rmi t$, we can find the initial conditions for the canonical coordinate
\begin{subequations}\label{eq:quantum initial cond}
    \begin{align}
            &\bra{H_0} \hat{\ell}(\tfrac{\beta}{2})\ket{H_0}=\ell_*~,\\
    \partial_t&\eval{\bra{H_0} \hat{\ell}\qty(\tfrac{\beta}{2}+\rmi t)\ket{H_0}}_{t=0}=0~.\label{eq:last ETW in cond}
    \end{align}
\end{subequations}
\paragraph{Proof of \eqref{eq:last ETW in cond}}Note from \eqref{eq:heisenberg 1} that
\begin{align}
-\rmi\partial_t\eval{\bra{H_0} \hat{\ell}\qty(\tfrac{\beta}{2}+\rmi t)\ket{H_0}}_{t=0}&=\lambda\bra{H_0}\rme^{-\frac{\beta}{2}\hH_{\rm ASC}}[\hH_{\rm ASC},~\hat{n}]\rme^{-\frac{\beta}{2}\hH_{\rm ASC}}\ket{H_0}~,\\
=\lambda\bigg(\prod_{i=1}^2\int_{\theta_{i}=0}^{\theta_{i}=\pi}\rmd\theta_i~\tilde{\mu}(\theta_i)\bigg)&E(\theta_1)\rme^{-\frac{\beta}{2}(E(\theta_1)+E(\theta_2))}\qty(\bra{\theta_1}\hat{n}\ket{\theta_2}-\bra{\theta_2}\hat{n}\ket{\theta_1})~,
\end{align}
where we applied the completeness relation \eqref{eq:completeness relation}, and $E(\theta)$ appears in \eqref{eq:energy spectrum}. Therefore (\ref{eq:last ETW in cond}) follows since we can express $\hat{n}=\sum_{n=0}^\infty n\ket{H_n}\bra{H_n}$ (\ref{eq:number op ASC}).

\subsection{Saddle points in the path integral}\label{ssec:saddle points}
We will now evaluate the classical phase space solutions for $\rme^{-\ell}$.\footnote{In fact, one could proceed differently, by first computing a geodesic distance that interpolates between an AdS$_2$ asymptotic boundary and the ETW brane, and then constructing a bulk Hamiltonian of the form (\ref{eq:ASC Hamiltonian}) from JT gravity \cite{Gao:2021uro} or sine dilaton gravity \cite{Blommaert:2025avl}. However, we will derive the result in a different (but equivalent way), since our starting point is the DSSYK model with constraints (instead of AdS$_2$ bulk geodesics).}
We consider a path integral for the constrained theory (\ref{eq:ASC Hamiltonian}) from:
\begin{equation}\label{eq:path int ASC}
    \int[\rmd P][\rmd \ell]\exp[\int\rmd\tau\qty(\frac{1}{\lambda}\rmi P\dv{\ell}{\tau}-H_{\rm ASC})]~.
\end{equation}
Here, $H_{\rm ASC}$ can be interpreted as the bulk Hamiltonian dual to the DSSYK after imposing constraints in the chord space $\mathcal{H}_1$ (which we derived in Sec. \ref{sec:ASC from DSSYK}). The saddle point equations become:
\begin{align}\label{eq:ell ETW 1}
    \frac{\lambda}{1-q}&\dv{\ell}{\tilde{t}}=\rmi\qty(2\cos\theta-2\rme^{-\rmi P}-(X+Y)\rme^{-\ell})~,\\
    &\dv{P}{\tilde{t}}=2\qty(\cos\theta-\cos P)+XY\rme^{\rmi P}\rme^{-2\ell}~,\label{eq:momentum eq ETW}
\end{align}
with $\tilde{t}=J\sqrt{\frac{\lambda}{1-q}}t$. Due to (\ref{eq:quantum initial cond}), we consider the following initial condition
\begin{equation}\label{eq:initial cond ETW}
    \ell(t=0)=\ell_{*}~,\quad \eval{\dv{\ell}{t}}_{t=0}=0~,
\end{equation}
corresponding to a HH preparation of state. (\ref{eq:ell ETW 1}) together with (\ref{eq:initial cond ETW}) imply 
\begin{equation}
    \rme^{-\rmi P(t=0)}=\cos\theta-\frac{X+Y}{2}\rme^{-\ell_{*}}~.
\end{equation}
Thus, the energy conservation relation for (\ref{eq:ASC H2}) can be expressed as
\begin{equation}\label{eq:initial length ETW}
    \qty(\cos\theta-\frac{X+Y}{2}\rme^{-\ell_{*}})^2=\qty(1-XY\rme^{-\ell_{*}})\qty(1-\rme^{-\ell_{*}})~.
\end{equation}
Since (\ref{eq:ASC H2}) has a conserved energy $\frac{2J}{\sqrt{\lambda(1-q)}}\cos\theta$, we can extract the canonical momentum as
\begin{equation}
    \rme^{-\rmi P(t)}=\cos\theta-\frac{X+Y}{2}\rme^{-\ell(t)}\pm\sqrt{\qty(\frac{X+Y}{2}\rme^{-\ell(t)}-\cos\theta)^2-(1-\rme^{-\ell})\qty(1-{XY}\rme^{-\ell(t)})}~.
\end{equation}
We now solve the EOM (\ref{eq:ell ETW 1}, \ref{eq:momentum eq ETW}) with the initial condition (\ref{eq:initial cond ETW}),\footnote{To solve the differential equation, it is useful to make a change of variables $\ell(t)=\log u(t)$.} resulting in:
\begin{equation}\label{eq:womrhole XY}
    \begin{aligned}
        \ell(t)=\log\biggl(\frac{ 1+X
   Y-(X+Y)\cos
   \theta+\cosh(2Jt\sin\theta)\prod_{X_i=X,Y}\sqrt{X_i^2-2 X_i
   \cos \theta +1}}{2 \sin^2\theta}\biggr)~,
    \end{aligned}
\end{equation}
where we have substituted $\ell_{*}$ from the relation (\ref{eq:initial length ETW}). 

\subsection{Disk thermal two-point functions}\label{ssec:correlation functions}
We can now derive the two-point correlation functions in the constrained model from the two-sided two-point functions in the DSSYK model (see Fig. \ref{fig:brane_from_DSSYK} for an illustration). The latter case has been worked out (differently) in \cite{Berkooz:2022fso,Aguilar-Gutierrez:2025pqp}. We remark that higher correlation functions proceed analogously, especially those relevant for deriving the switchback effect \cite{Aguilar-Gutierrez:2025mxf}; however, we will focus on the first non-trivial case in this work. The path integral (\ref{eq:partition function Faddeev form}) that evaluates the two-sided two-point correlation function in the HH state is given by
\begin{equation}\label{eq:two-point constrained}
    \frac{\bra{\Psi_{\Delta}(\tau_L,\tau_R)}q^{\Delta\hat{N}}\ket{\Psi_{\Delta}(\tau_L,\tau_R)}}{Z_\Delta(\beta_L,\beta_R)}=\int\prod_{i=L,R}[\rmd \ell_i][\rmd P_i]\rme^{-\Delta(\ell_L+\ell_R)}\rme^{I_E[\qty{\ell_i,~P_i}]}~,
\end{equation}
where $\lambda\hat{N}=-\hat{\ell}_L+\hat{\ell}_R$ is the total chord number, and the action is still given by (\ref{eq:path E action ETW}). Diagrammatically, the correlation functions in the bulk can be seen as shooting a probe particle between the asymptotic boundary of the bulk to the ETW brane, see Fig. \ref{fig:ETW brane interpretation}. From the boundary side it can be seen as in Fig. \ref{fig:brane_from_DSSYK} adding a probe correlator between the thermal circle and the particle.  

We can also immediately evaluate the classical limit of the thermal two-point correlation function in this system
\begin{align}\label{eq:correlator XY}
    &\eval{G^{(\Delta_w)}_{\rm disk}(\tau)}_{\tau=\frac{\beta(\theta)}{2}+\rmi t}=\frac{\bra{H_0}\rme^{-\qty(\frac{\beta(\theta)}{2}-\rmi t)\hH_{\rm ASC}}q^{\Delta\hat{n}}\rme^{-\qty(\frac{\beta(\theta)}{2}+\rmi t)\hH_{\rm ASC}}\ket{H_0}}{\bra{H_0}\rme^{-\beta(\theta)\hH_{\rm ASC}}\ket{H_0}}\\
&\eqlambda\qty(\frac{2\sin^2\theta}{1+XY-(X+Y)\cos\theta+\cosh(2J\sin\theta t)\prod_{X_i=X,Y}\sqrt{X_i^2-2X_i\cos\theta+1}})^{\Delta}~,\nonumber
\end{align}
where we took the expectation value with respect to the state (\ref{eq:HH ETW}). Note that the correlation function above also depends on $q^{\Delta}$ of the DSSYK operator insertion through $X$ an $Y$, following from (\ref{eq:XY 1s}, \ref{eq:Z2 sym condition}).

As mentioned in \cite{Aguilar-Gutierrez:2025pqp}, the $\lambda\rightarrow0$ limit corresponds to the semiclassical thermal two-point function for a given chord space $\mathcal{H}_m$ (with fixed total chord number). We could perform a chord evaluation that matches (\ref{eq:correlator XY}). However, (\ref{eq:correlator XY}) requires strong backreaction (i.e. $q^{\Delta}$ being finite as $\lambda\rightarrow0$). We currently lack the proper ansatz to confirm (\ref{eq:correlator XY}) from chord diagrams. We leave it for future directions. The reader can find more details on the thermodynamics from the ETW partition function in App. \ref{app:ETW brane partition function}.

\subsection{Spread complexity}\label{ssec:spread}
While the non-Hermitian representation of the constrained Hamiltonian (\ref{eq:ASC Hamiltonian}) is useful for path integral computations, we can also express in a symmetric form, which is {{more convenient for studying}} spread complexity. This follows from a shift in the canonical momentum
\begin{equation}\label{eq:momentum new}
\begin{aligned}
\rme^{\rmi \hat{P}}&\rightarrow \rme^{\rmi \hat{P}}\qty((1-XY q^{\hat{n}-1})(1-q^{\hat{n}}))^{-1/2}~,\\
\rme^{-\rmi \hat{P}}&\rightarrow\sqrt{(1-XY q^{\hat{n}-1})(1-q^{\hat{n}})}\rme^{-\rmi \hat{P}}~,
\end{aligned}
\end{equation}
so that the ASC Hamiltonian (\ref{eq:ASC Hamiltonian}) in the symmetric form becomes
\begin{equation}\label{eq:ETW brane H symmetrized}
\begin{aligned}
    \hH_{\rm ASC}\rightarrow\hH^{(\rm sym)}_{\rm ASC}=\frac{J}{\sqrt{\lambda(1-q)}}\biggl(&\rme^{\rmi\hat{P}}\sqrt{(1-XY q^{\hat{n}-1})(1-q^{\hat{n}})}\\
    &+\sqrt{(1-XY q^{\hat{n}-1})(1-q^{\hat{n}})}\rme^{-\rmi\hat{P}}+(X+Y)q^{\hat{n}}\biggr)~.
\end{aligned}
\end{equation}
The new form of the Hamiltonian amounts to a transformation of the basis $\qty{\ket{H_n}}$ into an orthonormal one $\qty{\ket{K_n}}$:
\begin{equation}\label{eq:Krylov basis ETW brane sym}
    \braket{\theta}{K_n}=\frac{Q_n(\cos\theta|X,Y;q)}{\sqrt{(q,XY;q)_n}}~,
\end{equation}
where the normalization in (\ref{eq:Krylov basis ETW brane sym}) follows from (\ref{eq:ortho relations ASC}), and $\ket{K_0}=\ket{H_0}$. In contrast to (\ref{eq:momentum new}), the momentum operator and the number operator act as
\begin{equation}\label{eq:shift in momentum canonical}
\begin{aligned}
    &\rme^{-\rmi\hat{P}}\ket{K_n}=\sqrt{[n+ 1]_q}\ket{K_{n+1}}~,\quad \rme^{-\rmi\hat{P}}\ket{K_n}=\sqrt{[n]_q}\ket{K_{n-1}}~,\\
&\hat{n}\ket{K_n}=n\ket{K_n}.
\end{aligned}
\end{equation}
Density matrices decomposed in the basis $\qty{\ket{H_n}}$ (\ref{eq:number op ASC}) are not altered by the change basis to $\qty{\ket{K_n}}$ (\ref{eq:Krylov basis ETW brane sym}), as seen from
\begin{equation}\label{eq:density element n}
\sum_nc_n\ket{H_n}\bra{H_n}=\sum_nc_n\ket{K_n}\bra{K_n}~,
\end{equation}
where $c_n$ is an arbitrary sequence. For instance, (\ref{eq:density element n}) becomes the identity operator when $c_n=1(\forall n\in\mathbb{Z}_{\geq0})$, and powers of the number operator, $\hat{n}^k$, when $c_n=n^k$ (\ref{eq:density element n}).

We will now work out the Krylov basis, Lanczos coefficients and spread complexity, which we can late match with a bulk minimal length geodesic.

From (\ref{eq:ETW brane H symmetrized}), we also deduce that, if we take the HH state (with $\tau=\frac{\beta}{2}+\rmi t$)
\begin{equation}\label{eq:HH ETW}
\ket{\Psi(\tau)}=\rme^{-\tau\hH_{\rm ASC}}\ket{K_0}=\sum_n\Psi_n(\tau)\ket{K_n}~,
\end{equation}
where $\Psi_n(\tau)=\bra{K_n}\ket{\Psi(\tau)}$. Here, the Krylov basis above $\qty{\ket{K_n}}$ are defined \cite{Balasubramanian:2022tpr}
\begin{equation}\label{eq:lanczos alg}
    \begin{aligned}
	&\ket{A_{n+1}}\equiv (\hH - a_n)\ket{K_n} - b_n \ket{K_{n-1}}~,\\
&\ket{K_n} \equiv b_n^{-1}\ket{A_n}~.
\end{aligned}
\end{equation}
Then, the amplitudes $\Psi_n(\tau)$ obey
\begin{equation}\label{eq:Lanczos ETW sym}
-\partial_\tau\Psi_n(\tau)=b_n \Psi_{n-1}(\tau)+b_{n+1}\Psi_{n+1}(\tau)+a_n\Psi_n(\tau)~,
\end{equation}
where the Lanczos coefficients from (\ref{eq:ETW brane H symmetrized}) are
\begin{equation}\label{eq:Lanczos ETW XY form}
\boxed{a_n=\frac{J}{\sqrt{\lambda(1-q)}}(X+Y)q^n~,\quad b_n=\frac{J}{\sqrt{\lambda}}\sqrt{(1-XYq^{n-1})[n]_q}~.}
\end{equation}
In particular:
\begin{itemize}
\item \textbf{One-sided branes}: When $Y=q^{\Delta}$, $X=0$,
\begin{equation}\label{eq:DSSYK with ETWB Hamiltonian}
a_n=\frac{J}{\sqrt{\lambda(1-q)}}q^{\Delta+n}~,\quad b_n=\frac{J}{\sqrt{\lambda}}\sqrt{[n]_q}~.
\end{equation}
Our results are in exact agreement with the Hermitian version of the ETW brane Hamiltonian (4.5) of \cite{Okuyama:2023byh}.

\item \textbf{Two-sided branes}: When $X=-Y=\rmi q^{(\Delta+1)/2}$, the Lanczos coefficients (\ref{eq:Lanczos ETW XY form}) become
\begin{equation}\label{eq:an bn Z2 case}
a_n=0~,\quad b_n=J\sqrt{\frac{1-q}{\lambda}}\sqrt{[n+\Delta]_q[n]_q}~.
\end{equation}
Note that the result is consistent with general arguments that quantum systems with a symmetric density of states should have $a_n=0$ when the initial state in the Lanczos algorithm {{is the}} infinite temperature {{HH}} state \cite{Nandy:2024zcd,Balasubramanian:2022dnj}. It is also interesting to {{compare our results with}} the Lanczos coefficients {{of the zero-particle chord Hilbert space}} (see e.g. \cite{Rabinovici:2023yex,Aguilar-Gutierrez:2025pqp}) {{which are now}} modified by a product of q-numbers, $[n+\Delta]_q[n]_q$ in (\ref{eq:an bn Z2 case}) due to the particle insertion. 
\end{itemize}
It is clear from  our results so far that the number operator $\hat{n}$ in (\ref{eq:shift in momentum canonical}) corresponds to the Krylov complexity operator 
\begin{equation}
    \hat{\mathcal{C}}\equiv\sum_n n\ket{K_n}\bra{K_n}~,
\end{equation}
for the reference state $\ket{\Psi(\tau)}$ in (\ref{eq:HH ETW}) since we have identified the Krylov basis (\ref{eq:Krylov basis ETW brane sym}), and the corresponding Lanczos algorithm (\ref{eq:Lanczos ETW sym}). It follows from the semiclassical expectation value of the chord number in the Hartle-Hawking state (\ref{eq:womrhole XY}) that the spread complexity is given by
\begin{equation}\label{eq:womrhole XY2}
    \begin{aligned}
        \mathcal{C}(t)=\frac{1}{\lambda}\log\biggl(\tfrac{ 1+X
   Y-(X+Y)\cos
   \theta}{2 \sin^2\theta}+\tfrac{\prod_{X_i=X,Y}\sqrt{X_i^2-2 X_i
   \cos \theta +1}}{2\sin^2\theta}\cosh(2J\sin(\theta)~t)\biggr)~.
    \end{aligned}
\end{equation}
Half of the spread and Krylov operator complexity for the two-sided HH state in \cite{Aguilar-Gutierrez:2025pqp} without constraints have exactly the same time dependence as (\ref{eq:womrhole XY}), although the overall coefficients are different. We provide further comments on this point in Sec. \ref{ssec:outlook}.

We now determine the spread complexity (\ref{eq:womrhole XY}) corresponding to the one-sided and two-sided symmetric ETW brane configurations
\begin{align}\label{eq:length ETW brane}
    &\mathcal{C}^{(1s)}(t)=\frac{1}{\lambda}\log(\frac{1-q^{\Delta}\cos\theta+\sqrt{1+q^{\Delta}(q^{\Delta}-2\cos\theta)}\cosh(2J\sin\theta~t)}{2\sin^2\theta})~,\\
    &\mathcal{C}^{(2s)}(t)=\frac{1}{\lambda}\log(\frac{1+q^{\Delta+1}+\cosh(2J\sin\theta t)\sqrt{(1-q^{\Delta+1})^2+4q^{\Delta+1}\cos^2\theta}}{2\sin^2\theta})~.\label{eq:length Z2 wormhole length}
\end{align}
In both cases the spread complexity experiences an early time parabolic and late time linear growth, similar to the spread complexity of the two-sided HH state in \cite{Aguilar-Gutierrez:2025pqp} evolving through the two-sided DSSYK Hamiltonian (\ref{eq:pair DSSYK Hamiltonians 1 particle}). Neither of these examples displays the scrambling features of the Krylov operator complexity \cite{Aguilar-Gutierrez:2025pqp,Ambrosini:2024sre}, which instead experiences a transition from exponential growth before the scrambling time. 

\subsection{The holographic dictionary, and the physical bulk Hilbert space}\label{ssec:spread bulk picture from ETW}

We can also translate (\ref{eq:womrhole XY}) in the more familiar language of JT gravity,
\begin{equation}\label{eq:total JT action ETW}
    I_{\rm total}=I_{\rm JT}+\frac{1}{\kappa^2}\int_{\mathcal{S}_{\rm ETW}}(\Phi_bK-m_{\rm ETW})~,
\end{equation}
where $I_{\rm JT}$ is the JT gravity action 
\begin{equation}
\begin{aligned}
    I_{\rm JT}=&-\frac{\Phi_0}{2\kappa^2}\qty(\int_{\mathcal{M}}\sqrt{g}\mathcal{R}+2\int_{\partial\mathcal{M}}\sqrt{h}K)\\
    &-\frac{1}{2\kappa^2}\qty(\int_{\mathcal{M}}\sqrt{g}\Phi(\mathcal{R}+2)+2\int_{\partial\mathcal{M}}\sqrt{h}\Phi_b(K-1))~,\label{eq:JT gravity}    
\end{aligned}
\end{equation}
while $\kappa^2=8\pi G_N$, $\Phi_0$ is a constant, $\Phi_b=\eval{\Phi}_{\partial\mathcal{M}}$, $\mathcal{S}_{\rm ETW}$ is the worldvolume of the ETW brane, $m_{\rm ETW}$ corresponds to the ETW brane mass (equivalent to its tension since it is a one-dimensional), and we denote $K_{\rm ETW}$ the extrinsic curvature at $\mathcal{S}_{\rm ETW}$. The geodesic distance connecting an asymptotic AdS boundary to the ETW brane in JT gravity is found as \cite{Blommaert:2025avl,Gao:2021uro},
\begin{equation}\label{eq:length ETW JT gravity}
    L_{\rm ETW}(u)=\log\qty(\frac{m_{\rm ETW}}{\Phi_h^2\sqrt{1-K_{\rm ETW}^2}}+\sqrt{1+\frac{m^2_{\rm AdS}}{\Phi_h^2{\qty(1-K_{\rm ETW}^2)}}}\frac{\cosh(\Phi_h u)}{\Phi_h})+L_{\rm reg}~,
\end{equation}
where, using the coordinates in Lorentzian signature, we consider AdS$_2$ black holes
\begin{equation}\label{eq:effective geometry}
\begin{aligned}
    &\rmd s_{\rm AdS}^2=-f(r)\rmd u^2+\frac{\rmd r^2}{f(r)}~,\\
    &f(r)=r^2-\qty(\Phi_h)^2~.
\end{aligned}
\end{equation}
Here $r=\Phi_h={2\pi}/\beta_{\rm AdS}$ is the location of the event horizon; $u_L=u_R\equiv u$ is the AdS$_2$ asymptotic boundary time; and $L_{\rm reg}$ is a scheme dependent regulator.

Furthermore, it was shown in \cite{Blommaert:2025avl} that (\ref{eq:length ETW JT gravity}) also corresponds to a wormhole distance in sine dilaton gravity with ETW branes, which is described by
\begin{equation}\label{eq:SDG with ETW brane}
\begin{aligned}
    I_{\rm SDG}=-\frac{1}{2\kappa^2}\bigg(&\int_{\mathcal{M}}\rmd^2x\sqrt{g}\qty(\Phi\mathcal{R}+2\sin\Phi)+2\int_{\partial\mathcal{M}}\rmd x\sqrt{h}\qty(\Phi_{b} K)\\
    &-\zeta\int_{\rm brane}\sqrt{h}~\rme^{-\rmi\Phi_b}-\overline{\zeta}\int_{\rm brane}\sqrt{h}~\rme^{\rmi\Phi_b}\bigg)~,
\end{aligned}
\end{equation}
where $\zeta$, $\overline{\zeta}$ are constants that can be mapped to the parameters in (\ref{eq:length ETW JT gravity}). 

\begin{figure}
    \centering
    \subfloat[One-sided ETW brane]{\includegraphics[width= 0.30\textwidth]{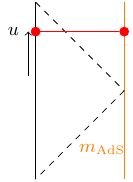}}\hfill\subfloat[Two-sided ETW brane]{\includegraphics[width= 0.59\textwidth]{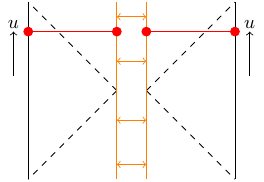}}
    \caption{Bulk dual of the DSSYK  model with constraints: (a) a one-sided ETW brane and (b) two-sided symmetric ETW brane (glued together through the orange arrows). An AdS$_2$ black hole background has an ETW brane (orange) with tension $m_{\rm ETW}$ (\ref{eq:dictionary 3}), dual to a heavy operator $\hat{\mathcal{O}}_\Delta$. The red line denotes a minimal length geodesic curve (which defines the distance $L_{\rm ETW}=\lambda~\mathcal{C}(t)$ (\ref{eq:dictionary 1})) starting the asymptotic boundary at time $u$ in the bulk coordinates (\ref{eq:effective geometry}) (with $\tau=\rmi u+\beta/2$) and reaching the ETW brane.}
    \label{fig:ETW brane interpretation}
\end{figure}

\paragraph{Holographic dictionary}We can now deduce a dictionary between sine dilaton gravity with the ETW branes and the DSSYK model by equating the AdS$_2$ wormhole distance (\ref{eq:length ETW JT gravity}) with the canonical variable $\ell$ (\ref{eq:womrhole XY}) (corresponding to spread complexity in the DSSYK side), as follows
\begin{subequations}
    \begin{align}
&\lambda~\mathcal{C}(t)=L_{\rm ETW}(u)~,\quad  \Phi_hu={2J\sin\theta~t}~,\label{eq:dictionary 1}\\
&L_{\rm reg}=\log\frac{\sqrt{\prod_{X_i=X,Y}(X_i^2-2X_i\cos\theta+1)-(1+XY-(X+Y)\cos\theta)^2}}{2\sin\theta}~,\label{eq:dictionary 2}\\
    &\frac{m_{\rm ETW}}{\sqrt{1-K_{\rm ETW}^2}}=\frac{(1+XY-(X+Y)\cos\theta)\sin\theta}{\sqrt{\prod_{X_i=X,Y}(X_i^2-2X_i\cos\theta+1)-(1+XY-(X+Y)\cos\theta)^2}}~.\label{eq:dictionary 3}
\end{align}
\end{subequations}
Therefore, the spread complexity of the DSSYK model after imposing constraints (\ref{eq:womrhole XY}) can be matched to a wormhole distance in the bulk (similar observation as other parts of the literature \cite{Rabinovici:2023yex,Xu:2024gfm,Ambrosini:2024sre,Heller:2024ldz}), where the matter chord corresponds to an ETW brane in the bulk, which in this case ends on an ETW brane. Our bulk interpretation is displayed in Fig. \ref{fig:ETW brane interpretation}. Table \ref{tab:holographic_dictionary} summarizes these results.

\paragraph{Physical bulk Hilbert space}
We now derive the bulk Hilbert space of the DSSYK with a given symmetry \eqref{eq:LR transformation}. We perform Dirac quantization in the bulk {{Hilbert space, which satisfies the chord symmetry constraints (\ref{eq:constraints Okuyama}, \ref{eq:constraints Xu case}), and they}} are further selected to also obey a Schrodinger equation with the same energy spectrum as the boundary theory (\ref{eq:energy spectrum}) and a momentum shift symmetry\footnote{The authors in \cite{Blommaert:2024whf} denote this as gauging the momentum shift symmetry. From the Dirac formulation, it is a second-class constraint as $[\hH_{\rm ASC},~{\sum_{-\infty}^{\infty}\rme^{\rmi m \hat{n}}}]\neq0$.}; this means:
\begin{equation}\label{eq:further constraints}
    \hH_{\rm ASC}^{\rm bulk}\ket{\psi(\theta)}=E(\theta)\ket{\psi(\theta)}~,\quad \sum_{m=-\infty}^{\infty}\qty(\rme^{\rmi \frac{2\pi m}{\lambda}\hat{L}_{L/R}^{\rm AdS}}-\mo)\ket{\psi(\theta)}=0~,
\end{equation}
where
\begin{equation}
    \hat{L}_{L/R}^{\rm AdS}=\lambda~\hat{n}~,
\end{equation}
is an entry in the holographic dictionary relating minimal geodesic lengths in the bulk (\ref{eq:length ETW brane}), and the chord number in the boundary theory (see Fig. \ref{fig:ETW brane interpretation}). The first constraint requires that the energy spectrum of the bulk and the boundary theory are the same {{which is motivated by \cite{Blommaert:2025avl}}}, namely
\begin{equation}\label{eq:energy spectrum}
    E(\theta)=\frac{2J}{\sqrt{\lambda(1-q)}}\cos\theta~,
\end{equation} 
while the second one requires that bulk geodesic lengths are discrete.

Solving the bulk constraints (\ref{eq:further constraints}) allowed \cite{Blommaert:2025avl} to find a Hamiltonian with the same form as (\ref{eq:ASC Hamiltonian}). Our work shows that the bulk model they solved is in fact dual to the DSSYK model with matter with either of the symmetry relations between left/right chord sectors in (\ref{eq:constraints Okuyama}) and (\ref{eq:constraints Xu case}).

\begin{figure}
    \centering
    \subfloat[]{\includegraphics[width=0.46\linewidth]{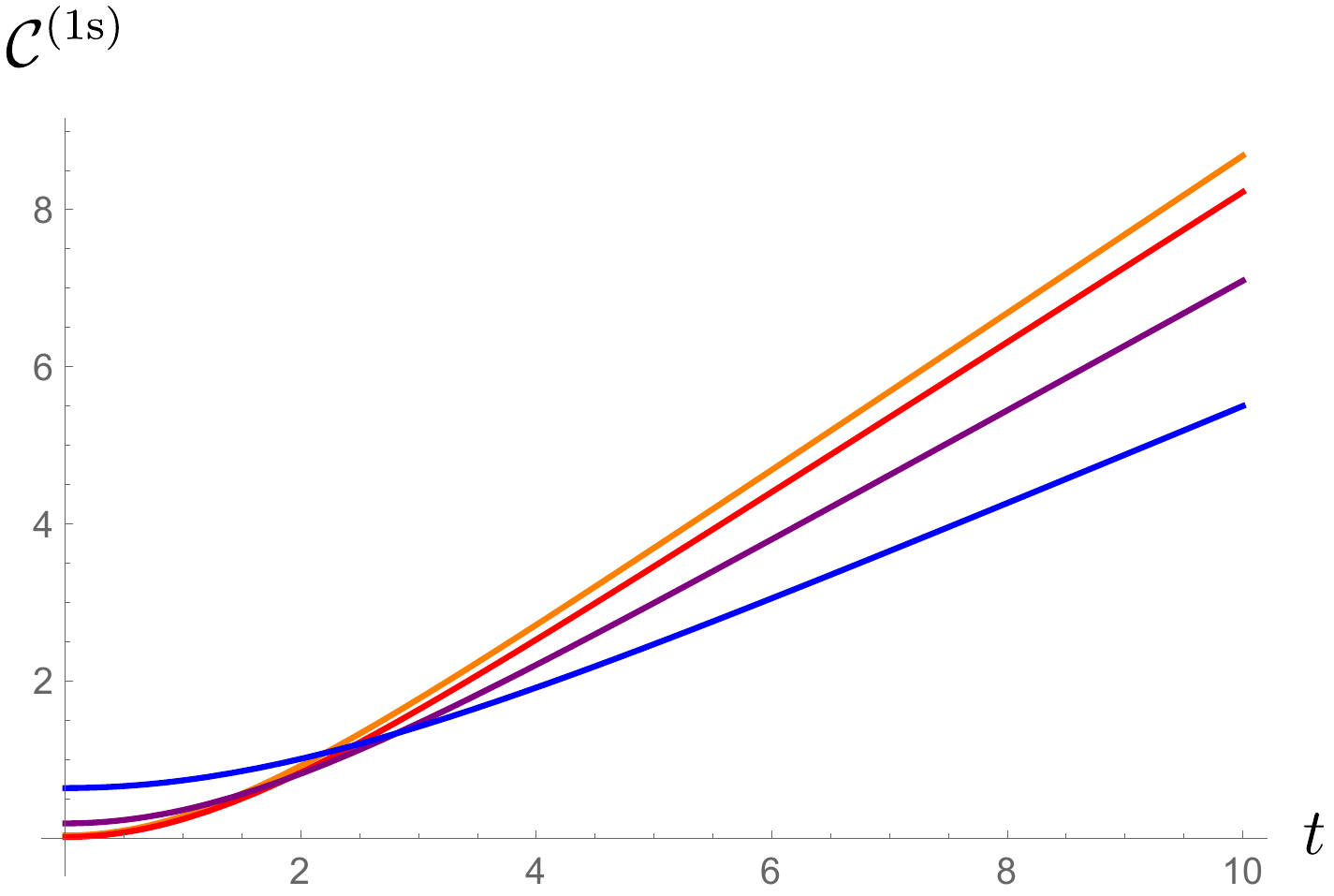}}\hfill\subfloat[]{\includegraphics[width=0.46\linewidth]{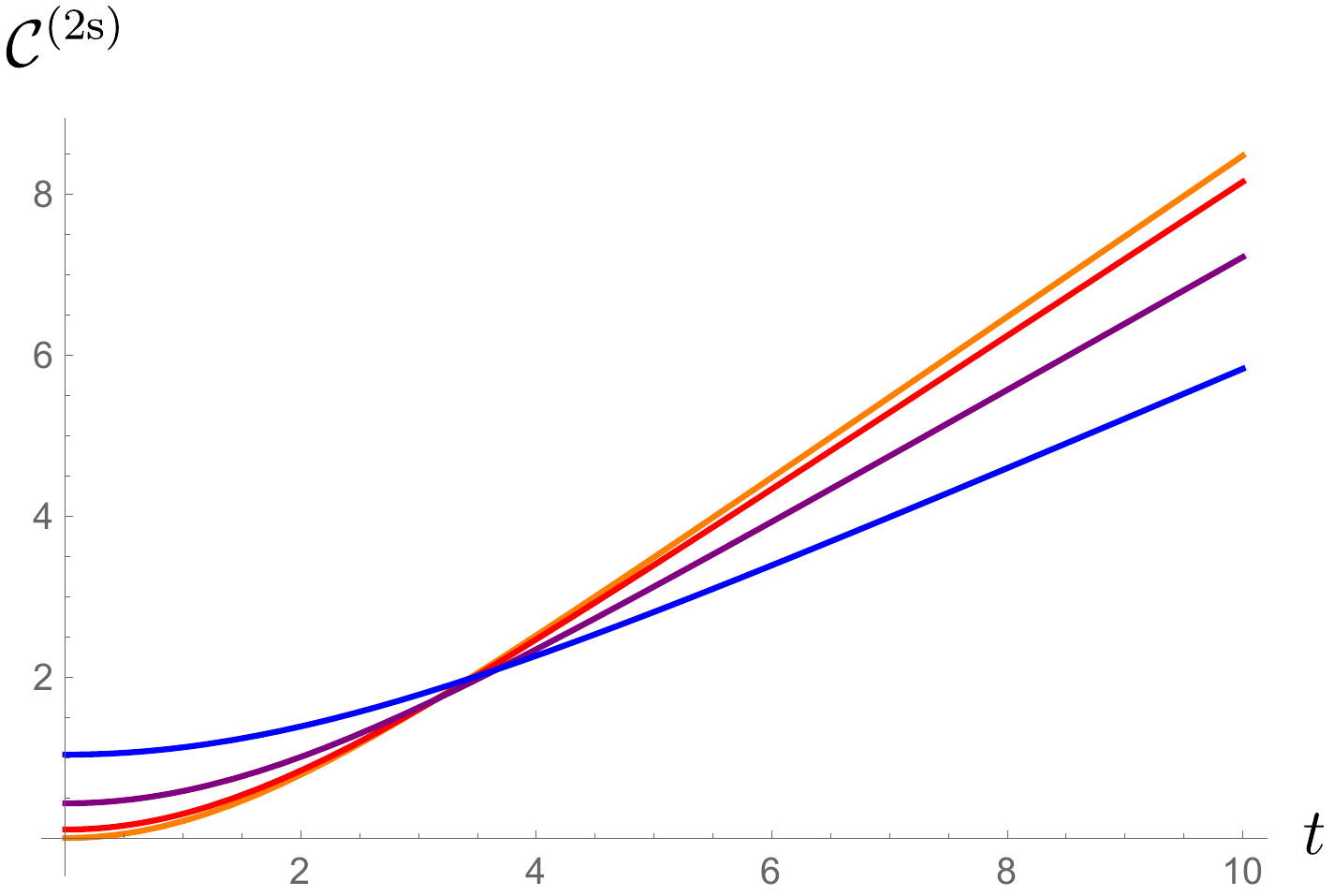}}
    \caption{Comparison plot between the spread complexity of the (a) one-sided brane system (\ref{eq:length ETW brane}), and (b) a two-sided symmetric brane system (\ref{eq:length Z2 wormhole length}). We are taking $\lambda=0.001$, $\lambda\Delta=1$, $J=1$, and $\theta$ decreases from $\theta=\pi/2$ (orange) in multiples of $0.3$ from top to bottom at late times. In both cases, there is a early time parabolic growth of Krylov complexity, and late time growth. Note the different rates of growth depending on the energy $E(\theta)$ of the solution.}
    \label{fig:plot_Krylov_complexities}
\end{figure}

\paragraph{Physical interpretation} 
The special cases investigated in (\ref{eq:length ETW brane}) and (\ref{eq:length Z2 wormhole length}) have a clear interpretation in the bulk as the Lorentzian wormhole distance intercepting with an ETW brane (Fig. \ref{fig:ETW brane interpretation}). We contrast the growth of spread complexity in unconstrained DSSYK model with the ETW brane systems in Fig. \ref{fig:plot_Krylov_complexities}.

We can also compare the holographic dictionary for the two types of ETW brane models. Since the regularization variable $L_{\rm reg}$ in (\ref{eq:dictionary 2}) is just a scheme dependent constant, we will be mostly interested in ratio between the brane tensions and extrinsic curvature on the ETW brane (\ref{eq:dictionary 3})
\begin{equation}
    {m_{\rm ETW}}=\begin{cases}
        (q^{-\Delta}-1)\sqrt{1-K_{\rm ETW}^2}~,&\text{for }Y=q^{\Delta},~~X=0~,\\
        \frac{1}{2}\qty(q^{\frac{{\Delta}+1}{2}}+q^{-\frac{{\Delta}+1}{2}})\sqrt{K_{\rm ETW}^2-1}~,&\text{for }X=-Y=\rmi q^{\frac{\Delta+1}{2}}~,
    \end{cases}
\end{equation}
We notice that for the tension to be real, there are bounds on the extrinsic curvature ($K_{\rm ETW}\leq1$ or $K_{\rm ETW}\geq1$ respectively), which acts as an arbitrary constant parameter in JT gravity \cite{Blommaert:2025avl}.

\section{Euclidean wormholes, perturbative stability and baby universes}\label{sec:wormhole topology}
In this section, we {{define}} Euclidean wormhole partition functions {{in the boundary theory}} and {{deduce a corresponding cylinder two-point}} correlation function {{by implementing}} constraints in the chord space (\ref{eq:pair DSSYK Hamiltonians 1 particle}). This essentially follows from connecting ETW branes in an appropriate way that changes the bulk topology, as seen in JT gravity \cite{Gao:2021uro} or sine dilaton gravity \cite{Blommaert:2025avl}. Our goal is to {{study}} these geometries from the boundary side, importantly, without making assumptions about generating a finite N completion of the DSSYK model through the ETH matrix model \cite{Jafferis:2022uhu,Jafferis:2022wez,Okuyama:2023yat}. {{However, we remark this section lacks a purely boundary interpretation of the wormhole partition functions since they are defined based on the corresponding bulk observable, translated in DSSYK terms. There are specific problems that need to be addressed to have a well-defined boundary interpretation of the wormhole observables in this section, which we elaborate on Sec. \ref{ssec:outlook}.}} 

Our results indicate that, even thought the DSSYK lives in a disk topology itself, the chord diagram Hamiltonians with constraints can be used to build Euclidean wormholes in the bulk. We specifically derive the baby universe Hilbert space associated with multi-boundary wormholes and study their perturbative stability. We will find that this simple one-dimensional model has a one-dimensional baby universe Hilbert space; and that the Euclidean wormhole is stable. This means that it mimics specific properties in higher dimensions, which could help us understand them better in this explicitly solvable case.

\paragraph{Outline} In Sec. \ref{ssec:partition functions wormholes} we derive the half-wormhole, trumpet and double trumpet partition functions from our boundary formulation of the ETW brane Hamiltonians. In Sec. \ref{ssec:two point Euclidean wormhole} we also deduce thermal two-point correlation functions in a cylinder topology from constraints in two-sided two-point functions in the one-particle space of the DSSYK model. Then, in Sec. \ref{ssec:stability} we show that the Euclidean wormhole with matter correlators is perturbatively stable. Later, in Sec. \ref{ssec:baby universes} we derive the multi-boundary partition functions in this system, and study the baby universe Hilbert space of the Euclidean wormholes. We show the Hilbert space is one-dimensional.

\subsection{Wormhole partition functions}\label{ssec:partition functions wormholes}
We will now {{present}} the trumpet and multi-boundary partition functions {{in \cite{Blommaert:2025avl}} in terms of} the DSSYK model.

\begin{figure}
\centering
    \subfloat[]{\includegraphics[height=0.33\textwidth,valign=t]{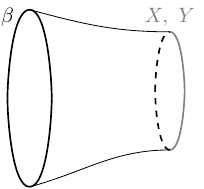}}\hfill\subfloat[]{\includegraphics[height=0.3\textwidth,valign=t]{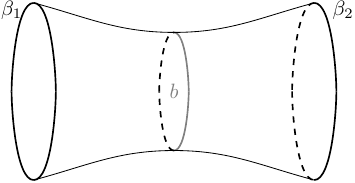}}
    \caption{(a) Half wormhole wavefunction (\ref{eq:half wormhole ETW}), and (b) double trumpet partition function (\ref{eq:double trumpet}). $b$ is the circumference of the curve in the boundary, representing the origin of the constrained Hamiltonian (\ref{eq:ASC Hamiltonian}). The boundary parameters $X$ and $Y$ are shown in (\ref{eq:XY 1s}) and (\ref{eq:Z2 sym condition}) for the one and two-sided brane models respectively.}
    \label{fig:trumpets}
\end{figure}

Using the orthogonality relation (\ref{eq:half wormhole ETW}), we can define a partition function on the constrained system, previously presented from a bulk analysis in \cite{Blommaert:2025avl}, as
\begin{equation}\label{eq:half wormhole ETW}
\begin{aligned}
Z_{X,Y}(\beta)&=\Tr[\rme^{-\beta \hH_{\rm ASC}}]\\
&=\sum_{n=0}^\infty\int_{0}^{\pi}\rmd\theta~\tilde{\mu}(\theta)\rme^{-\beta E(\theta)}\bra{K_n}\ket{\theta}\bra{\theta}\ket{K_n}~,
\end{aligned}
\end{equation}
where the Krylov basis is shown in (\ref{eq:Krylov basis ETW brane sym}), and $\tilde{\mu}(\theta)$ in (\ref{eq:completeness relation}).
In the bulk picture of \cite{Blommaert:2025avl}, the partition function (\ref{eq:half wormhole ETW}) represents a half-wormhole partition function (which is different from the disk partition function in (\ref{eq:general ETW partition function})) as it connects the asymptotic AdS boundary with the ETW brane in a cylinder topology. We illustrate this in Fig. \ref{fig:trumpets} (a). Moreover, given that the ASC polynomials are real $\bra{K_n}\ket{\theta}=\bra{\theta}\ket{K_n}$, and due to the identity $\mo=\sum_{n=0}^\infty\ket{K_n}\bra{K_n}$, then (\ref{eq:half wormhole ETW}) can be evaluated as
\begin{equation}
    Z_{X,Y}(\beta)=\int\rmd\theta~\delta(\theta-\theta)\rme^{-\beta E(\theta)}~.
\end{equation}
While the quantity above is formally divergent, one can define a finite wormhole partition function in terms of an integral transform of (\ref{eq:half wormhole ETW}), as we present next.

\paragraph{Trumpet wavefunctions}From the boundary side, we can then recover the corresponding trumpet wavefunction as an integral transform over the parameter $Y$,
\begin{equation}\label{eq:trumpet amplitude2}
\begin{aligned}
    \tilde{Z}_{b}(\beta)&=\frac{1}{2\pi\rmi}\oint\frac{\rmd Y}{Y^{(b+1)}}Z_{X,Y}(\beta)\\
    &=\frac{2}{(q;q)_\infty}\sum_{n=0}^\infty\int_0^\pi\rmd\theta~2\cos (b\theta)~\rme^{-\beta E(\theta)}q^{nb}\\
    &=\frac{2}{(q;q)_\infty(1-q^b)}I_b\qty(\frac{-2J\beta}{\sqrt{\lambda(1-q)}})~,
\end{aligned}
\end{equation}
where the integration contour is a disk for $Y\in\mathbb{C}$, and the last relation follows from (6.1) in \cite{Okuyama:2023byh}, (B.18) in \cite{Blommaert:2025avl}. Note that this \emph{does not} require an analytic continuation for the parameter $q^\Delta$ in (\ref{eq:XY 1s}, \ref{eq:Z2 sym condition}), instead one can simply shift the momentum $P_{L/R}\rightarrow P_{L/R}+$ const($\in\mathbb{C}$) in the constraints (\ref{eq:constraints Okuyama}), (\ref{eq:constraints Xu case}) respectively. The evaluation of the integral leads to the Modified Bessel function of the first kind $I_b(x)$ in (\ref{eq:trumpet amplitude}) \cite{Okuyama:2024eyf,Blommaert:2025avl}. Also, note that $X$ and $Y$ are mutually dependent in the cases where the ETW brane Hamiltonians are derived from the DSSYK two-sided Hamiltonian (i.e. either $X=-Y$ or $X=0$ for arbitrary $Y$), for this reason the trumpet partition function does not longer depend on $X$.

In order to simplify (\ref{eq:trumpet amplitude2}), we define a normalized trumpet partition function (which retains the same functional dependence on $\beta$), similar to \cite{Blommaert:2025avl,Okuyama:2023byh}
\begin{equation}
    \label{eq:trumpet amplitude}
    Z_b(\beta):=\frac{(q;q)_\infty(1-q^b)}{2}\tilde{Z}_b(\beta)=I_b\qty(\frac{-2J\beta}{\sqrt{\lambda(1-q)}})~.
\end{equation}
The trumpet partition function above agrees with the sine dilaton answer (4.1) in \cite{Blommaert:2025avl}, where the equivalence to our expression (\ref{eq:trumpet amplitude}) follows from their (B.9). Interestingly, (\ref{eq:trumpet amplitude}) also reproduces the ETH matrix model answer \cite{Okuyama:2024eyf}.

\paragraph{Double trumpet}
The double trumpet partition function (Fig. \ref{fig:trumpets} (b)) can then be defined from \cite{Jafferis:2022wez,Okuyama:2023byh,Blommaert:2025avl}
\begin{equation}\label{eq:double trumpet}
    Z(\beta_1,\beta_2)=\sum_{b=0}^\infty b~ Z_{b}(\beta_1)Z_{b}(\beta_2)~.
\end{equation}
An analytic expression can be found in (6.5) \cite{Okuyama:2023byh}. We represent this partition function in Fig. \ref{fig:trumpets} (b). 

\subsection{Cylinder two-point correlation functions}\label{ssec:two point Euclidean wormhole}
\begin{figure}
    \centering
    \includegraphics[width=0.6\textwidth]{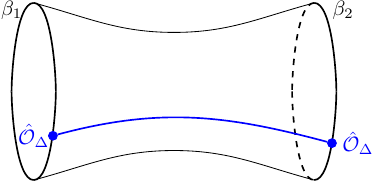}
    \caption{Cylinder two-point correlation function (\ref{eq:cylinder amplitude}) of the gauged DSSYK Hamiltonian (\ref{eq:ASC Hamiltonian}). $\hat{\mathcal{O}}_{\Delta}$ (blue) represent the matter chord insertions.}
    \label{fig:cylindercorrelatorETW}
\end{figure}

We now evaluate two-point correlation functions in the cylinder by imposing constraints in the two-sided two-point functions of the constrained model from two-sided two-point functions similar to (\ref{eq:two-point constrained}). We follow the same procedure as in the generalization from the disk to wormhole partition function, namely, we promote the trace to account for all $\ket{K_n}\in\mathcal{H}_{\rm phys}$. {{We remark that similar to the partition functions in the previous section, while modifying the trace has a natural bulk interpretation, its boundary interpretation needs further development.}} The unnormalized thermal two-point correlation function (\ref{eq:two-point constrained}) {{then}} becomes
\begin{equation}\label{eq:G cylinder}
    \begin{aligned}
   &G^{(\Delta_w)}_{\rm cylinder}=\Tr({{\hat{\mathcal{O}}_{\Delta_w}}\rme^{-\beta_1\hH_{\rm ASC}}{\hat{\mathcal{O}}_{\Delta_w}}\rme^{-\beta_2\hH_{\rm ASC}}})\\
    &={\sum_{n=0}^\infty\bra{K_n}\rme^{-\beta_2\hH_{\rm ASC}}q^{\Delta_w\hat{n}}\rme^{-\beta_1\hH_{\rm ASC}}\ket{K_n}}\\
    &=\sum_{n=0}^\infty\qty(\int_{0}^{\pi}\prod_{i=1}^2\rmd\theta_i\frac{(XY,\rme^{\pm2\rmi\theta_i}; q)_{\infty}}{(X\rme^{\pm \rmi\theta_i},Y\rme^{\pm \rmi\theta_i};q)_{\infty}}\rme^{-\beta_iE(\theta_i)})\bra{K_n}\ket{\theta_1}\bra{\theta_1}q^{\Delta_w\hat{n}}\ket{\theta_2}\bra{\theta_2}\ket{K_n}~.
    \end{aligned}
\end{equation}
We illustrate this correlation function in Fig. \ref{fig:cylindercorrelatorETW}. The result from the saddle point evaluation is shown in (\ref{eq:final cylinder amplitude}) below.

\subsection{Perturbative stability analysis}\label{ssec:stability}
We now show that both the one and two-sided Euclidean wormhole are \emph{perturbatively stable}. This is similar to recent findings in Euclidean wormholes in higher dimensions (e.g. \cite{Hertog:2024nys,Aguilar-Gutierrez:2023ril,Loges:2022nuw,Marolf:2021kjc,Jonas:2023ipa,Jonas:2023qle,Marolf:2025evo}) (while technical details about the boundary conditions remain under active debate). Using the orthogonality relations of the ASC polynomials (\ref{eq:ortho relations ASC}), and $\beta_1=\beta-\beta_2$, then (\ref{eq:G cylinder}) can be expressed as
\begin{equation}\label{eq:cylinder amplitude}
    G^{(\Delta_w)}_{\rm cylinder}(\beta)={\int_0^\pi \rmd\theta~\tilde{\mu}(\theta)\bra{\theta}q^{\Delta_w\hat{n}}\ket{\theta}\rme^{-\beta E(\theta)}}~,
\end{equation}
where $\tilde{\mu}(\theta)=\frac{(XY,\rme^{\pm2\rmi\theta}; q)_{\infty}}{(X\rme^{\pm \rmi\theta},Y\rme^{\pm \rmi\theta};q)_{\infty}}$. 

Moreover, inserting $\sum_n\ket{K_n}\bra{K_n}=\mathbb{1}$ and $\hat{n}\ket{K_n}=n\ket{K_n}$ in (\ref{eq:cylinder amplitude}) leads to
\begin{equation}\label{eq:correlator cylinder 2}
    G^{(\Delta_w)}_{\rm cylinder}(\beta)=\sum_{n=0}^\infty q^{\Delta_wn}{\int_0^\pi\rmd\theta~\tilde{\mu}(\theta)\abs{\bra{\theta}\ket{K_n}}^2\rme^{-\beta E(\theta)}}~.
\end{equation}
To carry on the evaluation of (\ref{eq:correlator cylinder 2}) one can use the Fourier kernel of the ASC polynomials. We take steps in this direction in App. \ref{app:alternative derivation} (similar to \cite{Okuyama:2023yat} in a different model). However, due to technical difficulties in the asymptotic analysis, we find more convenient to work in the Krylov basis to find explicit solutions of (\ref{eq:cylinder amplitude}). Given that in the semiclassical limit $\lambda n$ is held fixed, we may use the ASC Hamiltonian (\ref{eq:ETW brane H symmetrized})
\begin{equation}
    \begin{aligned}
        2\cos\theta\bra{\theta}\ket{K_n}=\bra{\theta}\biggl(&\rme^{\rmi\hat{P}}\sqrt{(1-XY q^{\hat{n}-1})[n]_q}\\
    &+\sqrt{(1-XY q^{\hat{n}-1})[n]_q}\rme^{-\rmi\hat{P}}+\frac{(X+Y)q^{\hat{n}}}{\sqrt{1-q}}\biggr)\ket{K_n}~.
    \end{aligned}
\end{equation}
The conjugate canonical variable to the chord number, $\hat{p}=\lambda\hat{k}$ has rescaled eigenvalues $k\sim\mathcal{O}(1)$, which indicates that  
\begin{equation}\label{eq:anticomm 1}
\begin{aligned}
    &\rme^{\rmi\hat{P}}\sqrt{(1-XY q^{\hat{n}-1})[\hat{n}]_q}+\sqrt{(1-XY q^{\hat{n}-1})[\hat{n}]_q}~\rme^{-\rmi\hat{P}}\\
    &\simeq2\sqrt{(1-XY q^{\hat{n}})[\hat{n}]_q}-\lambda\frac{q^{\hat{n}}XY\sqrt{[\hat{n}]_q}}{\sqrt{1-XY q^{\hat{n}}}}+\rmi\lambda\qty[\hat{k},~\sqrt{(1-XY q^{\hat{n}})[\hat{n}]_q}]+\mathcal{O}(\lambda^{3/2})~.
\end{aligned}
\end{equation}
Moreover, in this limit the last term in (\ref{eq:anticomm 1}) coincides with
\begin{equation}\label{eq:comm 2}
    \qty[\rme^{\rmi \hat{p}},~\sqrt{(1-XYq^{\hat{n}})[\hat{n}]_q}]=\rmi\lambda\qty[\hat{k},~\sqrt{(1-XY q^{\hat{n}})[\hat{n}]_q}]+\mathcal{O}(\lambda^{3/2})~.
\end{equation}
Next, using the constrained chord algebra (\ref{eq:shift in momentum canonical}), we can express (\ref{eq:comm 2}) as
\begin{equation}\label{eq:comm3}
\begin{aligned}
    &\qty[\rme^{\rmi \hat{p}},~\sqrt{(1-XYq^{\hat{n}})[\hat{n}]_q}]\ket{K_n}\\
    &=\qty(\sqrt{(1-XYq^{\hat{n}})[\hat{n}]_q}-\sqrt{(1-XYq^{\hat{n}+1})[\hat{n}+1]_q})\rme^{\rmi \hat{p}}\ket{K_n}~.
\end{aligned}
\end{equation}
We can then find the saddle point (s.p.) solution $\theta$ from (\ref{eq:ASC recurrence relation}) when $\lambda\rightarrow0$ and including one-loop corrections. This becomes
\begin{equation}\label{eq:sp theta}
    \cos\theta_{\rm s.p.}=\sqrt{[n]_q}\qty(\sqrt{1-XY q^n}+\frac{q^n(X+Y)}{2\sqrt{1-q^n}}-\lambda\frac{q^n\qty(1+\qty(3-4q^n)XY)}{4(1-q^n)\sqrt{1-XY q^n}})+\mathcal{O}(\lambda^2)~.
\end{equation}
An alternative derivation of (\ref{eq:sp theta}) is explained in App. \ref{app:alternative derivation}. Using the saddle point solution (\ref{eq:sp theta}) in (\ref{eq:correlator cylinder 2}),
\begin{equation}\label{eq:G cylinder saddle}
    G^{(\Delta_w)}_{\rm cylinder}(\beta)\eqlambda\sum_{n=0}^\infty q^{\Delta_wn}
    \rme^{-\beta E(\theta_{\rm s.p.})}\int_0^\pi\rmd\theta~\tilde{\mu}(\theta)\abs{\bra{\theta}\ket{K_n}}^2~.
\end{equation}
From (\ref{eq:ortho relations ASC} and \ref{eq:G cylinder saddle}) at first order non-trivial order in $\lambda$ we thus derive
\begin{equation}\label{eq:final cylinder amplitude}
\begin{aligned}
    G^{(\Delta_w)}_{\rm cylinder}&\simeq\sum_{n=0}^\infty \rme^{-I_{\rm WH}(n)}~,\\
    I_{\rm WH}(n)&:= \Delta_w \lambda+\frac{\beta J\sqrt{[n]_q}}{\sqrt{\lambda}}\qty(\sqrt{1-XY q^n}+\frac{q^n(X+Y)}{\sqrt{1-q^n}}-\lambda\frac{q^n\qty(1+\qty(3-4q^n)XY)}{4(1-q^n)\sqrt{1-XY q^n}})~.
\end{aligned}
\end{equation}
The sum (\ref{eq:final cylinder amplitude}) for $\beta=0$ reproduces the same solution as \cite{Okuyama:2023yat} (which instead assumes the ETH matrix model \cite{Jafferis:2022uhu,Jafferis:2022wez} for deriving the corresponding observables).

We proceed studying stability in the semiclassical limit for a fixed length wormhole, using similar techniques as \cite{Stanford:2020wkf,Okuyama:2023yat}. In bulk terms, this corresponds to imposing a constraint in the dilaton gravity equations of motion ({{as}} treated in \cite{Stanford:2020wkf}) which is reminiscent of gravitational instantons, e.g. \cite{Cotler:2020lxj,Cotler:2021cqa,Morvan:2022aon,Morvan:2022ybp}. 

\begin{figure}
    \centering
    \subfloat[]{\includegraphics[height=0.26\textwidth]{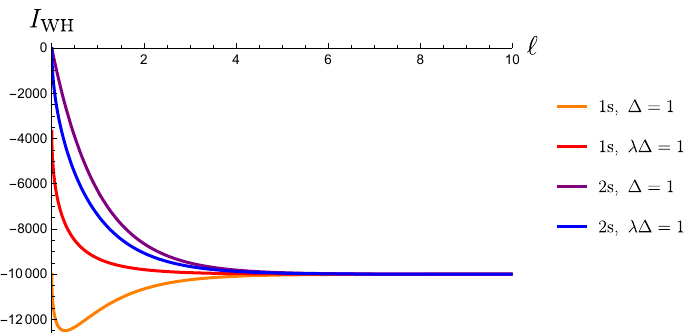}} \subfloat[]{\includegraphics[height=0.26\textwidth]{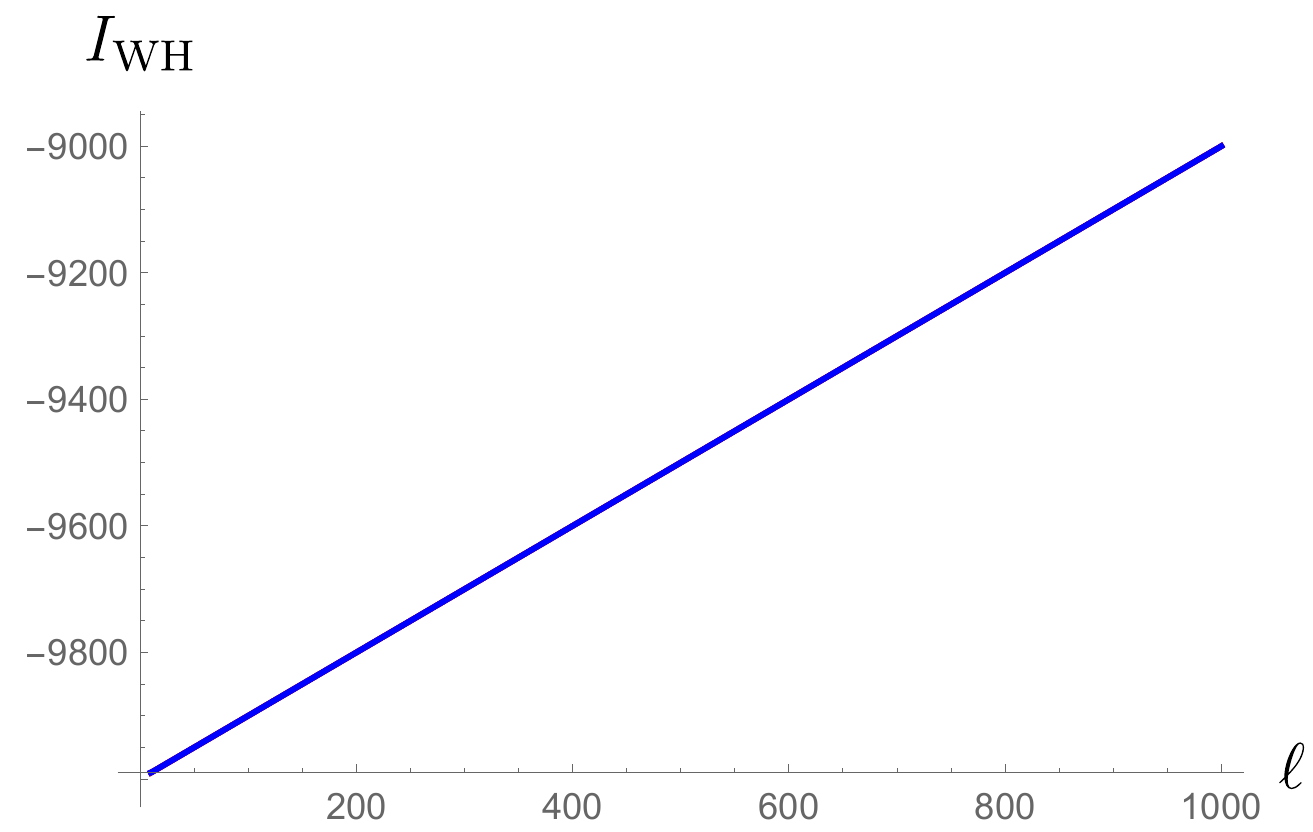}}
    \caption{Effective action in the sum (\ref{eq:final cylinder amplitude}) at fixed length $\ell=\lambda n$ for the one-sided (1s) (\ref{eq:XY 1s}), and the two-sided (2s) symmetric (\ref{eq:Z2 sym condition}) ETW brane models. $\ell$ takes values (a) $\sim\mathcal{O}(1)$, and (b) $\sim\mathcal{O}(10^2)$. The chosen parameters are $J\beta=-1$, $\Delta_w=1$, $\lambda=10^{-4}$, and the different values of $\Delta$ shown in the figure. Very similar results are obtained for other parameters. The asymptotic evolution shows that (\ref{eq:final cylinder amplitude}) always has a global minimum.}
    \label{fig:stability wormhole}
\end{figure}

In terms of the DSSYK model, we evaluate the minimum of the overall exponent in (\ref{eq:final cylinder amplitude}). We find analytic results for the minimum for the one-sided (\ref{eq:XY 1s}) and two-sided symmetric (\ref{eq:Z2 sym condition}) ETW brane solutions. However, the expressions are lengthy and not illuminating; for this reason, we show the numerical results instead in Fig. \ref{fig:stability wormhole}.\footnote{The effective action is bounded from below when $J$ or $\beta$ are negative. Both cases are allowed since both the inverse temperature $\beta$ (\ref{eq:temperature entropy ETW one sided}) and the coupling $J$ in $E(\theta)=2J\cos\theta/\sqrt{\lambda(1-q)}$ can be negative.}
It can be seen that in both (one and two-sided brane Hamiltonians) cases, there exists a non-zero minimum for the overall exponent, which indicates stability of the wormhole solution \cite{Okuyama:2023yat,Stanford:2020wkf} for a fixed geodesic length $\ell=\lambda n$. As seen in our numerical solutions, it is crucial for the non-vanishing total conformal dimension to find stability, since even if the operators are light it leads to a bounded effective action. Thus, we need additional matter to stabilize the fixed length wormholes, which is a similar observation as in \cite{Stanford:2020wkf}, and in recent discussions about the caterpillar wormholes in the SYK model \cite{Magan:2024aet}.

The reason for this is that the on-shell action has a negative minimum value in the on-shell solution, indicating that linearized perturbations around the minimum result in a bounded quadratic action. While our analysis is from the boundary perspective, we expect the bulk analysis to follow similar considerations as \cite{Stanford:2020wkf} for general dilaton gravity theories, albeit with one-loop quantum corrections in our case.

To conclude, our solutions for the Euclidean wormhole partition functions (\ref{eq:half wormhole ETW}, \ref{eq:trumpet amplitude}, \ref{eq:double trumpet}) and two-point correlation function (\ref{eq:G cylinder}) show that we can recover non-trivial topology from the constrained theory ({{albeit assuming the validity of the corresponding expressions in the bulk dual theory \cite{Blommaert:2025avl}}}). Given that the theory comes from a sum over chord diagrams in the DSSYK model, our results provide a way first steps in evaluate Euclidean wormhole observables from chord diagrams without assuming a direct relation with the ETH matrix model \cite{Jafferis:2022wez}. We stress the Krylov basis (\ref{eq:trumpet amplitude}) is a convenient technical tool in the derivation of (\ref{eq:half wormhole ETW}) and in the explicit evaluation of the two-point function (\ref{eq:G cylinder}), resulting in (\ref{eq:final cylinder amplitude}).\footnote{\label{fnt:spread in wormhole}It would be very interesting to associate this two-point correlation function (\ref{eq:final cylinder amplitude}) with a generating function of spread complexity of a given reference state associated with the sum over the original Krylov basis ${\ket{K_n}}$ in (\ref{eq:Krylov basis ETW brane sym}).}

\subsection{Baby universe Hilbert space}\label{ssec:baby universes}
\begin{figure}
    \centering
    \includegraphics[width=0.7\linewidth]{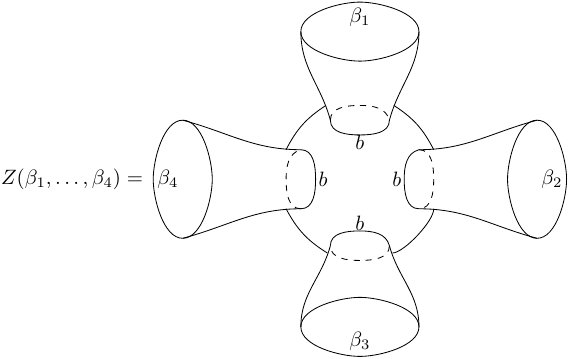}
    \caption{Representation of the multi-boundary wormhole (\ref{eq:multiboundary partition function}) for $m=4$ resulting from the inner product (\ref{eq:inner produt baby universe}) of baby universe states $\ket{Z(\beta_1,\dots,\beta_m;b)}$ (\ref{eq:baby universe state}) (where $0\leq m\leq4$).}
    \label{fig:multiboundary_wormhole}
\end{figure}

We now study the Hilbert space of the Euclidean wormhole topologies connecting multiple connected asymptotic boundaries. For this purpose, we find convenient to express the trumpet partition function (\ref{eq:trumpet amplitude}) in terms of self-adjoint operators, as in JT gravity or the ETH matrix model \cite{Okuyama:2024eyf,miyaji2025finitenbulkhilbert,Blommaert:2022ucs,Penington:2023dql,Saad:2019pqd}
\begin{equation}\label{eq:Zb beta}
    \begin{aligned}
        &Z_{b}(\beta)=\bra{K_0}\rme^{-\beta\hH_{\rm ASC}}\int_0^{\pi}\frac{\rmd\theta}{2\pi}2\cos(b\theta)\ket{\theta}\bra{\theta}\ket{K_0}~.
    \end{aligned}
\end{equation}
After replacing $\ket{0}\rightarrow\ket{K_0}$, (\ref{eq:Zb beta}) is exactly the same result that one recovers from the ETH matrix model approach in \cite{Okuyama:2024eyf}. Note that our derivation is independent from that. Based on the double-trumpet (\ref{eq:double trumpet}), and multi-boundary Euclidean wormhole partition function in JT gravity (see e.g. \cite{Bhattacharyya:2025gvd,Iliesiu:2021ari,Iliesiu:2024cnh}), and the ETH matrix model \cite{Jafferis:2022wez}, we define the multi-boundary Euclidean wormhole partition function without handles by\footnote{This should be checked explicitly in sine-dilaton gravity, including the generalizations with higher genus multi-boundary partition functions in (2.36) of \cite{miyaji2025finitenbulkhilbert}.}
\begin{equation}\label{eq:multiboundary partition function}
\begin{aligned}
    Z(\beta_1,\beta_2,\dots,\beta_m)&:=\sum_{b=1}^\infty b~Z_{b}(\beta_1)Z_{b}(\beta_2)\cdots Z_{b}(\beta_m)\\
    &=\sum_{b=1}^\infty b~I_b~\qty(\frac{-2J\beta_1}{\sqrt{\lambda(1-q)}})I_b\qty(\frac{-2J\beta_2}{\sqrt{\lambda(1-q)}})\cdots I_b\qty(\frac{-2J\beta_m}{\sqrt{\lambda(1-q)}})~,
\end{aligned}
\end{equation}
where the sum over $b$ glues the wormhole throats as in the double trumpet case (\ref{eq:double trumpet}). We illustrate the partition function (\ref{eq:multiboundary partition function}) in Fig. \ref{fig:multiboundary_wormhole}.
We interpret (\ref{eq:multiboundary partition function}) as the dimension of the symmetric DSSYK models living in multiple boundaries that are connected through the bulk Euclidean wormholes. As a consistency check, one can notice from the modified Bessel functions in (\ref{eq:multiboundary partition function}) that $Z(0,0,\dots,0)=0$ when the multi-boundary wormhole has vanishing size. 

Next, we study the baby universe Hilbert space $\mathcal{H}_{\rm BU}$, defined as the space of states of Euclidean wormholes with multiple asymptotic boundaries in the bulk. We generate the states by acting with operators, $\hat{Z}_b$, connecting a boundary with periodicity $\beta$ to an Euclidean wormhole geometry with other boundaries on the trace state, i.e.
\begin{equation}\label{eq:baby universe state}
\ket{Z(\beta_1,\beta_2,\dots,\beta_m)}=\hat{Z}_b(\beta_1)\hat{Z}_b(\beta_2)\cdots \hat{Z}_b(\beta_m)\ket{K_0}~.
\end{equation}
We need to specify the inner product of Hilbert space $\mathcal{H}_{\rm BU}$. We also stress that, unlike the previous literature, we do not rely on matrix models to derive the operators. Instead, our derivation follows directly from the ASC Hamiltonian (\ref{eq:ASC Hamiltonian}).

Therefore, we define the inner product on $\mathcal{H}_{\rm BU}$ that generates the wormhole partition function (\ref{eq:multiboundary partition function}) with $n+m$ boundaries as
\begin{equation}\label{eq:inner produt baby universe}
    Z(\beta_1,\beta_2,\dots,\beta_m)=\sum_{b=1}^\infty b\bra{Z(\beta_1,\dots\beta_m;b)}\ket{K_0}\bra{K_0}\ket{Z(\beta_1,\dots\beta_n;b)}
\end{equation}
We can now construct the baby universe operator, $\hat{Z}_b(\beta_i)$. We define it in such a way that the multi-boundary partition function factorizes into a product of the individual partition functions, i.e.
\begin{equation}\label{eq:baby univ op many}
    {Z}(\beta_1,\dots,\beta_m)=\sum_{b=0}^\infty b\bra{\Psi_{\rm HH}}\prod_{i=1}^m\hat{Z}_b(\beta_i)\ket{\Psi_{\rm HH}}~,
\end{equation}
where we denote $\ket{\Psi_{\rm HH}}$ the HH state, which in the constrained chord space is $\ket{\Psi(\tau)}$ (\ref{eq:HH ETW}). The multi-boundary partition function can be conveniently expressed in terms of the alpha-states of the theory, which are eigenstates of the baby universe operator
\begin{equation}\label{eq:alpha states eq}
    \hat{Z}_b(\beta)\ket{\alpha}={Z}_\alpha(\beta)\ket{\alpha}~.
\end{equation}
so that (\ref{eq:baby univ op many}) becomes
 \begin{equation}\label{eq:multiboundary alpha}
    Z(\beta_1,\dots\beta_m)=\sum_\alpha \abs{\bra{\Psi_{\rm HH}}\ket{\alpha}}^2\prod_{i=1}^m{Z}_\alpha(\beta_i)~.
\end{equation}
It can be seen that $\alpha$ is the charge for each superselection sector of the Euclidean wormhole. The alpha-states therefore provide a way for factorization of the theory, according to \cite{Coleman:1988cy} this causes a shift in the coupling constants of the theory, which we discussed in Sec. \ref{sec:intro}.

Using the definition of the baby universe states (\ref{eq:baby universe state}) and the inner product (\ref{eq:inner produt baby universe}), we identify the corresponding \emph{baby universe operator}\footnote{Note that this is a different definition from the baby universe operator appearing in \cite{Okuyama:2024eyf,miyaji2025finitenbulkhilbert}. First, the operators in our approach and those in \cite{Okuyama:2024eyf,miyaji2025finitenbulkhilbert}, which act on different Hilbert spaces. Moreover, we define the baby universe operator based on both the baby universe state (\ref{eq:baby univ op many}) and the multi-boundary wormhole amplitude (\ref{eq:inner produt baby universe}), as done in the literature (e.g. \cite{Blommaert:2022ucs,Marolf:2020rpm,Marolf:2020xie}) instead of a single wormhole partition function.}
\begin{equation}\label{eq:baby universe operator}
    \hat{Z}_b(\beta)=Z_b(\beta)\ket{K_0}\bra{K_0}~.
\end{equation}
Given that the infinite temperature HH state in this model corresponds to the empty open chord number state $\ket{\Psi_{\rm HH}}=\ket{K_0}$, the alpha-states (\ref{eq:alpha states eq}) and the eigenvalues of (\ref{eq:baby universe operator}) follow as
\begin{equation}\label{eq:alpha states}
\ket{\alpha}=\sqrt{b}\ket{K_0}~,\quad Z_\alpha(\beta)=Z_b(\beta)
\end{equation}
where the normalization ensures the equality of (\ref{eq:multiboundary alpha}) and (\ref{eq:inner produt baby universe}). Physically, baby universes are one-dimensional, even in this toy model, given that it is constructed from an ensemble averaged auxiliary Hamiltonian (\ref{eq:ASC Hamiltonian}). The result seems to agree with Coleman's interpretation \cite{Coleman:1988cy},{{ albeit in lower dimensions}}. The only coupling constant that can be shifted is $J$, so there is a single alpha state ($\ket{K_0}$). {However, the Hilbert space of the baby universe can be enlarged by including other type of sources in the path integral \cite{Chen:2025fwp}. We discuss about this in Sec. \ref{ssec:outlook}}. It is also possible that in a finite matrix model completion with more parameters there would be a non-trivial baby universe Hilbert space; so that higher genus multi-boundary partition functions might also modify the result. Regardless, we hope this model can offer some insights to explore baby universes holographically in more general settings.

\section{The doubled DSSYK models in dS\texorpdfstring{$_3$}{} holography from constraints in chord space}\label{sec:NV model}
In this section, we recover doubled DSSYK models, which take the role of wordline boundary theories in the static patch of dS$_3$ holographic proposals \cite{Narovlansky:2023lfz,Gaiotto:2024kze,Tietto:2025oxn,Verlinde:2024znh,Verlinde:2024zrh}, from constraints in the chord Hilbert space of a single DSYYK model with matter. This provides a concrete connection between the doubled DSSYK/dS$_3$ holographic proposals with sine dilaton gravity with matter in the previous section (also \cite{Aguilar-Gutierrez:2025pqp,Aguilar-Gutierrez:2025mxf}) in terms of symmetry sectors in the chord Hilbert space.

Using the perspective neutral approach to QRFs, we show that Krylov complexity is a relational observable in this particular model.
Furthermore, we introduce a notion of clock state with a finite temperature canonical ensemble to define expectation values for relational observables. This is a rather natural state from the holographic perspective, as an analytic continuation of the HH state with real time evolution; something so far unexplored in the QRF literature. Our work then highlights that both areas, QRFs and holography, would benefit from exchanging and merging concepts.

We stress that our calculations are performed in the boundary theory, {{and we do not attempt to provide new evidence for the bulk interpretation in the original proposal \cite{Narovlansky:2023lfz}}. While} we do not assume that the model must be dual to dS$_3$ space itself {{to derive our results;}} we do comment on the interpretation of the results for dS$_3$ holography assuming the validity in the series of works \cite{Narovlansky:2023lfz,Narovlansky:2025tpb,Gaiotto:2024kze,Tietto:2025oxn,Verlinde:2024znh,Verlinde:2024zrh}.

\paragraph{Outline:}In Sec. \ref{ssec:DDSSYKs} we implement constraints in the chord Hilbert space with matter to derive a condition for factorization in the left/right chord sectors. We compare with the results with sine-dilaton gravity. In Sec. \ref{ssec:path integral NV}, we study a path integral formulation of the original model proposed by \cite{Narovlansky:2023lfz}, and the one we construct based on the chord Hilbert space with matter. In Sec. \ref{ssec:gauge_inv_algebra} we study relational quantum dynamics from the perspective neutral approach. We derive the algebra of gauge invariant observables, and we discuss a dS holographic interpretation about the observables in this framework.

\subsection{Recovering the doubled DSSYK model}\label{ssec:DDSSYKs}
We now look for constraints to factorize the one-particle chord space into a pair of zero-particle states. This can be expressed
\begin{equation}\label{eq:const doubled DSSYK}
    \qty(\hF_\Delta-\mo)\ket{\psi}=0~,
\end{equation}
where $\ket{\psi}\in\mH_{0}\otimes\mH_{0}$, and $\hF_\Delta$ is the isometric linear map \cite{Xu:2024gfm,Xu:2024hoc,Okuyama:2024yya,Okuyama:2024gsn} acting on the chord number or energy basis respectively as
\begin{equation}\label{eq:isometric factorization}
\begin{aligned}
    \hF_\Delta\ket{\Delta;n_L,n_R}=&(q^\Delta \ha_L\otimes\ha_R;q)_\infty\ket{n_L}\otimes\ket{n_R}~,\\
    \hF_\Delta\ket{\Delta;\theta_L,\theta_R}=&\sqrt{\bra{\theta_L}q^{\Delta\hat{n}}\ket{\theta_R}}\ket{\theta_L}\otimes\ket{\theta_R}~,
\end{aligned}
\end{equation}
where $\hat{n}$ is the chord number in the zero-particle chord space. The factor in the square root can be expanded in the Krylov basis of the zero-particle space (e.g.\cite{Lin:2022rbf}) as
\begin{equation}\label{eq:matrix element 1}
    \bra{\theta_L}q^{\Delta\hat{n}}\ket{\theta_R}=\sum_{n=0}^\infty q^{\Delta n}\frac{H_n(\cos\theta_L|q)H_n(\cos\theta_R|q)}{(q;q)_n}~,
\end{equation}
where $H_n(\cos\theta_R|q)$ is the q-Hermite polynomial
\begin{equation}\label{eq:H_n def}
    H_n(\cos\theta|q)=\sum_{k=0}^n\begin{bmatrix}
        n\\
        k
    \end{bmatrix}_q\rme^{\rmi(n-2k)\theta}~,\quad
    \begin{bmatrix}
        n\\
        k
    \end{bmatrix}_{q}\equiv\frac{(q;~q)_{n}}{(q;~q)_{n-k}(q;~q)_{k}}~.
\end{equation}
Then, when $\Delta\rightarrow\infty$ for $q\in[0,1)$ leads to $\bra{\theta_L}q^{\Delta\hat{n}}\ket{\theta_R}\rightarrow1$. Thus, (\ref{eq:matrix element 1}) factorizes.
It can then be seen that (\ref{eq:const doubled DSSYK}) is satisfied for states with $\Delta\rightarrow\infty$ in either energy or chord number basis (\ref{eq:isometric factorization}) since $q\in[0,1)$.\footnote{This can also be seen by evaluating correlation functions with very heavy operators \cite{Berkooz:2020uly} (albeit the explicit results also require a triple scaling limit).} Moreover, the two-sided DSSYK Hamiltonian (\ref{eq:pair DSSYK Hamiltonians 1 particle}) for physical states (\ref{eq:const doubled DSSYK}) becomes
\begin{equation}\label{eq:factorized HLR}
    \hH_{L/R}\ket{\psi}=\frac{J}{\sqrt{\lambda(1-q)}}\qty(\rme^{-\rmi\hP_{L/R}}+\rme^{\rmi\hP_{L/R}}\qty(1-\rme^{-\hel_{L/R}}))\ket{\psi}~,
\end{equation}
given that $q^{\Delta}\rightarrow0$. In fact, the same argument above can be applied to the two-sided Hamiltonian (\ref{eq:two-sided new}) with $m$ particles when we enforce
\begin{equation}
\lim_{\Delta_1,\dots,\Delta_m\rightarrow\infty}\mH_m=\mH_0\otimes\mH_0~,
\end{equation}
which recovers (\ref{eq:factorized HLR}).\footnote{A related interpretation in terms of two-point correlation functions and the zero-particle DSSYK partition function was previously noticed in \cite{Okuyama:2023byh}.} However, in contrast to (\ref{eq:const doubled DSSYK}), we do not currently have an explicit factorization map with more than one-particle in the chord Hilbert space, which might allow to abstractly enforce the condition $\Delta_i\rightarrow\infty$. We expect this can be realized based on \cite{Aguilar-Gutierrez:2025mxf}. Regardless, in the relevant case ($i=1$) $\lim_{\Delta\rightarrow\infty}\mH_1=\mH_0\otimes\mH_0$. 
Thus, the one-particle chord space can be constrained into a two copy of the DSSYK without matter. {{Furthermore, the two copy system has a $\mathbb{Z}_2$ symmetry that can be gauged by implementing constraints identifying symmetric states of the two-copy system, as noticed in \cite{Maldacena:2001kr}\footnote{{{In the context of the information paradox, it was discussed that by gauging the $\mathbb{Z}_2$ symmetry in a doubled CFT without wormholes connecting them, this procedure can lead to orbifold two-sided black hole spacetime with an ETW brane in the interior (in a pure global state). We thank the anonymous referee for pointing this out. It would be interesting to relate this with our earlier construction in Sec. \ref{sec:ASC from DSSYK}.}}} for CFTs instead of SYK systems. This shares some similarities to the constraints that we implement below in Sec. \ref{ssec:gauge_inv_algebra} from a boundary perspective.}}


We can now develop the above construction in a path integral formulation to compare with the previous literature, particularly \cite{Narovlansky:2023lfz}, which we develop next.

\subsection{A path integral approach and time translations}
In this subsection, we briefly review the path integral formulation of the doubled DSSYK model, before performing an alternative formulation, in terms of the path integral in canonical variables, similar to (\ref{eq:path int ASC}). Moreover, we gauge relative time automorphisms between the pair of decoupled DSSYK as in \cite{Narovlansky:2023lfz,Narovlansky:2025tpb} to study dynamical aspects of the model.

In the original doubled DSSYK model \cite{Narovlansky:2023lfz}, one considers a decoupled pair of SYK models
\begin{equation}\label{eq:SYK theory}
    I_{\rm SYK}=\int\rmd t\qty(\sum_{i=1}^N\frac{\rmi}{2}\qty(\psi_i^L\dot{\psi}_i^L+\psi_i^R\dot{\psi}_i^R)-e(t)\qty(H^{(\rm SYK)}_L-H^{(\rm SYK)}_R))~,
\end{equation}
where $N$ is the total number of Majorana fermions, while
\begin{equation}
H^{(\rm SYK)}_{L/R}=\rmi^{p/2}\sum_{i_1\dots i_p}J_{i_1\dots i_p}{\psi}_{i_1}^{L/R}\dots{\psi}_{i_p}^{L/R}    ~,
\end{equation}
and $e$ is a dynamical Lagrange multiplier. The total {{action}} is expressed
\begin{equation}\label{eq:total theory}
    I_{\rm tot}=I_{\rm SYK}+\int\rmd t\qty(g\dot{e}+\dot{b}\dot{c})~.
\end{equation}
Here $b$, $c$ are ghost fields, and $g$ a constant Lagrange multiplier. One can then perform the path integral quantization of the theory considering
\begin{equation}\label{eq:PI NV original}
  Z=\int[\rmd \psi_i^L][\rmd \psi_i^R][\rmd b][\rmd c][\rmd e]\rme^{-I_{\rm tot}}  ~.
\end{equation}
Then, by taking the double-scaling limit of the system (\ref{eq:total theory}) and studying ensemble averaged Hamiltonian moments or correlation functions, \cite{Narovlansky:2023lfz} finds the DSSYK auxiliary Hamiltonian, and its associated observables.

Below we recover a doubled DSSYK model by implementing specific constraints in the chord Hilbert space with matter. Similarly to \cite{Narovlansky:2025tpb}, we do not need to use a maximal entropy (infinite temperature) state of the DSSYK.

\paragraph{Gauging time translations}\label{ssec:path integral NV}
As a result of dividing the original DSSYK into a decoupled pair of DSSYKs without matter, we note that the system acquires a notion of relative evolution, given by the Hamiltonian of each DSSYK. This can be seen by imposing the first-class constraint generating \emph{time automorphisms} due to internal evolution, both at the level of the physical Hilbert space and the Dirac observables:
    \begin{equation}\label{eq:constraint NV}
        \qty(\hH_L-\hH_R)\ket{\psi}=0~,\quad [\hH_L-\hH_R,\hO_\Delta]=0~,
    \end{equation}
where $\hO_{\Delta}(t)=\int\rmd t'~\hO_{\Delta_L}^L(t')\hO^R_{\Delta_R}(t-t')$ \cite{Narovlansky:2023lfz}.
    
Equivalently to (\ref{eq:constraint NV}), one can impose a constraint of the form \cite{Narovlansky:2025tpb}
\begin{equation}\label{eq:constraints others}
    \qty(J\int_{0}^{2\pi}\rmd\eta~\rme^{\rmi\eta\qty(\hH_L-\hH_R)}-\mo)\ket{\psi}=0~,\quad\qty[J\int_{0}^{2\pi}\rmd\eta~\rme^{\rmi\eta\qty(\hH_L-\hH_R)},~\hO_{\Delta}]=0~.
\end{equation}
(\ref{eq:constraint NV}) or (\ref{eq:constraints others}) can be seen as analogous the model on an observer in dS space \cite{Chandrasekaran:2022cip} with $\hat{q}=-\hH_L$, so that the constraint above takes the form $\hat{q}+\hH_{R}=0$. On the other hand, the interpretation of gauging the relative evolution (\ref{eq:constraints others}), generated by the total Hamiltonian $\hH_L+\hH_R$, which we will not discuss in more detail as it does not accommodate for perspective neutral observables.

\paragraph{Path integral formulation}We can also formulate the above constraints in the chord Hilbert space with the double-scaled two-sided auxiliary Hamiltonians (\ref{eq:two-sided new}) in a path integral approach, to compare with the original formulation in \cite{Narovlansky:2023lfz}. Moreover, this allows us to work with the premise that the auxiliary chord Hilbert space describes the holographic dual theory, which help us to test the ideas put forward in \cite{Lin:2022rbf}. 

For instance, in the FP approach, we may define
\begin{equation}
    {\varphi}=\xi\qty(H_L-H_R)~,
\end{equation}
where $\xi$ is the Lagrange multiplier imposing $\varphi=0$ for the solutions of the path integral, 
\begin{subequations}\label{eq:main path integral}
     \begin{align}\label{eq:partition function Faddeev form}
&Z=\int\prod_{i=L,R}[\rmd \ell_i][\rmd P_i][\rmd \xi]~\det(\qty{\varphi,~\chi})~\delta(\chi)\rme^{I_E[\qty{\ell_i,~P_i,~\xi}]}~,\\
         \label{eq:path E action ETW}
    I_E=\int&\rmd\tau_L\rmd\tau_R\qty(\sum_{i}\qty(\frac{\rmi}{\lambda} P_i(\partial_{\tau_L}+\partial_{\tau_R})\ell_i+\varphi[\xi,\ell_i,P_i])-(H_L- H_R))~,
     \end{align}
 \end{subequations}
and $\chi$ a gauge fixing condition. This path integral prepares double-scaled wormhole density matrices \cite{Aguilar-Gutierrez:2025mxf} (based on previous work \cite{Berkooz:2022fso}).

More generally, given $M$ constraints $\varphi^a(\ell_i,P_i)=0$ one has to find $N$ auxiliary constraints $\chi^a(\ell_i,P_i)=0$ such that $\qty{\varphi^a,~\chi_b}$ (the FP determinant \cite{Kunstatter:1991ds}) is \emph{invertible} for the determinant to exist. Since in the case (\ref{eq:main path integral}) we impose a single constraint and an auxiliary gauge fixing condition, the only requirement is that $\qty{\varphi,~\chi}\neq0$, which can be satisfied for instance by the gauge-fixing condition
\begin{equation}\label{eq:condition chi}
   {\chi}=P_L-P_R~.
\end{equation}
The path integral (\ref{eq:main path integral}) then becomes
\begin{equation}\label{eq:Hamilton PI2}
\begin{aligned}
    &\int\prod_{i=0}^m[\rmd \ell_i][\rmd P_i][\rmd \xi]\exp\qty[\int\rmd\tau_L\rmd\tau_R\qty(\frac{\rmi}{\lambda}\sum_{i=0}^mP_i\qty(\partial_{\tau_L}+\partial_{\tau_R})\ell_i-(\xi+1)(H_L-H_R))]\\
    &\qquad\times\delta(\chi)\text{det}\qty(\qty{H_L-H_R,~\chi})~.
\end{aligned}
\end{equation}
We can use the commutation relations in the chord algebra \cite{Lin:2023trc} to write
\begin{equation}
    \qty[\hH_L-\hH_R,\rme^{-\rmi \hat{P}_{L}}-\rme^{-\rmi \hat{P}_{R}}]=J\sqrt{\tfrac{1-q}{\lambda}}\qty(\rme^{-2\rmi \hat{P}_{L}}-\rme^{-2\rmi \hat{P}_{R}})-(1-q)\qty(\rme^{-\rmi \hat{P}_{L}}\hH_{L}-\rme^{-\rmi \hat{P}_{R}}\hH_{R})~.
\end{equation}
Thus, the determinant (\ref{eq:Hamilton PI2}) for the condition (\ref{eq:condition chi}) is invertible, since it is non-vanishing, even when $\lambda\rightarrow0$. Then, (\ref{eq:Hamilton PI2}) implies that the representation of the Hamiltonian for the physical states is
\begin{equation}\label{eq:NV model}
    \hH_{L/R}=\frac{J}{\sqrt{\lambda(1-q)}}(\rme^{-\rmi \hat{P}_{L/R}}-\rme^{\rmi \hat{P}_{L/R}}(1-q^{\hat{n}_{L/R}}))~.
\end{equation}
Thus, the combined system, after {{implementing}} the constraints in the path integral is isomorphic to the single DSSYK model without matter, albeit with the important difference of being probed by a paired observer. Given that we do not need to differentiate between left/right systems, we will drop this label from now on. The commutation relations in the constrained Hilbert space is isomorphic to those of a single DSSYK
\begin{equation}\label{eq:commutators rels}
    [\hat{n},\hat{a}^\dagger]=\hat{a}^\dagger~,\quad [\hat{n},\hat{a}]=-\hat{a}~,\quad [\hat{a},~\hat{a}^\dagger]_q=1~,
\end{equation}
where we expressed
\begin{equation}\label{eq:H DSSYK}
    \hH=\frac{J}{\sqrt{\lambda}}\qty(\hat{a}^\dagger+\hat{a})~,\quad \hat{a}^\dagger=\frac{1}{\sqrt{1-q}}\rme^{-\rmi \hat{P}}~,\quad \hat{a}=\sqrt{1-q}~\rme^{\rmi \hat{P}}~,
\end{equation}
in (\ref{eq:NV model}), similar to (\ref{eq:number op ASC}).

As seen from the above, the doubled DSSYK model is very well-adapted to the language of relational quantum dynamics (see e.g. \cite{Hoehn:2019fsy,Hoehn:2020epv,Hoehn:2023ehz,Casali:2021ewu}). Either of the DSSYK models is just a clock observer (i.e. a QRF), and the other one takes the role of system probed by its pair. The total constraint $(\hH_L-\hH_R)\ket{\Psi}=0$ $\forall\ket{\Psi}\in\mathcal{H}_{\rm phys}$ in Dirac quantization describes the evolution according to either of the clock Hamiltonians $\hH_{L/R}$ \cite{DeVuyst:2024pop,DeVuyst:2024uvd,Hoehn:2019fsy,Hoehn:2020epv,Hoehn:2023ehz,Casali:2021ewu}. We develop this in more detail in the following.

\subsection{Gauge invariant algebra and its holographic interpretation}\label{ssec:gauge_inv_algebra}
We consider the algebra of observables for the decoupled pair of DSSYKs from the perspective neutral approach of relational operator algebra in \cite{DeVuyst:2024pop,DeVuyst:2024uvd}. Since the kinematical Hilbert space (after applying the constraint (\ref{eq:const doubled DSSYK}) and before gauging time translations) is a tensor factor of zero-particle states,  {{which means}}
\begin{equation}\label{eq:kinematica physical Hilbert space}
    \mathcal{H}_{\rm kin}=\mathcal{H}_0\otimes\mathcal{H}_0~.
\end{equation}
By imposing a gauge constraint (\ref{eq:constraint NV}) we can describe the evolution according to either of the clock Hamiltonians $\hH_{L/R}$, while only after applying a gauge fixing condition, we choose the perspective of either clock DSSYK observer in the physical Hilbert space \cite{Vanrietvelde:2018pgb}.

We therefore proceed constructing the so-called clock states of the gauge invariant algebra (e.g. \cite{DeVuyst:2024pop,DeVuyst:2024uvd}). We allow for an analytic continuation where time can be complex (the real part being the inverse temperature, and the imaginary the real time), so that the clock state is just a HH state of the form
\begin{equation}\label{eq:HH clock state}
    \ket{\psi(\tau)}=\frac{\rme^{-\rmi g(\hH)-\tau\hH}}{\sqrt{Z(\beta)}}\ket{0}=\frac{1}{\sqrt{Z(\beta)}}\int_{0}^{\pi}\frac{\rmd\theta}{2\pi}(q,~\rme^{\pm 2 \rmi \theta};q)_\infty\rme^{-\rmi g(E(\theta))}\rme^{-\tau E(\theta)}\ket{\theta}~,
\end{equation}
where $g(E(\theta))\in\mathbb{R}$, {{with}} $\tau=\beta/2+\rmi t$; $\hH$ indicates the Hamiltonian of the system (either $\hH_{L/R}$); the energy spectrum, $E(\theta)$ takes the form (\ref{eq:energy spectrum}); and the measure of integration in energy basis corresponds to the DSSYK without matter, see e.g. \cite{Berkooz:2018jqr,Berkooz:2018qkz}. 

Note that the DSSYK is a so-called non-ideal clock (i.e. $\bra{\psi(\rmi t')}\ket{\psi(\rmi t)}\cancel{\propto}\delta(t-t')$) since its energy spectrum is bounded; and in fact
\begin{equation}
    \bra{\psi(\tfrac{\beta}{2}+\rmi t')}\ket{\psi(\tfrac{\beta}{2}+\rmi t)}=Z(\beta+\rmi(t'-t))/Z(\beta)~.
\end{equation}
This basis has the properties
\begin{subequations}
\begin{align}
    &\int_{\mathbb{R}}\rmd t\eval{\ket{\psi(\tau)}\bra{\psi(\tau)}}_{\tau=\frac{\beta}{2}+\rmi t}=\frac{\rme^{-\beta\hH}}{Z(\beta)}~,\\
    & \rme^{-\tau_2\hH}\ket{\psi(\tau_1)}=\ket{\psi(\tau_1+\tau_2)}~,
    \end{align}
\end{subequations}
where the $\beta\rightarrow0$ {{limit gives 
\begin{equation}
    \lim_{\beta\rightarrow0}\int_{\mathbb{R}}\rmd t\eval{\ket{\psi(\tau)}\bra{\psi(\tau)}}_{\tau=\frac{\beta}{2}+\rmi t}=\mathbb{1}_C~,
\end{equation}
with the subindex}} C denoting the reference clock subspace.

This formalism can be useful to build the gauge invariant algebra of observables \cite{DeVuyst:2024pop,DeVuyst:2024uvd}, which in this case is defined as\footnote{Note that this relational algebra does not correspond to the double-scaled algebra in \cite{Lin:2022nss,Xu:2024gfm}.}
\begin{equation}\label{eq:invariant}
    \mathcal{A}_{\rm inv}=\qty(\mathcal{B}{(\mathcal{H}_0^C)}\otimes\mathcal{B}{(\mathcal{H}_0^S)})^{\hat{\varphi}}=\expval{\hat{O}_C^\tau(\hat{A}),~\mathbb{1}_S\otimes \rme^{-\rmi t_C \hH}|\hat{A}\in\mathcal{B}({\mathcal{H}_0^S}),t\in\mathbb{R}}~,
\end{equation}
where, for notational simplicity, we include subindices $S/C$ to denote the system and the clock observer. $\mathcal{B}$ represents the set of bounded operators acting on each factor  Hilbert space; while the notation $\expval{\cdot,\cdot|\cdot}$ indicates an algebra generated by the corresponding operators acting on the system/clock Hilbert spaces with the constraint $\hat{\varphi}\ket{\psi}=0\forall\ket{\psi}\in\mathcal{H}_{\rm kin}$. Since (\ref{eq:invariant}) is a type I$_\infty$ algebra it has natural notions of traces in the chord basis $\ket{m}_C\otimes\ket{m}_S$. This means that we can trace out one of the clocks in $\mathcal{H}_{\rm kin}$, e.g.
\begin{equation}\label{eq:reduced DM clock}
    \hat{\rho}_S=\sum_{n=0^\infty}\bra{n}_C\otimes\mathbb{1}_S\ket{\psi}\bra{\psi}\ket{n}_C\otimes\mathbb{1}_S~,\quad\forall\ket{\psi}\in\mathcal{H}_{\rm kin}~,
\end{equation}
{{and also}} use the energy basis $\ket{\theta}_C\otimes\ket{\theta}_S$. In the dS context, it might be more natural to take the later case, \cite{Narovlansky:2023lfz,Verlinde:2024znh} the dual Bunch-Davies  state; which is not required for our remaining arguments.

Closure of the algebra is defined with respect to the norm of states $\in\mathcal{H}_{\rm kin}=\mathcal{H}^C_0\otimes\mathcal{H}^S_0$.
Here $\hat{O}_C^\tau(\hat{A})$ is a relational operator $\hat{A}\in\mathcal{B}(\mathcal{H}_0^S)$ and conditioned on a reading $\tau_0$ according to the clock $C$; which is defined by an incoherent group averaging (G-twirl \cite{DeVuyst:2024pop,DeVuyst:2024uvd}) over the time translations,
\begin{equation}\label{eq:conditional operator}
\hat{O}_C^{\tau_0}(\hat{A})=\frac{1}{{Z(\beta)}}\int_{\mathbb{R}}\rmd t_1\ket{\psi(\tau_1)}\bra{\psi(\tau_1)}\otimes \eval{\sum_{n=0}^\infty\frac{(\tau_1-\tau_0)^n}{n!}[\hat{A},~\hH]_n}_{\tau_{0,1}=\frac{\beta}{2}-\rmi t_{0,1}}~,
\end{equation}
with $[\hat{A},~\hH]_n$ being a nested commutator, i.e.
\begin{equation}\label{eq:nested}
    [\hat{A},~\hH]_n:=[[[[A,\hH],\hH],H],\dots] \quad \text{where }\hH\text{ appears n-times}~,
\end{equation}
and again we are allowing for a canonical ensemble through analytic continuation in the time coordinate. 

For instance, consider that $\hat{A}=\hat{n}$, which is the Krylov complexity operator in $\mathcal{H}_0$. We can then evaluate (\ref{eq:conditional operator}) explicitly using the chord algebra. Denoting $\hH=\frac{J}{\sqrt{\lambda}}(\hat{a}+\hat{a}^\dagger)$ as in (\ref{eq:H DSSYK}), we have that
\begin{equation}\label{eq:rel chord velocity operator}
    [\hat{n},\hat{a}^\dagger+\hat{a}]=\rmi\hat{\pi}_n~,\quad [a^\dagger-a,a^\dagger+a]=-2(1+(q-1)\hat{a}^\dagger \hat{a})~,
\end{equation}
where $\hat{\pi}_n\equiv \rmi{(\hat{a}-\hat{a}^\dagger)}$ is called the chord velocity operator \cite{miyaji2025finitenbulkhilbert}. Then, (\ref{eq:conditional operator}) becomes
\begin{equation}
\begin{aligned}
    \hat{O}_C^\tau(\hat{n})=&\frac{1}{Z(\beta)}\int_{\mathbb{R}}\rmd t_1\ket{\psi(\tau_1)}\bra{\psi(\tau_1)}\otimes\qty(\hat{n}+\qty(\tT_0-\tT_1)\hat{\pi}_n-\frac{J^2}{\lambda}(\tT-\tT_0)^2\mathbb{1}_S+\mathcal{O}(\lambda))\\
    =&\frac{\rme^{-\beta\hH}}{Z(\beta)} \bigg(\mathbb{1}_S\otimes\qty(\hat{n}+\tT_0\hat{\pi}_n-\tT_0^2\mathbb{1}_S)+ \frac{J^2}{\lambda}\qty(\rmi g''(\hH)-\qty(g'(\hH)+\rmi\frac{\beta}{2}\mathbb{1}_C)^2)\otimes \mathbb{1}_S\\
    &\quad\qquad+\frac{J}{\sqrt{\lambda}}\qty(\rmi g'(\hH)+\frac{\beta}{2}\mathbb{1}_C)\otimes\qty(\hat{\pi}_n+2\tT_0\mathbb{1}_S)+\mathcal{O}(\lambda)\bigg)~,
\end{aligned}
\end{equation}
where $\tT:=Jt/\sqrt{\lambda}$ and we applied the nested commutator \eqref{eq:conditional operator} to recover the final expression. Thus, spread complexity in the DSSYK model, $\expval{\hat{n}}$, is a \emph{relational observable}, as it takes part of the gauge invariant algebra (\ref{eq:invariant}).\footnote{A related work found a different reason for spread complexity to be observer dependent in a different setting \cite{Li:2025fqz} -- albeit not in the framework of relational quantum dynamics.}

The result can then be interpreted with the Page-Wootters formalism \cite{page1983evolution,Hoehn:2019fsy,Hoehn:2020epv,Hoehn:2023ehz,Casali:2021ewu}. The expectation value of these observables $\bra{\Psi}\hat{O}_C^\tau(\hat{A})\ket{\Psi}$ $\forall\ket{\Psi}\in\mathcal{H}_1$
correspond to measurements where one of the reference observer DSSYK has a clock reading $t={\rm Im}(\tau)$, while $\beta={\rm Re}(\tau)$ accounts for the microcanonical inverse temperature.

\paragraph{De Sitter interpretation}
The subalgebra for each factor (\ref{eq:invariant}), i.e. $\mathcal{B}(\mathcal{H}_0^S)$, as well as for $\mathcal{A}_{\rm inv}$ (\ref{eq:invariant}) {{which acts}} on the kinematical Hilbert space corresponds to a type I$_\infty$ Von Neumann algebra since they are isomorphic to those of the DSSYK model without matter \cite{Xu:2024hoc}, in contrast with the double-scaled algebra \cite{Lin:2022rbf}, which instead only considers operators $\expval{\hH_{L/R},~\hO^{L/R}_{\Delta}}$, with $\hO^{L/R}_{\Delta}$ being matter chord operators. For this reason, there is no notion of generalized horizon entropy associated with (\ref{eq:invariant}) at this level yet. However, there is no discrepancy with respect to the type II$_1$ algebra of dS space \cite{Chandrasekaran:2022cip},\footnote{It has been recently questioned whether some of the assumptions in \cite{Chandrasekaran:2022cip} leading to the type II$_1$ algebra are realized in microscopic models.} which assumes that $G_N\rightarrow0$ and that the mass of the observer is infinite, as remarked by \cite{Kolchmeyer:2024fly}. In order to recover the emergent notion of cosmological horizon and the type II$_1$ algebra in \cite{Chandrasekaran:2022cip} according to one of the observers, one has to take the thermodynamic limit. In fact, the algebra (\ref{eq:invariant}) is of the same type as \cite{Kolchmeyer:2024fly}, which instead considers the algebra of observables in dS$_2$ space {{with}} two quantum mechanical observers, similar to our setting. There are noticeable differences between the set-up of \cite{Kolchmeyer:2024fly} and ours; however, we expect that they lead to similar conclusions regarding the emergence of the cosmological horizon. Namely, in the semiclassical limit $\lambda\rightarrow0$, each of the DSSYKs {{gets}} localized to a worldline trajectory, with an algebra of operators of the type (\ref{eq:conditional operator}). It would be interesting to study this explicitly by carrying out the canonical quantization of dS$_3$ space in \cite{Verlinde:2024znh} and studying the transition to the semiclassical regime and emergence of the type II$_1$ dS algebra of observables \cite{Chandrasekaran:2022cip} with two antipodal observers.

\section{Discussion}\label{sec:disc}
In this work, we derived constraints in the chord Hilbert space with matter that reproduce ETW branes and Euclidean wormhole geometries in sine dilaton gravity, as well as for the doubled DSSYK models in dS$_3$ holographic proposals \cite{Verlinde:2024znh,Verlinde:2024zrh,Gaiotto:2024kze,Tietto:2025oxn,Narovlansky:2023lfz,Narovlansky:2025tpb}. {{While we do not claim that}} both of these holographic proposals {{are simultaneously correct, they are compatible with the existence}} of different types of symmetry sectors within the chord Hilbert space.

We specify the main findings for each of the cases:
\begin{itemize}
    \item We first recovered the ETW brane systems, we gauged chord symmetries between the left/right chord sectors of the DSSYK model total Hamiltonian by imposing constraints in terms of the quantum phase space of the total Hamiltonian. We carried out the evaluation of the partition function, the semiclassical thermodynamics (including the thermal stability) {{in}} App. \ref{app:ETW brane partition function}, and thermal correlation functions. The saddle point solutions of the path integral allowed us to evaluate the spread complexity of the HH state (which includes finite temperature effects). We matched it with the geodesic distance of an Einstein-Rosen bridge (where the geodesic curve lands on the ETW brane). This {{leads to}} a concrete identification of entries of holographic dictionary (seen in Tab. \ref{tab:holographic_dictionary}). Thus, spread complexity allows for a precise identification of entries in the holographic dictionary {{in this system}}. We also discussed the quantization of the bulk dual theory.
\item Meanwhile, we {{introduced}} Euclidean wormhole {{partition functions and correlation functions in the DSSYK model}} by implementing appropriate traces in chord space {{based on results}} in JT gravity and sine dilaton gravity with ETW branes \cite{Gao:2021uro,Blommaert:2025avl}. {{In the following subsection we comment about possible issues that need to be accounted for to interpret the corresponding observables from the boundary perspective.}} We stress we do not need to assume a matrix model completion in our analysis; and in fact, it could be used to determine the models that are compatible with correlation functions of the DSSYK model in the large $N$ limit, within a symmetry sector. We also studied the perturbative stability (i.e. including one-loop corrections) of spacetime wormholes at fixed length \cite{Stanford:2020wkf,Okuyama:2023yat}, {{which}} happens to be stable. After that, we constructed multi-boundary wormhole partition functions to derive the baby universe Hilbert space, {{which}} turns out to be one-dimensional, unless it contains correlated matter operators.
\item On the other hand, to recover the doubled DSSYK model we imposed a constraint that factorizes the chord Hilbert space. This allows for a notion of internal relative evolution between the DSSYKs. Either of them plays the role of a QRF performing measurements on its copy. We also highlighted that this model is well-adapted to describe relational quantum dynamics. We then built its algebra of gauge invariant observables in the perspective neutral approach. We also discussed about the interpretation of the algebraic findings in terms of dS$_3$ holography.
\end{itemize}
Furthermore, the examples above are complemented by our recent studies \cite{Aguilar-Gutierrez:2025pqp,Aguilar-Gutierrez:2025mxf}, where no additional constraints are imposed on the boundary theory with matter chords. These works provide a concrete link between two common approaches to holography in the DSSYK model through symmetry sectors in the chord Hilbert space, and on developing  the tools of relational quantum dynamics in an explicit solvable and UV finite model in holography.

We now conclude with some future directions.

\subsection{Outlook}\label{ssec:outlook}
\paragraph{{{Boundary interpretation of the wormhole observables and }}higher genus topology}
The Euclidean wormhole geometries, which we developed in Sec. \ref{sec:wormhole topology}, {{are based on bulk models \cite{Blommaert:2025avl,Gao:2021uro} where the ASC Hamiltonian \eqref{eq:ASC Hamiltonian} and the corresponding traces play a crucial role. It is an important future prospect to understand their boundary interpretation. To be more precise, it is an intricate problem to distinguish between Euclidean wormhole contributions from certain non-perturbative contributions in physical observables. This can be illustrated by computing double-trace Hamiltonian moments (i.e. $\expval{\tr H^{k_1}\tr H^{k_2}}/\expval{(\tr H^2)^{k_1+k_2}}$ in \cite{Cotler:2016fpe}, see also \cite{Pluma:2019pnc,Berkooz:2020fvm}). There are additional contractions with respect to those in the standard chord rules of the DSSYK model \cite{Berkooz:2018jqr} that contribute to the double-trace moments as $N^{-p}$ \cite{Cotler:2016fpe} (with $N$ the total number of fermions, $p$ the all-to-all interactions in the SYK model). This becomes non-perturbative in the double scaling limit, and thus it cannot be treated with the perturbative moment method in \cite{Berkooz:2020fvm}. Our calculations in this section do not take into account possible non-perturbative corrections that should be distinguished from the wormhole amplitudes in Sec. \ref{sec:wormhole topology}. One would need a boundary parameter in the DSSYK model (or its matrix model extension \cite{Jafferis:2022uhu,Jafferis:2022wez}) controlling the topological expansion of the bulk theory. We expect that the parameter controlling the topological expansion is related to the overall normalization constant of the partition function. This should be studied further to confirm the wormhole calculables have a well-defined boundary interpretation independent of the putative bulk dual theory.}} Moreover, the half wormholes in (\ref{eq:half wormhole ETW}), seem to be unrelated with the restoration of the factorization of partition functions in other settings \cite{Saad:2021uzi,Saad:2021rcu,Mukhametzhanov:2021nea,Yang:2025kgs,Garcia-Garcia:2021squ,Peng:2022pfa}. To improve this, it might be useful to search for more evidence relating the holographic Euclidean wormhole model in Sec. \ref{sec:wormhole topology} with respect to the ETH matrix model \cite{Jafferis:2022uhu,Jafferis:2022wez}, or other matrix model associated with the DSSYK with matter. Note for instance that the ETH matrix model does not reproduce the same multi-boundary connected correlation functions of the DSSYK model \cite{Berkooz:2020fvm} in the large $N$ regime. One might check if an extension of the Euclidean wormhole model in Sec. \ref{sec:wormhole topology} has similar features. Furthermore, it would also be interesting to treat the path integral (\ref{eq:path E Z ETW}) in terms of $G\Sigma$ variables as in other parts of the literature (e.g. \cite{Berkooz:2024evs,Berkooz:2024ofm,Cotler:2016fpe,Maldacena:2016hyu,kitaevTalks}).

\paragraph{More particle insertions}Incorporating matter in the chord Hilbert space \cite{Lin:2022rbf} was indispensable in this work. However, the physical Hilbert space of the DSSYK model within a symmetry sector, such as Sec. \ref{sec:ASC from DSSYK}, is analogous to the zero-particle chord Hilbert space. A natural future direction is to construct its Hilbert space with additional matter content.
Since we recovered matter correlation functions \eqref{eq:cylinder amplitude} by applying constraints in the crossed four-correlation functions, it should be possible to formulate an extended Hilbert space that includes one or more particle states in a similar way as \cite{Lin:2022rbf}.

Another natural extension in this direction is to study symmetry sectors in the two-sided DSSYK Hamiltonians with arbitrarily many particles (\ref{eq:pair DSSYK Hamiltonians 1 particle}). We found that specific types of constraints reproduce a family of ETW brane Hamiltonians (\ref{eq:ASC Hamiltonian}). As a concrete case, it would be interesting to learn if there are even more kinds of constraints that generalize the ASC Hamiltonians from the DSSYK model with matter. A natural case to consider is a generalization of the two-sided brane Hamiltonian (\ref{eq:H + ASC}) (assuming $\Delta_1=\cdots=\Delta_m:=\Delta$):
\begin{equation}\label{eq:all symmetric ETW H}
    \hH_L+\hH_R=\frac{2J}{\sqrt{\lambda(1-q)}}\qty(\rme^{-\rmi \hat{P}}+\rme^{\rmi \hat{P}}\qty(1-\rme^{-\hat{\ell}})\sum_{j=0}^m q^{\Delta}\rme^{-j\hat{\ell}})~,
\end{equation}
which follows directly from imposing $2m$ linear momentum constraints $\qty(\hat{P}-\hat{P}_i)\ket{\psi}=0$, and equal chord number constraints $(\hel_{0\leq i\leq m}-\hel)\ket{\psi}=0$ (\ref{eq:pair DSSYK Hamiltonians 1 particle}) $\forall\ket{\psi}\in\mathcal{H}_{\rm phys}$. On the other hand, to recover one-sided ETW brane Hamiltonians with one or more particle, one may implement e.g. $(\hat{P}-\hat{P}_0)\ket{\psi}=0$, $\hat{P}_{i\neq0}\ket{\psi}=0$, with $(\hel_0-\hel)\ket{\psi}=0$, $(\hel_{i\neq0}-\tilde{N}\mo)\ket{\psi}=0$ and take the limit $\tilde{N}\rightarrow\infty$. However, this does not modify the answer for the one-particle space in (\ref{eq:ASC Hamiltonian}) with the conditions (\ref{eq:XY 1s}), indicating that this is just a composite particle representation of the case we have developed in this work. It {{might be possible that}} there is no generalization to the one-sided ETW brane case, e.g. the number of chord sectors has to equal number of chord Hamiltonians.
It would also be useful to gain some physical intuition for why the specific constraints in Sec. \ref{ssec:constraints ETW} {{implementing}} chord symmetry generates ETW branes in the bulk (which we confirmed by computing disk partition functions in Sec. \ref{ssec:constraints ETW}, App. \ref{app:ETW brane partition function}).

\begin{figure}
    \centering
    \includegraphics[width=0.5\linewidth]{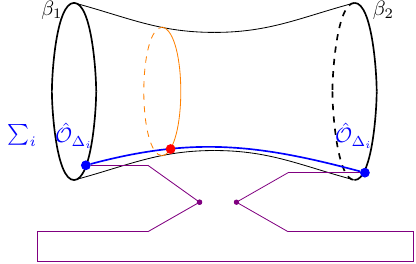}
    \caption{A closed universe S$^1$ slice (orange) in the Euclidean wormhole geometry, based on \cite{Harlow:2018tqv,Chen:2025fwp}, where the observer (red dot) corresponds to the matter correlator traversing the S$^1$ slice. In the quantum circuit representation, an entangled non-gravitating bath generates multiple correlated operator insertions, $\hO_{\Delta_i}$ (blue) on the disk boundaries. The number of two-point functions determines the Hilbert space dimension of the closed universe with the observer \cite{Chen:2025fwp,Harlow:2025pvj}. There are more topologies in the path integral; however, we display the only relevant one for Sec. \ref{sec:ASC from DSSYK}.}
    \label{fig:closed_universe}
\end{figure}

\paragraph{Relational holography in closed universes}There have been very interesting discussions in the literature \cite{Almheiri:2019hni,Penington:2019kki,Balasubramanian:2023xyd,Harlow:2025pvj,Abdalla:2025gzn} {{regarding}} the interpretation of the one-dimensional Hilbert space of closed universes as predicted by Euclidean path integrals (see also \cite{Akers:2025ahe,Chen:2025fwp,Nomura:2025whc,Blommaert:2025rgw,Blommaert:2025bgd}). How do we confront this with the non-trivial evolution of our universe? To recover non-trivial Hilbert spaces, under these conditions, one {{can introduce an observer in}} the closed universe; as for instance, by inserting a clock in de Sitter space \cite{Chandrasekaran:2022cip}. However, {{threating the observer as a fixed worldline trajectory breaks diffeomorphism invariance, and}} it is less clear how to introduce observers in closed universes, which have been prepared through a path integral. In particular, it was realized in \cite{Chen:2025fwp}, that one can construct the state of closed universes with a non-trivial Hilbert space by slicing the path integral of Euclidean wormhole generating baby universes by inserting ``partial sources’’ in the language of \cite{Chen:2025fwp}. In this picture, the correlation functions across the Euclidean wormhole (Fig. \ref{fig:cylindercorrelatorETW}) can be used to reproduce a closed universe with a non-trivial Hilbert space in  \cite{Harlow:2025pvj}, as we illustrate in Fig. \ref{fig:closed_universe}. Here, the bulk is governed by a WDW Hamiltonian constraint \cite{DeWitt:1967yk,Wheeler:1968iap}: $\hH_{\rm WDW}=0$. This can be related with the boundary theory through T$\overline{\text{T}}$ deformations (e.g. \cite{Gross:2019ach,Gross:2019uxi,Hartman:2018tkw}) based on finite cutoff holography \cite{McGough:2016lol,Iliesiu:2020zld}. This has been recently revisited for wormhole topologies in JT gravity in \cite{Bhattacharyya:2025gvd}.

In terms of dilaton-gravity with a matter stress tensor $T_{\mu\nu}$, the WDW Hamiltonian constraint, $\hH_{\rm WDW}=0$, for general dilaton-gravity theories must be put by hand \cite{Gross:2019ach,Iliesiu:2020zld}. It would be very interesting to deduce how the matter contribution appears in the WDW Hamiltonian constraint from first principles in our specific Euclidean wormhole models generalized by the DSSYK with constraints, to model a closed universe with observers as matter fields. This might require some differences compared to finite cutoff holography \cite{McGough:2016lol}.\footnote{We are aware other group is working on it at the moment in the context of sine-dilaton gravity.}

For instance, following the procedure in \cite{Chen:2025fwp,Harlow:2025pvj}, one can generate a closed universe model with an observer by slicing the wormhole shown in Fig. \ref{fig:closed_universe} and analytically continuing to Lorentzian signature the S$^1$ slice obtained by cutting the path integral in Fig. \ref{fig:closed_universe}. The matter chord operators $\hO_{\Delta_i}$, with $i=1,\dots, d$, generating a matter stress tensor $T_{\mu\nu}$ in the bulk takes the role of the observer. 
This corresponds to the canonical variable conjugate to the Hamiltonian of the dilaton-gravity system. 
Based on leading order analysis, the dimension of the Hilbert space corresponding to the configuration in Fig. \ref{fig:closed_universe}, where one sums over all operator contractions $\expval{\hO_{\Delta_i}\hO_{\Delta_i}}$ in the HH state \eqref{eq:HH ETW}, corresponding to a sum over the cylinder two-point correlation functions in \eqref{eq:cylinder amplitude} $\sum_{i}G^{(\Delta_i)}_{\rm cylinder}(\beta)$, which is given by \cite{Harlow:2025pvj,Chen:2025fwp} $\text{min}(d,\rme^{2S_0})$, where ${S_0}$ is a topological term in the dilaton gravity action (corresponding to an overall constant factor in the partition function), meaning that the closed universe, as produced by a HH preparation of state from a slice of the baby universe (Fig. \ref{fig:closed_universe}) with multiple sources can have a non-trivial Hilbert space. 
Alternatively, one might identify the disk edges of the cylinder to generate a periodic universe \cite{Abdalla:2025gzn}. In either case, it would an improvement with respect to the existing literature to develop all the steps above explicitly from the boundary theory. This would allow a first principles derivation of the backreaction caused by the observer in the closed universe through the WDW constraint, and justify under what conditions one can neglect it. An exciting prospect in this direction is to deduce the perspective neutral gauge invariant operator algebra, seen as the bulk matter field excitation in the closed universe. Following on this, it would be interesting to work on conditioned operators from the Page-Wootters reduction \cite{page1983evolution,Chataignier:2024eil,Hoehn:2020epv,Hoehn:2019fsy,DeVuyst:2024uvd}) without neglecting interactions.
This might provide advantages in model building more general cosmological backgrounds in two-dimensional quantum gravity. Based on the definitions above, the kinematical Hilbert space corresponds to each alpha sector in the baby universe Hilbert space (which in this particular model is a single one) with partial sources $\hO_{\Delta_j}$; while the physical Hilbert space is constructed after applying the WDW constraint in the baby universe. It would be worth to develop the details in this construction.

\paragraph{Relational path integral}
The approach we implemented to study the gauge invariant algebra of observables relies heavily on the chord Hilbert space. However, imposing constraints can also be done in a path integral formulation, and indeed this can be advantageous to derive the semiclassical interpretation of observables computed as saddle point solutions. which allowed us to derive the holographic dictionary in Tab. \ref{tab:holographic_dictionary}. Since there are proposals for relational path integrals \cite{Wei:2025guh}, we would like to extend our results in this direction, which might have some technical advantages, such as using the tools in classical dynamical reference frame \cite{Goeller:2022rsx} to derive relational observables in quantum gravity/gauge theories. In particular, we saw that Krylov complexity is part of the algebra of relational observables of the doubled DSSYK. Moreover, it would be interesting to show that quantum complexity is relational, in the sense of being QRF dependent, in a broader variety of systems that the ones in this work.

\paragraph{Subsystem relativity with multiple DSSYK models}It would be worth developing the notion of subsystem relativity in relational quantum dynamics \cite{Hoehn:2023ehz,DeVuyst:2024pop,DeVuyst:2024uvd,AliAhmad:2021adn} in a solvable holographic model, as in Sec. \ref{sec:NV model}. To do this, one needs to consider at least two clock observers and a target system in the configuration. This allows to describe what are the mutually accessible operators according to the different QRFs in the system. For instance, it might be possible to start with a pair of decoupled DSSYKs with one-particle insertion each and then study the limit where the particle becomes infinitely heavy. This could have a holographic interpretation in terms multiple observers in dS space, each detecting a different generalized horizon entropy \cite{DeVuyst:2024pop,DeVuyst:2024uvd,Witten:2023xze}.

\paragraph{Other symmetry sectors/bulk holograms}
Generally, it would be interesting to probe how vast is the landscape of bulk holograms dual to chord symmetry sectors. As a concrete next step, one could derive other types of ETW boundary conditions e.g. in sine dilaton gravity, since we only found ETW brane Hamiltonians for semi open channels in \cite{Blommaert:2025avl}. It might be worth investigating if there are other constraints in the boundary side that reproduce the closed and full open channel in \cite{Blommaert:2025avl}. If this were not possible, it would indicate limitations on the type of bulk theories that can be derived by gauging symmetries in the chord Hilbert space.

We hope to report progress on these exciting future directions.

\section*{Acknowledgments}
I am grateful to Goncalo Araujo, Andreas Blommaert, Xuchen Cao, Hong Zhe (Vincent) Chen, Ping Gao, Philipp Hoehn, Don Marolf, Masamichi Miyaji, Kazumi Okuyama and Tim Schuhmann for insightful discussions; especially Jiuci Xu for comments on an earlier version of the draft. It is a pleasure to thank the High Energy Theory group at UC Santa Barbara for hospitality in during the development of this work, and the QISS consortium for travel support. Earlier results in the manuscript were presented in the ``\textit{Workshop on Low-dimensional Gravity and SYK Model 2025}'' in Shinshu University. The author also acknowledges the Yukawa Institute for Theoretical Physics at Kyoto University, KU Leuven, and Perimeter institute where this work was developed and completed during the YITP-I-25-01 ``Black Hole, Quantum Chaos and Quantum Information'', ``The holographic universe'' and ``QIQG 2025'' workshops respectively. SEAG is supported by the Okinawa Institute of Science and Technology Graduate University. This project/publication was also made possible through the support of the ID\#62312 grant from the John Templeton Foundation, as part of the ‘The Quantum Information Structure of Spacetime’ Project (QISS), as well as Grant ID\# 62423 from the John Templeton Foundation. The opinions expressed in this project/publication are those of the author(s) and do not necessarily reflect the views of the John Templeton Foundation.

\appendix

\section{Notation}\label{app:notation}
\paragraph{Acronyms}
\begin{itemize}[noitemsep]
\item (A)dS: Anti-de Sitter
\item ASC: Al-Salam Chihara
\item CFT: Conformal field theory
\item (DS)SYK: Double-scaled Sachhev-Ye-Kitaev
\item ETW: End-Of-The-World
\item FJ: Faddeev-Jackiw
\item HH: Hartle-Hawking
\item JT: Jackiw-Teitelboim
\item NV: Narovlansky-Verlinde
\item WDW: Wheeler–DeWitt
\end{itemize}

\paragraph{Definitions}

\begin{itemize}
\item $\hvarphi_i$ (\ref{eq:symmetries general2}): General constraints acting on the states in the physical Hilbert space $\ket{\psi}\in{\mathcal{H}}_{\rm phys}$, where the observables obey $\qty[\hat{\varphi}_i,~\hO_\Delta]=0$.

\item $\hvarphi_A$, $\hvarphi_B$: Constraints for chord symmetric ETW brane models, where \eqref{eq:constraints Okuyama} corresponds to one-sided branes, and \eqref {eq:constraints Xu case} for two-sided branes.

\item $\hH_{\rm ASC}$ (\ref{eq:ASC Hamiltonian}): ASC Hamiltonian.

\item $E(\theta)$ (\ref{eq:energy spectrum}): Energy spectrum.

\item $\tilde{\mu}(\theta)$ \eqref{eq:completeness relation}: Integration measure factor in energy basis.

\item $\ket{\Psi(\tau)}$ (\ref{eq:HH ETW}): HH state.

\item $X$, $Y$: Parameters in the ASC Hamiltonian (\ref{eq:ASC Hamiltonian}) that depend on the total conformal dimension of the DSSYK operator $\hO_\Delta$ and the energy parametrization $\theta$ for the one-sided (\ref{eq:XY 1s}), and two-sided (\ref{eq:Z2 sym condition}) ETW brane systems.

\item $Q_l(\cos\theta|X,Y;q)$ (\ref{eq:ASC pol}): ASC polynomials \cite{al1976convolutions}.

\item $L_{\rm ETW}$ (\ref{eq:length ETW JT gravity}): Distance between an ETW brane and an asymptotic AdS$_2$ boundary (for the disk topology).

\item $\ell_*\eqlambda\bra{\Psi(\beta/2)}\hat{n}\ket{\Psi(\beta/2)}$ (\ref{eq:initial length ETW}): Initial rescaled total chord number.

\item $\mathcal{S}_{\rm ETW}$ (\ref{eq:total JT action ETW}): Worldvolume of the ETW brane.

\item $m_{\rm ETW}$, $K_{\rm ETW}$, $\Phi_h$, $L_{\rm reg}$ (\ref{eq:length ETW JT gravity}): ETW brane tension (equal to its ADM mass), its extrinsic curvature, the dilaton value at the AdS$_2$ black horizon (\ref{eq:effective geometry}), and a regularization scale, respectively.

\item $Z_{\rm ETW}(\beta|X,Y)$ (\ref{eq:general ETW partition function}): Disk partition function with ETW brane.

\item $G^{(\Delta_w)}(\tau)$ (\ref{eq:correlator XY}): Disk two-point correlation function with ETW brane.

\item $\ket{K_n}$ (\ref{eq:Krylov basis ETW brane sym}): Krylov basis for the ETW brane system.

\item $\ket{H_n}$ (\ref{eq:number op ASC}): Auxiliary basis for the ETW brane system.

\item $a_n$, $b_n$ (\ref{eq:Lanczos ETW XY form}): Lanczos coefficients.

\item $\Psi_n(\tau)=\bra{K_n}\ket{\Psi(\tau)}$ (\ref{eq:Lanczos ETW sym}): Amplitudes obeying the Lanczos algorithm:
$$-\partial_\tau\Psi_n(\tau)=b_n \Psi_{n-1}(\tau)+b_{n+1}\Psi_{n+1}(\tau)+a_n\Psi_n(\tau)~.$$

\item $\mathcal{C}(t)$ (\ref{eq:womrhole XY}), $\mathcal{C}^{(1s)}$ (\ref{eq:length ETW brane}), $\mathcal{C}^{(2s)}$ (\ref{eq:length Z2 wormhole length}): Spread complexity for general ETW branes; and for one-sided and two-sided sided ETW branes respectively.

\item $Z_{X,Y}(\beta)$ (\ref{eq:half wormhole ETW}), $Z_{b}(\beta)$ (\ref{eq:trumpet amplitude}), $Z(\beta_1,\beta_2)$ (\ref{eq:double trumpet}): Half wormhole, trumpet and double trumpet partition functions respectively.

\item $\hat{Z}_b(\beta_m)$ \eqref{eq:baby universe operator}: Baby universe operator.

\item $Z(\beta_1,\beta_2,\dots,\beta_m)$ \eqref{eq:multiboundary partition function}: Multi-boundary partition function.

\item $G^{(\Delta_w)}_{\rm disk}(\tau)$ (\ref{eq:cylinder amplitude}): Cylinder two-point correlation function.

\item $I_{\rm WH}(n)$ (\ref{eq:final cylinder amplitude}): Effective action for the cylinder correlation function with a matter correlation function \eqref{eq:G cylinder}.

\item $\ket{\alpha}$ (\ref{eq:alpha states}): Alpha states, i.e. a span of the baby universe Hilbert space.

\item $f_a(\qty{\ell_i,P_i})$ (\ref{eq:constraints path integral}): Second order constraints in the path integral.

\item $\omega_{ab}$ (\ref{eq:phase space}): Symplectic two-form.

\item $\hat{\pi}_n$ (\ref{eq:rel chord velocity operator}): Chord velocity operator.

\item $H_n(\cos\theta|q)$ (\ref{eq:H_n def}): q-Hermite polynomials.

\item $\hF_\Delta$ (\ref{eq:isometric factorization}): Isometric factorization linear map.

\item $\mathcal{A}_{\rm inv}$ (\ref{eq:invariant}): Gauge-invariant algebra of observables.

\item $\mathcal{B}(\mH)$: Set of bounded operators acting on the Hilbert space $\mH$.

\item $\qty[\hat{A},~\hH]_n$ (\ref{eq:nested}): Nested commutator.

\item $\ket{\psi(\tau)}$ (\ref{eq:HH clock state}): Clock states with temperature dependence.

\item $\hat{O}^\tau_C(\hat{A})$ (\ref{eq:conditional operator}): Relational operator for a characteristic $\hat{A}\in\mathcal{B}(\mathcal{H}_0^S)$, with a clock state reading $\tau$.

\item $\hat{\rho}_S$ (\ref{eq:reduced DM clock}): Reduced density matrix in chord Hilbert space with respect to the clock basis.
\end{itemize}

\section{ETW brane partition function}\label{app:ETW brane partition function}
We will evaluate the partition function for the ETW brane theories resulting from the constraints in the one-particle chord space (in Sec. \ref{ssec:spread bulk picture from ETW}). In general, the measure of integration in energy basis follows from the orthogonality relation of the ASC polynomials (\ref{eq:ortho relations ASC}), so that the partition function corresponds to\footnote{There is a different way to define the partition function where one instead evaluates a path integral connecting the AdS boundary with the ETW brane as in a trumpet case \cite{Blommaert:2025avl,Okuyama:2023byh} (similar to JT gravity \cite{Gao:2021uro}). We present the cylinder partition function evaluated directly from the (constrained) DSSYK Hamiltonian in (\ref{eq:half wormhole ETW}).}
\begin{equation}\label{eq:general ETW partition function}
    \begin{aligned}
        Z_{\rm ETW}(\beta&|X,Y)=\bra{K_0}\rme^{-\beta\hH_{\rm ASC}}\ket{K_0}=\int_0^\pi\rmd\theta\frac{(XY,\rme^{\pm2\rmi\theta};q)_\infty}{(X\rme^{\pm\rmi\theta},Y\rme^{\pm\rmi\theta};q)_\infty}\rme^{-\beta E(\theta)}
        \\
        &\eqlambda\int_0^\pi\rmd\theta\sqrt{\frac{8\pi(1-XY)\sin^2\theta}{\lambda(1+X^2-2X\cos\theta)(1+Y^2-2Y\cos\theta)}}\rme^{S_\Delta(\theta)-\beta E(\theta)}~,
    \end{aligned}
\end{equation}
where the thermodynamic entropy, and microcanonical temperature for the saddle point in $\theta\in[0,\pi]$ are respectively:
\begin{equation}\label{eq:temperature entropy ETW one sided}
    \begin{aligned}
        S_{\rm ETW}(\theta)&=S(\theta)+\frac{1}{\lambda}\qty({\rm Li}_2(XY)+\frac{\pi^2}{6}+\sum_{\varepsilon=\pm}\qty({\rm Li}_2(X\rme^{\rmi\varepsilon\theta})+{\rm Li}_2(Y\rme^{\rmi\varepsilon\theta})))~,\\
    \beta_{\rm ETW}(\theta)&=\dv{S_\Delta}{E}=\beta(\theta)+\frac{\rmi}{2J\sin\theta}\log\frac{\qty(1-X\rme^{i\theta})\qty(1-Y\rme^{i\theta})}{\qty(1-X\rme^{-\rmi\theta})\qty(1-Y\rme^{-\rmi\theta})}~.
    \end{aligned}
\end{equation}
where $S(\theta)$, $\beta(\theta)$ correspond to the saddle point solutions \cite{Aguilar-Gutierrez:2025pqp},
\begin{align}\label{eq:saddles 1 particle S}
    &S(\theta_{L/R})=-\frac{2}{\lambda}\qty(\theta_{L/R}+n\pi-\frac{\pi}{2})^2~,\quad n\in\mathbb{Z}~,\\\label{eq:saddles 1 particle beta}
    &\beta_{L/R}=\pdv{S(\theta_{L/R})}{E_{L/R}}=\frac{2\theta_{L/R}+(2n-1)\pi}{J\sin\theta_{L/R}}~.
\end{align}

\subsection{Comparison with the one-particle space partition function}\label{sapp:comparison 1p partition function}
We will compare the partition function defined in (\ref{eq:general ETW partition function}) with the DSSYK partition function with a one-particle insertion (i.e. two matter chord operators $\hat{\mathcal{O}}_\Delta$) and inverse temperatures $\beta_{L/R}$ \cite{Aguilar-Gutierrez:2025pqp}
\begin{equation}\label{eq:factorized EPI DSSYK}
\begin{aligned}
    Z_{\Delta}(\beta_L,\beta_R)&=(q^{2\Delta};q)_\infty\int\rmd\mu(\theta_R)\rme^{-\beta_RE_R}\int\rmd\mu(\theta_L)\frac{\rme^{-\beta_LE_L}}{(q^{\Delta}\rme^{\pm\rmi\theta_L\pm\rmi\theta_R};q)_\infty}~.
\end{aligned}
\end{equation}
where $\mu(\theta)\equiv\frac{\rmd\theta}{2\pi}(q,~\rme^{\pm 2 \rmi \theta};q)_\infty$. We first show that the one-sided ETW brane agrees with the interpretation in \cite{Goel:2023svz} (for a different notion of PETS). This follows from the DSSYK partition function (\ref{eq:factorized EPI DSSYK}) when $\beta_{L/R}\rightarrow\infty$ while $\beta_{R/L}$ is kept finite in (\ref{eq:factorized EPI DSSYK}) (as suggested in \cite{Goel:2023svz} to recover a bulk ETW brane from a PETS), which leads to $\theta_{L/R}\rightarrow0$ in the evaluation of the integral, and it means that\footnote{Note that the corresponding density of states, $\rme^{S_\Delta(\theta)}$, is not symmetric with respect to $E=0$. {{This is related to the non-vanishing}} of the Lanczos coefficient $a_n$ in (\ref{eq:DSSYK with ETWB Hamiltonian}), as discussed for the DSSYK model without matter in \cite{Nandy:2024zcd}.}
\begin{equation}\label{eq:part integral ETW}
    Z_{\Delta}(\beta_L,\beta_R)\eqbetaR\mathcal{N}\int\rmd\mu(\theta_L)\frac{\rme^{-\beta_LE_L}}{(q^{\Delta}\rme^{\pm\rmi\theta_L};q)_\infty}~.
\end{equation}
This agrees with (\ref{eq:general ETW partition function}) under the conditions derived in Sec. \ref{ssec:spread bulk picture from ETW}, $X=0$, and $Y=q^\Delta$. This also shows that the coherent state in \eqref{eq:B Delta} (first appearing in \cite{Okuyama:2023byh} (4.17)) comes from the integration measure in the one-particle partition function.

\subsection{Heat capacity}\label{sapp:heat capacity}
\begin{figure}
    \centering
    \subfloat[]{\includegraphics[width=0.48\textwidth]{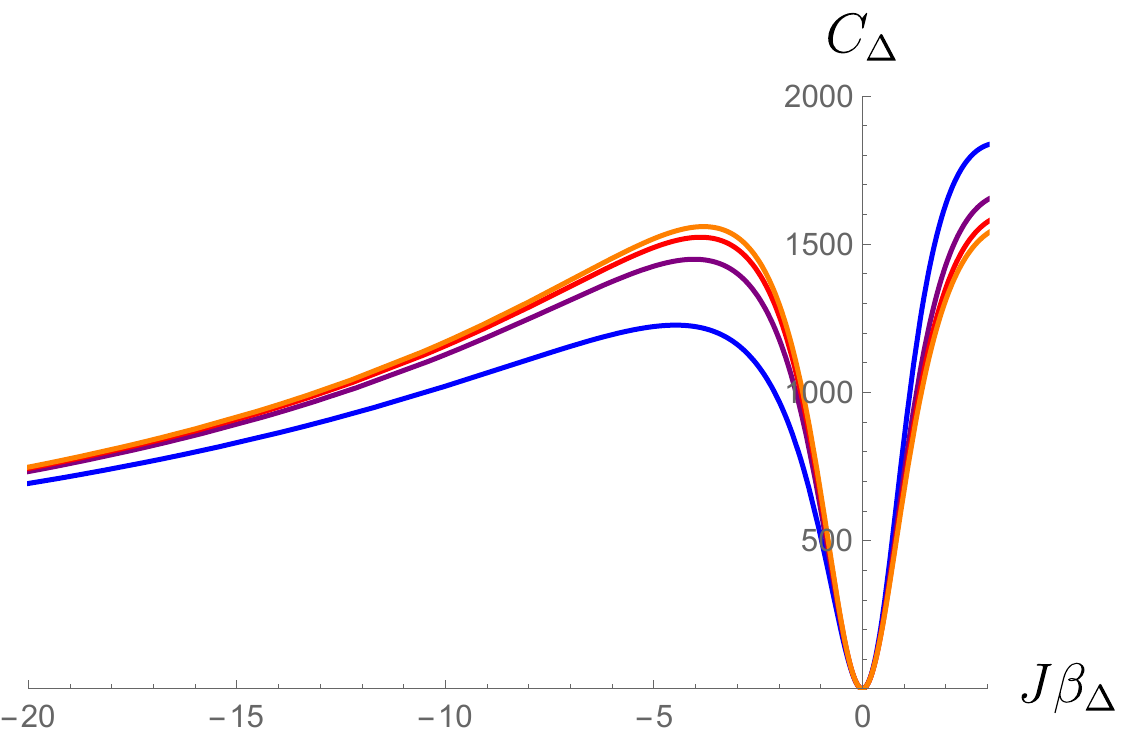}}\subfloat[]{\includegraphics[width=0.48\textwidth]{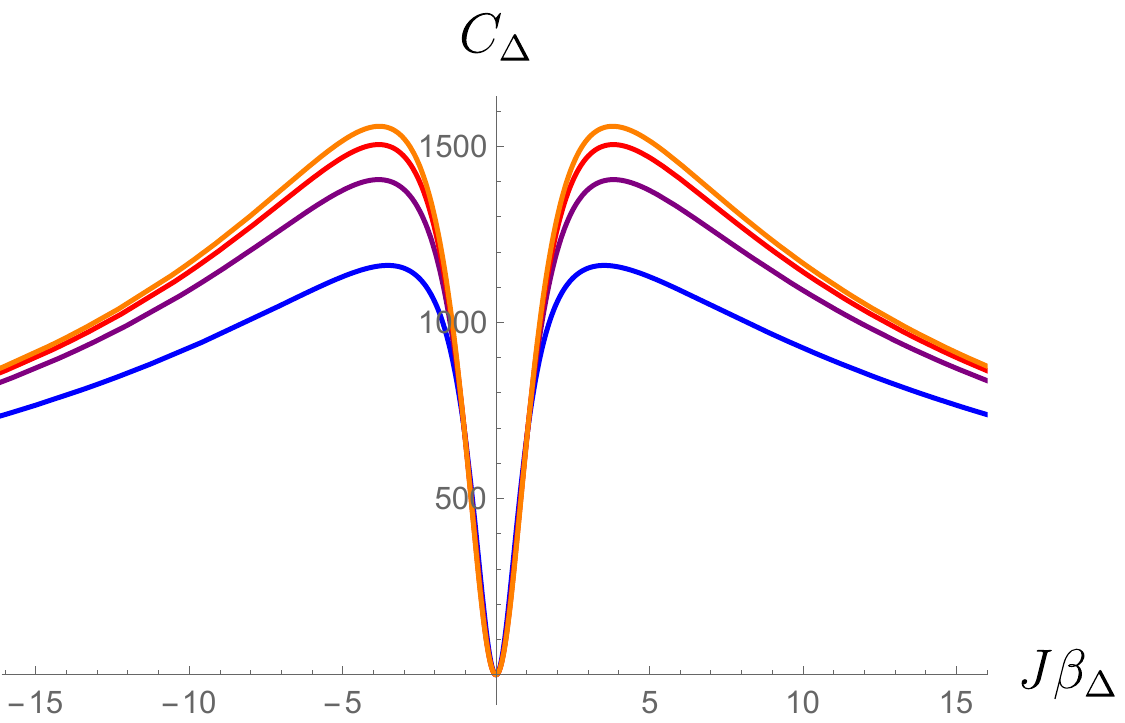}}
    \caption{Heat capacity for (a) the one-sided and (b) two-sided ETW brane described by constraints in the chord space. There is a finite range of real inverse temperatures for case (a), while it extends to all the real line in case (b). In both cases, we notice that the heat capacity is positive definite. The blue, red, purple and orange curves correspond to $\lambda\Delta=\qty{1,2,3,5}$ respecetively, where we take $\lambda=10^{-3}$. Similar results are recovered for other values of $\Delta$.}
    \label{fig:ETW Heat capcity}
\end{figure}

We consider the heat capacity for the saddle points with $n=0$ in (\ref{eq:temperature entropy ETW one sided}) where we label the one and two-sided branes as (1s) and (2s) respectively, i.e.
\begin{align}
        & 
        C^{(\rm 1s)}_\Delta=\frac{\left(4(\theta +n)+\rmi \log \left(\frac{1-e^{\rmi \theta }
   q^{\Delta }}{1-\rme^{-\rmi \theta } q^{\Delta }}\right)-2 \pi
   \right)^2}{\lambda  \left(\cot (\theta ) \left(-4(\theta +n)-\rmi
   \log \left(\frac{1-\rme^{\rmi \theta } q^{\Delta }}{1-e^{-\rmi \theta }
   q^{\Delta }}\right)+2 \pi \right)+\frac{2-2 \cos (\theta )
   q^{\Delta }}{q^{2 \Delta }-2 \cos (\theta ) q^{\Delta
   }+1}+2\right)}~,\label{eq:Heat capacity 1s}\\
       &
        C^{(\rm 2s)}_\Delta=\frac{\left(4(\theta +n)+\rmi \log \left(\frac{e^{2 \rmi \theta }-\rme^{4
   \rmi \theta } q^{\Delta +1}}{-q^{\Delta +1}+\rme^{2 \rmi \theta
   }}\right)-2 \pi \right)^2}{\lambda  \left(\frac{4-4 \cos (2
   \theta ) q^{\Delta +1}}{-2 \cos (2 \theta ) q^{\Delta +1}+q^{2
   \Delta +2}+1}+\cot (\theta ) \left(-4(\theta +n)-\rmi \log
   \left(\frac{\rme^{2 \rmi \theta }-\rme^{4 \rmi \theta } q^{\Delta
   +1}}{-q^{\Delta +1}+\rme^{2 \rmi \theta }}\right)+2 \pi
   \right)\right)}~.\label{eq:Heat capacity 2s}
    \end{align}
We find that in both cases the semiclassical heat capacity is positive (and hence the system is thermodynamically stable) for the saddle points with $n=0$ in (\ref{eq:Heat capacity 1s}, \ref{eq:Heat capacity 2s}), which we illustrate in Fig. \ref{fig:ETW Heat capcity}.

\section{Morse potential from the Al-Salam Chihara (ASC) Hamiltonian}\label{app:Morse from ASC}
As a consistency check, we discuss the triple-scaling limit of the ETW brane Hamiltonian (\ref{eq:ASC Hamiltonian}) to verify that one can recover JT gravity Hamiltonian with a Morse potential derived in \cite{Gao:2021uro}. The result will not be required elsewhere in the text. We first recast the Hamiltonian (\ref{eq:ASC Hamiltonian}) in a different basis, which has recently appeared in \cite{Blommaert:2025avl}, by doing a momentum shift
\begin{equation}
\begin{aligned}
\rme^{\rmi \hat{P}}&\rightarrow \rme^{\rmi \hat{P}}\qty(1+\rmi\qty(\sqrt{XY}+\frac{1}{\sqrt{XY}})q^{\hat{n}}-q^{2\hat{n}})^{-1/2}~,\\
\rme^{-\rmi \hat{P}}&\rightarrow\rme^{-\rmi \hat{P}}\sqrt{1+\rmi\qty(\sqrt{XY}+\frac{1}{\sqrt{XY}})q^{\hat{n}}-q^{2\hat{n}}}~.
\end{aligned}
\end{equation}
This results in a non-Hermitian Hamiltonian 
\begin{equation}
    \begin{aligned}\label{eq:ETW brane H different}
    \hH_{\rm ASC}\rightarrow \frac{J}{\sqrt{\lambda(1-q)}}\biggl(&\cos\hat{P}\sqrt{1+\rmi\qty(\sqrt{XY}+\frac{1}{\sqrt{XY}})q^{\hat{n}}-q^{2\hat{n}}}-\rmi\frac{X+Y}{\sqrt{XY}}q^{\hat{n}}\biggr)~.
\end{aligned}
\end{equation}
When considering the triple-scaling limit, we take
\begin{equation}\label{eq:relation triple scaling}
    q^{\hat{n}}=2\lambda\rme^{-\hat{L}}~,\quad \hat{P}=\lambda \hat{k}~,
\end{equation}
with $\hat{L}$ and $\hat{k}$ held fixed as $\lambda\rightarrow0$. 
The result for (\ref{eq:ETW brane H different}) is then
\begin{equation}\label{eq:morse potential}
    \eval{\hH_{\rm ASC}}_{\rm triple-scaled}-\frac{J}{\sqrt{\lambda(1-q)}}=-2\frac{J}{\sqrt{\lambda(1-q)}}\lambda^2\qty(\frac{\hat{k}^2}{4}+\qty[\rmi\frac{X+Y-XY-1}{2\lambda\sqrt{XY}}]\rme^{-\hat{L}}+\rme^{-2\hat{L}})~.
\end{equation}
Here, the constant in the left-hand side represents a zero-point energy. Also, note that $J<0$ for the Hamiltonian to be bounded from below. (\ref{eq:morse potential}) agrees with the ETW brane Hamiltonian in JT gravity of \cite{Gao:2021uro} where the ETW brane tension corresponds to the term in square brackets. Since we are using a non-Hermitian representation of the Hamiltonian, one should consider $X,~Y\in\mathbb{C}$ to recover a real ETW brane tension in the triple-scaling limit. There are other ways to implement the triple-scaling limit that also recover the Morse potential in \cite{Gao:2021uro}. They have been explicitly studied for the one-sided and two-sided symmetric branes in \cite{Okuyama:2023byh} and \cite{Xu:2024hoc} respectively.

We emphasize that the triple-scaling limit above is \emph{not required} for our study, but it serves as a consistency check that we can recover the expected JT gravity physics from the constrained DSSYK Hamiltonian of the ASC form in (\ref{eq:ASC Hamiltonian}).

\section{Faaddev-Jackiw (FJ) path integral approach}\label{app:FJ approach}
The FJ approach was introduced in \cite{Faddeev:1988qp} to handle canonical quantization in second-class constrained systems; and it has been developed in the path integral formulation in \cite{Huang:2009zzc,PhysRevD.75.025025,Toms:2015lza}.\footnote{However, the previous literature formulates the path integral in a gauge-dependent matter, which is not compatible with the perspective neutral approach to relational observables. A gauge invariant formulation of the path integral in other systems has been treated in e.g. \cite{Parker:2009uva,DeWitt:1964mxt,DeWitt:1967yk,DeWitt:1967ub,DeWitt:1975ys,DeWitt:2003pm,Kunstatter:1991ds}. It would be worth developing the previous formulations with a relational path integral, as we mention in Sec. \ref{ssec:outlook}.} As an advantage over other methods, there is no classification of primary and secondary constraints as in Dirac's approach, so they can be treated the same way. This method agrees with Dirac quantization with appropriate ordering (similar to reduced phase space quantization, e.g. \cite{Kunstatter:1991ds,faddeev1969feynman}). Instead of performing quantization in the boundary theory, we will use it to implement constraints which lead to gauging the chord symmetry (\ref{eq:constraints Okuyama}, \ref{eq:constraints Xu case}) there is an intuitive understanding about the canonical phase space constraints in terms of a particle moving in a constrained surface in phase space, found in more general theories by \cite{Toms:2015lza}.

To be explicit, we impose second-order constraints in the canonical coordinates $\qty{\ell_i,~P_i}$ of the form
\begin{equation}\label{eq:constraints path integral}
    f_a(\qty{\ell_i,P_i})=0~.
\end{equation}
After imposing (\ref{eq:constraints path integral}) through Lagrange multipliers and integrating them out, the path integral, similar to (\ref{eq:main path integral}), can be written \cite{Toms:2015lza,Jackiw:1993in}
\begin{equation}\label{eq:path E Z ETW}
    Z=\int\prod_{i=L,R}[\rmd \ell_i][\rmd P_i]{\rm Pff}[\omega_{ab}]{\rm Pff}[\qty{f_a,~f_b}]\delta(f_a)\rme^{I_E[\qty{\ell_i,~P_i}]}~,
\end{equation}
where ${\rm Pff}[\cdot]$ is a Pfaffian functional, $\qty{\cdot,~\cdot}$ denotes the Poisson brackets, and $\omega_{ab}$ is the symplectic two-form, which is defined by,
\begin{equation}\label{eq:phase space}
    \dv{x^a}{t_{L/R}}=(\omega^{-1})^{ba}\partial_b H_{L/R}~,\quad\text{where}\quad x^a=\qty{\ell_i,~P_i}~.
\end{equation}
We can find conditions on the canonical variables such that the two-sided one-particle DSSYK Hamiltonian matches the JT and sine dilaton gravity Hamiltonians with ETW branes (\ref{eq:ASC Hamiltonian}) by implementing
\begin{equation}\label{eq:FJ constraints}
   f=f(\ell_{L/R})~,\quad \tilde{f}(\ell_{L/R},P_{L/R})=P_L\partial_{\ell_L}f+P_R\partial_{\ell_R}f~,
\end{equation}
where $\tilde{f}(\ell_{L/R},P_{L/R})$ is a required to regularize the path integral (to eliminate zero modes) due to $f(\ell_{L/R})$. Physically, the second constraint imposes that the canonical momenta is tangential to the constraint surface, $f(\ell_{L/R})=0$. This describes a particle moving a surface in phase space \cite{Toms:2015lza}. The path integral then simplifies to \eqref{eq:path E Z ETW}
\begin{equation}
    Z=\int\prod_{i=a,b}[\rmd \ell_i][\rmd P_i]\abs{\partial_{\ell_i}f}^2\delta(f(\ell_{L/R}))\delta(P_i\partial_{\ell_i}f(\ell_{L/R}))\rme^{I_E[\qty{\ell_i,~P_i}]}~,
\end{equation}
Then, the one-sided and two-sided ETW brane systems in Sec. \ref{ssec:constraints ETW} arise naturally from (\ref{eq:FJ constraints}) by requiring
\begin{equation}
    \begin{aligned}
       &{\rm One-sided}:&&f=\ell_L-\tilde{N}~, \quad \tilde{f}=P_L~,\\
&{\rm Two-sided}:&&f=\ell_L-\ell_R~,\quad \tilde{f}=P_L-P_R~,
    \end{aligned}
\end{equation}
which leads to the constraints (\ref{eq:constraints Okuyama}, \ref{eq:XY 1s}). Thus, this form of implementing the constraints agrees {{with our results}} in Sec. \ref{sec:ASC from DSSYK}.

\section{Alternative derivation of the saddle point energy (\ref{eq:sp theta})}\label{app:alternative derivation}
An alternative but equivalent way to derive the saddle point energy parametrization (\ref{eq:sp theta}), is to directly evaluate (\ref{eq:correlator cylinder 2}) {{using a}} saddle point {{approximation for the corresponding}} integral, {{ in a similar way as \cite{Okuyama:2023yat}}}. 

Let us use some results in the math literature. It is known that the ASC polynomials satisfy a the Fourier kernel \cite{askey1996general} (14.8):
\begin{equation}\label{eq:explicit kernel}
    \begin{aligned}
        &\sum_{l=0}^\infty q^{{\Delta}l}\frac{Q_l(\cos\theta_1|q^{\Delta_w}\rme^{\pm\rmi\theta_3};q)Q_l(\cos\theta_2|q^{\Delta_w}\rme^{\pm\rmi\theta_4};q)}{(q,q^{2\Delta_w};q)_l}\\
    &=\frac{(q^{\Delta}\rme^{-\rmi(\theta_3+\theta_4)},q^{{\Delta}+\Delta_w}\rme^{\rmi(\theta_4\pm\theta_1)},q^{{\Delta}+\Delta_w}\rme^{\rmi(\theta_3\pm\theta_2)};q)_\infty}{(q^{{\Delta}+\Delta_w}\rme^{\rmi(\theta_3+\theta_4)},q^{\Delta}\rme^{\rmi\qty(\pm\theta_1\pm\theta_2)};q)_\infty} \cdot\\
    &\quad \cdot{}_8W_7\qty(q^{{\Delta}+2\Delta_w-1}\rme^{\rmi\qty(\theta_3+\theta_4)},q^{\Delta}\rme^{\rmi\qty(\theta_3+\theta_4)},q^{\Delta_w}\rme^{\rmi\qty(\theta_1\pm\theta_3)},q\rme^{\rmi\qty(\theta_4\pm\theta_2)};q,q^{\Delta}\rme^{-\rmi(\theta_3+\theta_4)})~,
    \end{aligned}
\end{equation}
where $_8W_7$ is a very-well-poised basic hypergeometric series
\begin{equation}\label{eq:very-well-poised basic hypergeometric series}
    {}_{r+1}W_r(a_1;a_4,\ldots,a_{r+1};q,z) =
\sum_{j=0}^\infty  \frac{1-a_1q^{2j}}{1-a_1}
\frac{(a_1,a_4,\ldots,a_{r+1};q)_jz^j }{(q,qa_1/a_4,\ldots,
qa_1/a_{r+1};q)_j}~.
\end{equation}
Using the explicit Krylov basis (\ref{eq:Krylov basis ETW brane sym}), we recognize one of the terms is the Fourier kernel of the ASC polynomials in (\ref{eq:explicit kernel})
\begin{equation}\label{eq:Very well posed kernel}
    \begin{aligned}
        \bra{\theta_1}&q^{\Delta_w\hat{n}}\ket{\theta_2}=\sum_{n=0}^\infty q^{\Delta_wn}\bra{\theta_1}\ket{K_n}\bra{K_n}\ket{\theta_2}\\
        &=\sum_nq^{\Delta_wn}\frac{Q_n(\cos\theta_1|X,Y;q)Q_n(\cos\theta_2|X,Y;q)}{(q,XY;q)_\infty}\\
        &=\frac{\qty(\tfrac{Y}{X}q^{\Delta_w},Xq^{\Delta_w}\rme^{\pm\rmi\theta_1},Xq^{\Delta_w}\rme^{\pm\rmi\theta_2};q)_\infty}{(X^2q^{\Delta_w},q^{\Delta_w}\rme^{\rmi(\pm\theta_1\pm\theta_2)};q)_\infty}{}_8W_7\qty(X^2q^{\Delta_w},\tfrac{X}{Y}q^{\Delta_w},X\rme^{\pm\rmi\theta_1},X\rme^{\pm\rmi\theta_2};q,\tfrac{Y}{X}q^{\Delta_w})~,
    \end{aligned}
\end{equation}
where ${}_8W_7$ in (\ref{eq:very-well-poised basic hypergeometric series}) is a very-well-poised basic hypergeometric series. 

We can therefore express
\begin{equation}
\begin{aligned}\label{eq:substitution qn}
    &\abs{\bra{K_{n}}\ket{\theta}}^2=\int_0^{2\pi}\frac{\rmd s}{2\pi}\rme^{-\rmi ns}\sum_{m=0}^\infty\rme^{\rmi ms}\abs{\bra{K_{m}}\ket{\theta}}^2\\
    &=\int_0^{2\pi}\frac{\rmd s}{2\pi}\rme^{-\rmi ns}\frac{\qty(\tfrac{Y}{X}\rme^{\rmi s},X\rme^{\rmi s\pm\rmi\theta},X\rme^{\rmi s\pm\rmi\theta};q)_\infty}{(X^2\rme^{\rmi s},\rme^{\rmi s},\rme^{\rmi s},\rme^{\rmi(s\pm2\theta)};q)_\infty}{}_8W_7\qty(X^2\rme^{\rmi s},\tfrac{X}{Y}\rme^{\rmi s},X\rme^{\pm\rmi\theta},X\rme^{\pm\rmi\theta};q,\tfrac{Y}{X}\rme^{\rmi s})~,
\end{aligned}
\end{equation}
where in the second line we applied (\ref{eq:Very well posed kernel}) for $\theta_1=\theta_2$ and $q^{\Delta_w}\rightarrow\rme^{-\rmi\phi}$.

We can therefore express the cylinder two-point correlation function (\ref{eq:correlator cylinder 2}) as
\begin{equation}\label{eq:double int Gtau}
\begin{aligned}
    &G^{(\Delta_w)}_{\rm cylinder}(\beta)=\\
    &\int_0^{2\pi}\frac{\rmd s}{2\pi}\rme^{-\rmi ns}\frac{\qty(\tfrac{Y}{X}\rme^{\rmi s},X\rme^{\rmi s\pm\rmi\theta},X\rme^{\rmi s\pm\rmi\theta};q)_\infty}{(X^2\rme^{\rmi s},\rme^{\rmi s},\rme^{\rmi s},\rme^{\rmi(s\pm2\theta)};q)_\infty}{}_8W_7\qty(X^2\rme^{\rmi s},\tfrac{X}{Y}\rme^{\rmi s},X\rme^{\pm\rmi\theta},X\rme^{\pm\rmi\theta};q,\tfrac{Y}{X}\rme^{\rmi s})~.
\end{aligned}
\end{equation}
To make further simplifications, we one may use the asymptotic expansion for $0<\lambda\ll1$ \cite{Goel:2023svz}
\begin{equation}
\begin{aligned}\label{eq:approx (x;q)infty}
    (x;q)_\infty&=\exp\qty(-\frac{{\rm Li}_2(x)}{\lambda}+\frac{1}{2}\log(1-x)+\mathcal{O}(\lambda))~,\\
    (q;q)_\infty&=\sqrt{\frac{2\pi}{\lambda}}\exp\qty(-\frac{\pi^2}{6\lambda}+\mathcal{O}(\lambda))~,\\
    &{\rm Li}_2(z)+{\rm Li}_2(z^{-1})=-\frac{\pi^2}{6}-\frac{1}{2}\qty(\log(-z))^2~,
\end{aligned}
\end{equation}
In order to carry out the evaluation of (\ref{eq:double int Gtau}) with the leading order terms above, one may also use an integral representation of ${}_8W_7$ (as (B.23) in \cite{Berkooz:2018jqr}){{ for}} the asymptotic analysis. However, due to the technical complications, we have not verified if this alternative procedure leads to the same saddle point solution as (\ref{eq:sp theta}).

\bibliographystyle{JHEP}
\bibliography{references.bib}
\end{document}